% File jytex.tex, for jyTeX version 2.6M (June 1992)
% Copyright (c) 1991, 1992 by Jonathan P. Yamron
% For full documentation, "get jydoc" from hep-ph@xxx.lanl.gov
%   Problems?  Contact brahm@theory3.caltech.edu.

\catcode`\@=11

%*****************************************************************************

\message{Loading jyTeX fonts...}

%************************************************************
%*
%*             Available fonts
%*
%************************************************************

%************** 5-point fonts *******************************

\font\vptrm=cmr5 \font\vptmit=cmmi5 \font\vptsy=cmsy5 \font\vptbf=cmbx5

\skewchar\vptmit='177 \skewchar\vptsy='60 \fontdimen16
\vptsy=\the\fontdimen17 \vptsy

\def\vpt{\ifmmode\err@badsizechange\else
     \@mathfontinit
     \textfont0=\vptrm  \scriptfont0=\vptrm  \scriptscriptfont0=\vptrm
     \textfont1=\vptmit \scriptfont1=\vptmit \scriptscriptfont1=\vptmit
     \textfont2=\vptsy  \scriptfont2=\vptsy  \scriptscriptfont2=\vptsy
     \textfont3=\xptex  \scriptfont3=\xptex  \scriptscriptfont3=\xptex
     \textfont\bffam=\vptbf
     \scriptfont\bffam=\vptbf
     \scriptscriptfont\bffam=\vptbf
     \@fontstyleinit
     \def\rm{\vptrm\fam=\z@}%
     \def\bf{\vptbf\fam=\bffam}%
     \def\oldstyle{\vptmit\fam=\@ne}%
     \rm\fi}

%************** 6-point fonts *******************************

\font\viptrm=cmr6 \font\viptmit=cmmi6 \font\viptsy=cmsy6 \font\viptbf=cmbx6

\skewchar\viptmit='177 \skewchar\viptsy='60 \fontdimen16
\viptsy=\the\fontdimen17 \viptsy

\def\vipt{\ifmmode\err@badsizechange\else
     \@mathfontinit
     \textfont0=\viptrm  \scriptfont0=\vptrm  \scriptscriptfont0=\vptrm
     \textfont1=\viptmit \scriptfont1=\vptmit \scriptscriptfont1=\vptmit
     \textfont2=\viptsy  \scriptfont2=\vptsy  \scriptscriptfont2=\vptsy
     \textfont3=\xptex   \scriptfont3=\xptex  \scriptscriptfont3=\xptex
     \textfont\bffam=\viptbf
     \scriptfont\bffam=\vptbf
     \scriptscriptfont\bffam=\vptbf
     \@fontstyleinit
     \def\rm{\viptrm\fam=\z@}%
     \def\bf{\viptbf\fam=\bffam}%
     \def\oldstyle{\viptmit\fam=\@ne}%
     \rm\fi}

%************** 7-point fonts *******************************

\font\viiptrm=cmr7 \font\viiptmit=cmmi7 \font\viiptsy=cmsy7
\font\viiptit=cmti7 \font\viiptbf=cmbx7

\skewchar\viiptmit='177 \skewchar\viiptsy='60 \fontdimen16
\viiptsy=\the\fontdimen17 \viiptsy

\def\viipt{\ifmmode\err@badsizechange\else
     \@mathfontinit
     \textfont0=\viiptrm  \scriptfont0=\vptrm  \scriptscriptfont0=\vptrm
     \textfont1=\viiptmit \scriptfont1=\vptmit \scriptscriptfont1=\vptmit
     \textfont2=\viiptsy  \scriptfont2=\vptsy  \scriptscriptfont2=\vptsy
     \textfont3=\xptex    \scriptfont3=\xptex  \scriptscriptfont3=\xptex
     \textfont\itfam=\viiptit
     \scriptfont\itfam=\viiptit
     \scriptscriptfont\itfam=\viiptit
     \textfont\bffam=\viiptbf
     \scriptfont\bffam=\vptbf
     \scriptscriptfont\bffam=\vptbf
     \@fontstyleinit
     \def\rm{\viiptrm\fam=\z@}%
     \def\it{\viiptit\fam=\itfam}%
     \def\bf{\viiptbf\fam=\bffam}%
     \def\oldstyle{\viiptmit\fam=\@ne}%
     \rm\fi}

%************** 8-point fonts *******************************

\font\viiiptrm=cmr8 \font\viiiptmit=cmmi8 \font\viiiptsy=cmsy8
\font\viiiptit=cmti8
%\font\viiiptsl=cmsl8
\font\viiiptbf=cmbx8
%\font\viiipttt=cmtt8
%\font\viiiptss=cmss8

\skewchar\viiiptmit='177 \skewchar\viiiptsy='60 \fontdimen16
\viiiptsy=\the\fontdimen17 \viiiptsy

\def\viiipt{\ifmmode\err@badsizechange\else
     \@mathfontinit
     \textfont0=\viiiptrm  \scriptfont0=\viptrm  \scriptscriptfont0=\vptrm
     \textfont1=\viiiptmit \scriptfont1=\viptmit \scriptscriptfont1=\vptmit
     \textfont2=\viiiptsy  \scriptfont2=\viptsy  \scriptscriptfont2=\vptsy
     \textfont3=\xptex     \scriptfont3=\xptex   \scriptscriptfont3=\xptex
     \textfont\itfam=\viiiptit
     \scriptfont\itfam=\viiptit
     \scriptscriptfont\itfam=\viiptit
     \textfont\bffam=\viiiptbf
     \scriptfont\bffam=\viptbf
     \scriptscriptfont\bffam=\vptbf
     \@fontstyleinit
     \def\rm{\viiiptrm\fam=\z@}%
     \def\it{\viiiptit\fam=\itfam}%
     \def\bf{\viiiptbf\fam=\bffam}%
     \def\oldstyle{\viiiptmit\fam=\@ne}%
     \rm\fi}

%************** Optional 9-point fonts **********************

\def\getixpt{%
     \font\ixptrm=cmr9
     \font\ixptmit=cmmi9
     \font\ixptsy=cmsy9
     \font\ixptit=cmti9
%     \font\ixptsl=cmsl9
     \font\ixptbf=cmbx9
%     \font\ixpttt=cmtt9
%     \font\ixptss=cmss9
     \skewchar\ixptmit='177 \skewchar\ixptsy='60
     \fontdimen16 \ixptsy=\the\fontdimen17 \ixptsy}

\def\ixpt{\ifmmode\err@badsizechange\else
     \@mathfontinit
     \textfont0=\ixptrm  \scriptfont0=\viiptrm  \scriptscriptfont0=\vptrm
     \textfont1=\ixptmit \scriptfont1=\viiptmit \scriptscriptfont1=\vptmit
     \textfont2=\ixptsy  \scriptfont2=\viiptsy  \scriptscriptfont2=\vptsy
     \textfont3=\xptex   \scriptfont3=\xptex    \scriptscriptfont3=\xptex
     \textfont\itfam=\ixptit
     \scriptfont\itfam=\viiptit
     \scriptscriptfont\itfam=\viiptit
     \textfont\bffam=\ixptbf
     \scriptfont\bffam=\viiptbf
     \scriptscriptfont\bffam=\vptbf
     \@fontstyleinit
     \def\rm{\ixptrm\fam=\z@}%
     \def\it{\ixptit\fam=\itfam}%
     \def\bf{\ixptbf\fam=\bffam}%
     \def\oldstyle{\ixptmit\fam=\@ne}%
     \rm\fi}

%************** 10-point fonts ******************************

\font\xptrm=cmr10 \font\xptmit=cmmi10 \font\xptsy=cmsy10 \font\xptex=cmex10
\font\xptit=cmti10 \font\xptsl=cmsl10 \font\xptbf=cmbx10 \font\xpttt=cmtt10
\font\xptss=cmss10 \font\xptsc=cmcsc10 \font\xptbfs=cmb10
\font\xptbmit=cmmib10

\skewchar\xptmit='177 \skewchar\xptbmit='177 \skewchar\xptsy='60 \fontdimen16
\xptsy=\the\fontdimen17 \xptsy

\def\xpt{\ifmmode\err@badsizechange\else
     \@mathfontinit
     \textfont0=\xptrm  \scriptfont0=\viiptrm  \scriptscriptfont0=\vptrm
     \textfont1=\xptmit \scriptfont1=\viiptmit \scriptscriptfont1=\vptmit
     \textfont2=\xptsy  \scriptfont2=\viiptsy  \scriptscriptfont2=\vptsy
     \textfont3=\xptex  \scriptfont3=\xptex    \scriptscriptfont3=\xptex
     \textfont\itfam=\xptit
     \scriptfont\itfam=\viiptit
     \scriptscriptfont\itfam=\viiptit
     \textfont\bffam=\xptbf
     \scriptfont\bffam=\viiptbf
     \scriptscriptfont\bffam=\vptbf
     \textfont\bfsfam=\xptbfs
     \scriptfont\bfsfam=\viiptbf
     \scriptscriptfont\bfsfam=\vptbf
     \textfont\bmitfam=\xptbmit
     \scriptfont\bmitfam=\viiptmit
     \scriptscriptfont\bmitfam=\vptmit
     \@fontstyleinit
     \def\rm{\xptrm\fam=\z@}%
     \def\it{\xptit\fam=\itfam}%
     \def\sl{\xptsl}%
     \def\bf{\xptbf\fam=\bffam}%
     \def\tt{\xpttt}%
     \def\ss{\xptss}%
     \def\sc{\xptsc}%
     \def\bfs{\xptbfs\fam=\bfsfam}%
     \def\bmit{\fam=\bmitfam}%
     \def\oldstyle{\xptmit\fam=\@ne}%
     \rm\fi}

%************** Optional 11-point fonts *********************

\def\getxipt{%
     \font\xiptrm=cmr10  scaled\magstephalf
     \font\xiptmit=cmmi10 scaled\magstephalf
     \font\xiptsy=cmsy10 scaled\magstephalf
     \font\xiptex=cmex10 scaled\magstephalf
     \font\xiptit=cmti10 scaled\magstephalf
     \font\xiptsl=cmsl10 scaled\magstephalf
     \font\xiptbf=cmbx10 scaled\magstephalf
     \font\xipttt=cmtt10 scaled\magstephalf
     \font\xiptss=cmss10 scaled\magstephalf
     \skewchar\xiptmit='177 \skewchar\xiptsy='60
     \fontdimen16 \xiptsy=\the\fontdimen17 \xiptsy}

\def\xipt{\ifmmode\err@badsizechange\else
     \@mathfontinit
     \textfont0=\xiptrm  \scriptfont0=\viiiptrm  \scriptscriptfont0=\viptrm
     \textfont1=\xiptmit \scriptfont1=\viiiptmit \scriptscriptfont1=\viptmit
     \textfont2=\xiptsy  \scriptfont2=\viiiptsy  \scriptscriptfont2=\viptsy
     \textfont3=\xiptex  \scriptfont3=\xptex     \scriptscriptfont3=\xptex
     \textfont\itfam=\xiptit
     \scriptfont\itfam=\viiiptit
     \scriptscriptfont\itfam=\viiptit
     \textfont\bffam=\xiptbf
     \scriptfont\bffam=\viiiptbf
     \scriptscriptfont\bffam=\viptbf
     \@fontstyleinit
     \def\rm{\xiptrm\fam=\z@}%
     \def\it{\xiptit\fam=\itfam}%
     \def\sl{\xiptsl}%
     \def\bf{\xiptbf\fam=\bffam}%
     \def\tt{\xipttt}%
     \def\ss{\xiptss}%
     \def\oldstyle{\xiptmit\fam=\@ne}%
     \rm\fi}

%************** 12-point fonts ******************************

\font\xiiptrm=cmr12 \font\xiiptmit=cmmi12 \font\xiiptsy=cmsy10
scaled\magstep1 \font\xiiptex=cmex10  scaled\magstep1 \font\xiiptit=cmti12
\font\xiiptsl=cmsl12 \font\xiiptbf=cmbx12
%\font\xiipttt=cmtt12
\font\xiiptss=cmss12 \font\xiiptsc=cmcsc10 scaled\magstep1
\font\xiiptbfs=cmb10  scaled\magstep1 \font\xiiptbmit=cmmib10 scaled\magstep1

\skewchar\xiiptmit='177 \skewchar\xiiptbmit='177 \skewchar\xiiptsy='60
\fontdimen16 \xiiptsy=\the\fontdimen17 \xiiptsy

\def\xiipt{\ifmmode\err@badsizechange\else
     \@mathfontinit
     \textfont0=\xiiptrm  \scriptfont0=\viiiptrm  \scriptscriptfont0=\viptrm
     \textfont1=\xiiptmit \scriptfont1=\viiiptmit \scriptscriptfont1=\viptmit
     \textfont2=\xiiptsy  \scriptfont2=\viiiptsy  \scriptscriptfont2=\viptsy
     \textfont3=\xiiptex  \scriptfont3=\xptex     \scriptscriptfont3=\xptex
     \textfont\itfam=\xiiptit
     \scriptfont\itfam=\viiiptit
     \scriptscriptfont\itfam=\viiptit
     \textfont\bffam=\xiiptbf
     \scriptfont\bffam=\viiiptbf
     \scriptscriptfont\bffam=\viptbf
     \textfont\bfsfam=\xiiptbfs
     \scriptfont\bfsfam=\viiiptbf
     \scriptscriptfont\bfsfam=\viptbf
     \textfont\bmitfam=\xiiptbmit
     \scriptfont\bmitfam=\viiiptmit
     \scriptscriptfont\bmitfam=\viptmit
     \@fontstyleinit
     \def\rm{\xiiptrm\fam=\z@}%
     \def\it{\xiiptit\fam=\itfam}%
     \def\sl{\xiiptsl}%
     \def\bf{\xiiptbf\fam=\bffam}%
     \def\tt{\xiipttt}%
     \def\ss{\xiiptss}%
     \def\sc{\xiiptsc}%
     \def\bfs{\xiiptbfs\fam=\bfsfam}%
     \def\bmit{\fam=\bmitfam}%
     \def\oldstyle{\xiiptmit\fam=\@ne}%
     \rm\fi}

%************** Optional 13-point fonts *********************

\def\getxiiipt{%
     \font\xiiiptrm=cmr12  scaled\magstephalf
     \font\xiiiptmit=cmmi12 scaled\magstephalf
     \font\xiiiptsy=cmsy9  scaled\magstep2
     \font\xiiiptit=cmti12 scaled\magstephalf
     \font\xiiiptsl=cmsl12 scaled\magstephalf
     \font\xiiiptbf=cmbx12 scaled\magstephalf
     \font\xiiipttt=cmtt12 scaled\magstephalf
     \font\xiiiptss=cmss12 scaled\magstephalf
     \skewchar\xiiiptmit='177 \skewchar\xiiiptsy='60
     \fontdimen16 \xiiiptsy=\the\fontdimen17 \xiiiptsy}

\def\xiiipt{\ifmmode\err@badsizechange\else
     \@mathfontinit
     \textfont0=\xiiiptrm  \scriptfont0=\xptrm  \scriptscriptfont0=\viiptrm
     \textfont1=\xiiiptmit \scriptfont1=\xptmit \scriptscriptfont1=\viiptmit
     \textfont2=\xiiiptsy  \scriptfont2=\xptsy  \scriptscriptfont2=\viiptsy
     \textfont3=\xivptex   \scriptfont3=\xptex  \scriptscriptfont3=\xptex
     \textfont\itfam=\xiiiptit
     \scriptfont\itfam=\xptit
     \scriptscriptfont\itfam=\viiptit
     \textfont\bffam=\xiiiptbf
     \scriptfont\bffam=\xptbf
     \scriptscriptfont\bffam=\viiptbf
     \@fontstyleinit
     \def\rm{\xiiiptrm\fam=\z@}%
     \def\it{\xiiiptit\fam=\itfam}%
     \def\sl{\xiiiptsl}%
     \def\bf{\xiiiptbf\fam=\bffam}%
     \def\tt{\xiiipttt}%
     \def\ss{\xiiiptss}%
     \def\oldstyle{\xiiiptmit\fam=\@ne}%
     \rm\fi}

%************** 14-point fonts ******************************

\font\xivptrm=cmr12   scaled\magstep1 \font\xivptmit=cmmi12  scaled\magstep1
\font\xivptsy=cmsy10  scaled\magstep2 \font\xivptex=cmex10  scaled\magstep2
\font\xivptit=cmti12  scaled\magstep1 \font\xivptsl=cmsl12  scaled\magstep1
\font\xivptbf=cmbx12  scaled\magstep1
%\font\xivpttt=cmtt12  scaled\magstep1
\font\xivptss=cmss12  scaled\magstep1 \font\xivptsc=cmcsc10 scaled\magstep2
\font\xivptbfs=cmb10  scaled\magstep2 \font\xivptbmit=cmmib10 scaled\magstep2

\skewchar\xivptmit='177 \skewchar\xivptbmit='177 \skewchar\xivptsy='60
\fontdimen16 \xivptsy=\the\fontdimen17 \xivptsy

\def\xivpt{\ifmmode\err@badsizechange\else
     \@mathfontinit
     \textfont0=\xivptrm  \scriptfont0=\xptrm  \scriptscriptfont0=\viiptrm
     \textfont1=\xivptmit \scriptfont1=\xptmit \scriptscriptfont1=\viiptmit
     \textfont2=\xivptsy  \scriptfont2=\xptsy  \scriptscriptfont2=\viiptsy
     \textfont3=\xivptex  \scriptfont3=\xptex  \scriptscriptfont3=\xptex
     \textfont\itfam=\xivptit
     \scriptfont\itfam=\xptit
     \scriptscriptfont\itfam=\viiptit
     \textfont\bffam=\xivptbf
     \scriptfont\bffam=\xptbf
     \scriptscriptfont\bffam=\viiptbf
     \textfont\bfsfam=\xivptbfs
     \scriptfont\bfsfam=\xptbfs
     \scriptscriptfont\bfsfam=\viiptbf
     \textfont\bmitfam=\xivptbmit
     \scriptfont\bmitfam=\xptbmit
     \scriptscriptfont\bmitfam=\viiptmit
     \@fontstyleinit
     \def\rm{\xivptrm\fam=\z@}%
     \def\it{\xivptit\fam=\itfam}%
     \def\sl{\xivptsl}%
     \def\bf{\xivptbf\fam=\bffam}%
     \def\tt{\xivpttt}%
     \def\ss{\xivptss}%
     \def\sc{\xivptsc}%
     \def\bfs{\xivptbfs\fam=\bfsfam}%
     \def\bmit{\fam=\bmitfam}%
     \def\oldstyle{\xivptmit\fam=\@ne}%
     \rm\fi}

%************** 17-point fonts ******************************

\font\xviiptrm=cmr17 \font\xviiptmit=cmmi12 scaled\magstep2
\font\xviiptsy=cmsy10 scaled\magstep3 \font\xviiptex=cmex10 scaled\magstep3
\font\xviiptit=cmti12 scaled\magstep2 \font\xviiptbf=cmbx12 scaled\magstep2
\font\xviiptbfs=cmb10 scaled\magstep3

\skewchar\xviiptmit='177 \skewchar\xviiptsy='60 \fontdimen16
\xviiptsy=\the\fontdimen17 \xviiptsy

\def\xviipt{\ifmmode\err@badsizechange\else
     \@mathfontinit
     \textfont0=\xviiptrm  \scriptfont0=\xiiptrm  \scriptscriptfont0=\viiiptrm
     \textfont1=\xviiptmit \scriptfont1=\xiiptmit \scriptscriptfont1=\viiiptmit
     \textfont2=\xviiptsy  \scriptfont2=\xiiptsy  \scriptscriptfont2=\viiiptsy
     \textfont3=\xviiptex  \scriptfont3=\xiiptex  \scriptscriptfont3=\xptex
     \textfont\itfam=\xviiptit
     \scriptfont\itfam=\xiiptit
     \scriptscriptfont\itfam=\viiiptit
     \textfont\bffam=\xviiptbf
     \scriptfont\bffam=\xiiptbf
     \scriptscriptfont\bffam=\viiiptbf
     \textfont\bfsfam=\xviiptbfs
     \scriptfont\bfsfam=\xiiptbfs
     \scriptscriptfont\bfsfam=\viiiptbf
     \@fontstyleinit
     \def\rm{\xviiptrm\fam=\z@}%
     \def\it{\xviiptit\fam=\itfam}%
     \def\bf{\xviiptbf\fam=\bffam}%
     \def\bfs{\xviiptbfs\fam=\bfsfam}%
     \def\oldstyle{\xviiptmit\fam=\@ne}%
     \rm\fi}

%************** 21-point fonts ******************************

\font\xxiptrm=cmr17  scaled\magstep1
%\font\xxiptmit=cmmi12 scaled\magstep3
%\font\xxiptsy=cmsy10 scaled\magstep4
%\font\xxiptex=cmex10 scaled\magstep4
%\font\xxiptbf=cmbx12 scaled\magstep3

%\skewchar\xxiptmit='177 \skewchar\xxiptsy='60
%\fontdimen16 \xxiptsy=\the\fontdimen17 \xxiptsy

\def\xxipt{\ifmmode\err@badsizechange\else
     \@mathfontinit
%     \textfont0=\xxiptrm  \scriptfont0=\xivptrm  \scriptscriptfont0=\xptrm
%     \textfont1=\xxiptmit \scriptfont1=\xivptmit \scriptscriptfont1=\xptmit
%     \textfont2=\xxiptsy  \scriptfont2=\xivptsy  \scriptscriptfont2=\xptsy
%     \textfont3=\xxiptex  \scriptfont3=\xivptex  \scriptscriptfont3=\xptex
%     \textfont\bffam=\xxiptbf
%     \scriptfont\bffam=\xivptbf
%     \scriptscriptfont\bffam=\xptbf
     \@fontstyleinit
     \def\rm{\xxiptrm\fam=\z@}%
     \rm\fi}

%************** 25-point fonts ******************************

\font\xxvptrm=cmr17  scaled\magstep2
%\font\xxvptmit=cmmi12 scaled\magstep4
%\font\xxvptsy=cmsy10 scaled\magstep5
%\font\xxvptex=cmex10 scaled\magstep5
%\font\xxvptbf=cmbx12 scaled\magstep4

%\skewchar\xxvptmit='177 \skewchar\xxvptsy='60
%\fontdimen16 \xxvptsy=\the\fontdimen17 \xxvptsy

\def\xxvpt{\ifmmode\err@badsizechange\else
     \@mathfontinit
%     \textfont0=\xxvptrm  \scriptfont0=\xviiptrm  \scriptscriptfont0=\xiiptrm
%     \textfont1=\xxvptmit \scriptfont1=\xviiptmit \scriptscriptfont1=\xiiptmit
%     \textfont2=\xxvptsy  \scriptfont2=\xviiptsy  \scriptscriptfont2=\xiiptsy
%     \textfont3=\xxvptex  \scriptfont3=\xviiptex  \scriptscriptfont3=\xiiptex
%     \textfont\bffam=\xxvptbf
%     \scriptfont\bffam=\xviiptbf
%     \scriptscriptfont\bffam=\xiiptbf
     \@fontstyleinit
     \def\rm{\xxvptrm\fam=\z@}%
     \rm\fi}

%************** Other fonts *********************************

%\font\dummy=dummy

%******************************************************************************

\message{Loading jyTeX macros...}

%************************************************************
%*
%*              Simple modifications to plain
%*
%************************************************************
\message{modifications to plain.tex,}

% The "\outer" qualifier is removed from the definitions of \newcount through
% \newif so that they may be used in definitions.  \newif is also changed to
% make \if commands globally defined.

\def\newcount{\alloc@0\count\countdef\insc@unt}
\def\newdimen{\alloc@1\dimen\dimendef\insc@unt}
\def\newskip{\alloc@2\skip\skipdef\insc@unt}
\def\newmuskip{\alloc@3\muskip\muskipdef\@cclvi}
\def\newbox{\alloc@4\box\chardef\insc@unt}
\def\newtoks{\alloc@5\toks\toksdef\@cclvi}
\def\newhelp#1#2{\newtoks#1\global#1\expandafter{\csname#2\endcsname}}
\def\newread{\alloc@6\read\chardef\sixt@@n}
\def\newwrite{\alloc@7\write\chardef\sixt@@n}
\def\newfam{\alloc@8\fam\chardef\sixt@@n}
\def\newinsert#1{\global\advance\insc@unt by\m@ne
     \ch@ck0\insc@unt\count
     \ch@ck1\insc@unt\dimen
     \ch@ck2\insc@unt\skip
     \ch@ck4\insc@unt\box
     \allocationnumber=\insc@unt
     \global\chardef#1=\allocationnumber
     \wlog{\string#1=\string\insert\the\allocationnumber}}
\def\newif#1{\count@\escapechar \escapechar\m@ne
     \expandafter\expandafter\expandafter
          \xdef\@if#1{true}{\let\noexpand#1=\noexpand\iftrue}%
     \expandafter\expandafter\expandafter
          \xdef\@if#1{false}{\let\noexpand#1=\noexpand\iffalse}%
     \global\@if#1{false}\escapechar=\count@}

%************** Some parameter changes **********************

\newlinechar=`\^^J
\overfullrule=0pt

%************** Font-related modifications ******************

% The plain fonts are mapped onto the corresponding jyTeX fonts

% Some control sequences are disabled.

\let\itfam=\undefined

\let\bffam=\undefined

\count18=3

% German sharp s is given a new name (\ss is already taken)

\chardef\sharps="19

% The mathcode assignments of characters in the math italic font are changed to
% allow for switching to boldface.

\mathchardef\alpha="710B \mathchardef\beta="710C \mathchardef\gamma="710D
\mathchardef\delta="710E \mathchardef\epsilon="710F \mathchardef\zeta="7110
\mathchardef\eta="7111 \mathchardef\theta="7112 \mathchardef\iota="7113
\mathchardef\kappa="7114 \mathchardef\lambda="7115 \mathchardef\mu="7116
\mathchardef\nu="7117 \mathchardef\xi="7118 \mathchardef\pi="7119
\mathchardef\rho="711A \mathchardef\sigma="711B \mathchardef\tau="711C
\mathchardef\upsilon="711D \mathchardef\phi="711E \mathchardef\chi="711F
\mathchardef\psi="7120 \mathchardef\omega="7121 \mathchardef\varepsilon="7122
\mathchardef\vartheta="7123 \mathchardef\varpi="7124
\mathchardef\varrho="7125 \mathchardef\varsigma="7126
\mathchardef\varphi="7127 \mathchardef\imath="717B \mathchardef\jmath="717C
\mathchardef\ell="7160 \mathchardef\wp="717D \mathchardef\partial="7140
\mathchardef\flat="715B \mathchardef\natural="715C \mathchardef\sharp="715D

%************** Miscellaneous changes ***********************

% The dimension \p@ (1pt) is replaced with \rp@ (relative pt, defined below),
% whose size is determined by the base type size of the document.

\def\angle{{\vbox{\ialign{$\m@th\scriptstyle##$\crcr
     \not\mathrel{\mkern14mu}\crcr
     \noalign{\nointerlineskip}
     \mkern2.5mu\leaders\hrule height.34\rp@\hfill\mkern2.5mu\crcr}}}}
\def\vdots{\vbox{\baselineskip4\rp@ \lineskiplimit\z@
     \kern6\rp@\hbox{.}\hbox{.}\hbox{.}}}
\def\ddots{\mathinner{\mkern1mu\raise7\rp@\vbox{\kern7\rp@\hbox{.}}\mkern2mu
     \raise4\rp@\hbox{.}\mkern2mu\raise\rp@\hbox{.}\mkern1mu}}
\def\overrightarrow#1{\vbox{\ialign{##\crcr
     \rightarrowfill\crcr
     \noalign{\kern-\rp@\nointerlineskip}
     $\hfil\displaystyle{#1}\hfil$\crcr}}}
\def\overleftarrow#1{\vbox{\ialign{##\crcr
     \leftarrowfill\crcr
     \noalign{\kern-\rp@\nointerlineskip}
     $\hfil\displaystyle{#1}\hfil$\crcr}}}
\def\overbrace#1{\mathop{\vbox{\ialign{##\crcr
     \noalign{\kern3\rp@}
     \downbracefill\crcr
     \noalign{\kern3\rp@\nointerlineskip}
     $\hfil\displaystyle{#1}\hfil$\crcr}}}\limits}
\def\underbrace#1{\mathop{\vtop{\ialign{##\crcr
     $\hfil\displaystyle{#1}\hfil$\crcr
     \noalign{\kern3\rp@\nointerlineskip}
     \upbracefill\crcr
     \noalign{\kern3\rp@}}}}\limits}
\def\big#1{{\hbox{$\left#1\vbox to8.5\rp@ {}\right.\n@space$}}}
\def\Big#1{{\hbox{$\left#1\vbox to11.5\rp@ {}\right.\n@space$}}}
\def\bigg#1{{\hbox{$\left#1\vbox to14.5\rp@ {}\right.\n@space$}}}
\def\Bigg#1{{\hbox{$\left#1\vbox to17.5\rp@ {}\right.\n@space$}}}
\def\@vereq#1#2{\lower.5\rp@\vbox{\baselineskip\z@skip\lineskip-.5\rp@
     \ialign{$\m@th#1\hfil##\hfil$\crcr#2\crcr=\crcr}}}
\def\rlh@#1{\vcenter{\hbox{\ooalign{\raise2\rp@
     \hbox{$#1\rightharpoonup$}\crcr
     $#1\leftharpoondown$}}}}
\def\bordermatrix#1{\begingroup\m@th
     \setbox\z@\vbox{%
          \def\cr{\crcr\noalign{\kern2\rp@\global\let\cr\endline}}%
          \ialign{$##$\hfil\kern2\rp@\kern\p@renwd
               &\thinspace\hfil$##$\hfil&&\quad\hfil$##$\hfil\crcr
               \omit\strut\hfil\crcr
               \noalign{\kern-\baselineskip}%
               #1\crcr\omit\strut\cr}}%
     \setbox\tw@\vbox{\unvcopy\z@\global\setbox\@ne\lastbox}%
     \setbox\tw@\hbox{\unhbox\@ne\unskip\global\setbox\@ne\lastbox}%
     \setbox\tw@\hbox{$\kern\wd\@ne\kern-\p@renwd\left(\kern-\wd\@ne
          \global\setbox\@ne\vbox{\box\@ne\kern2\rp@}%
          \vcenter{\kern-\ht\@ne\unvbox\z@\kern-\baselineskip}%
          \,\right)$}%
     \null\;\vbox{\kern\ht\@ne\box\tw@}\endgroup}
\def\endinsert{\egroup
     \if@mid\dimen@\ht\z@
          \advance\dimen@\dp\z@
          \advance\dimen@12\rp@
          \advance\dimen@\pagetotal
          \ifdim\dimen@>\pagegoal\@midfalse\p@gefalse\fi
     \fi
     \if@mid\bigskip\box\z@
          \bigbreak
     \else\insert\topins{\penalty100 \splittopskip\z@skip
               \splitmaxdepth\maxdimen\floatingpenalty\z@
               \ifp@ge\dimen@\dp\z@
                    \vbox to\vsize{\unvbox\z@\kern-\dimen@}%
               \else\box\z@\nobreak\bigskip
               \fi}%
     \fi
     \endgroup}

% \normalbaselines is removed from \cases and \matrix.

\def\cases#1{\left\{\,\vcenter{\m@th
     \ialign{$##\hfil$&\quad##\hfil\crcr#1\crcr}}\right.}
\def\matrix#1{\null\,\vcenter{\m@th
     \ialign{\hfil$##$\hfil&&\quad\hfil$##$\hfil\crcr
          \mathstrut\crcr
          \noalign{\kern-\baselineskip}
          #1\crcr
          \mathstrut\crcr
          \noalign{\kern-\baselineskip}}}\,}

% \raggedbottom modified slightly

\newif\ifraggedbottom

\def\raggedbottom{\ifraggedbottom\else
     \advance\topskip by\z@ plus60pt \raggedbottomtrue\fi}%
\def\normalbottom{\ifraggedbottom
     \advance\topskip by\z@ plus-60pt \raggedbottomfalse\fi}

%************************************************************
%*
%*              Miscellaneous definitions
%*
%************************************************************
\message{hacks,}

%************** Hack registers ******************************

\toksdef\toks@i=1 \toksdef\toks@ii=2

%************** Basic macros ********************************

\def\TeX{T\kern-.1667em \lower.5ex \hbox{E}\kern-.125em X\null}
\def\jyTeX{{\leavevmode
     \raise.587ex \hbox{\it\j}\kern-.1em \lower.048ex \hbox{\it y}\kern-.12em
     \TeX}}

\let\then=\iftrue
\def\ifnoarg#1\then{\def\hack@{#1}\ifx\hack@\empty}
\def\ifundefined#1\then{%
     \expandafter\ifx\csname\expandafter\blank\string#1\endcsname\relax}
\def\useif#1\then{\csname#1\endcsname}
\def\usename#1{\csname#1\endcsname}
\def\useafter#1#2{\expandafter#1\csname#2\endcsname}

% Modify so that I can have \loop's within \loop's?
\long\def\loop#1\repeat{\def\@iterate{#1\expandafter\@iterate\fi}\@iterate
     \let\@iterate=\relax}
%\long\def\loop#1\repeat{\def\@loopbody{#1}\@iterate}
%\def\@iterate{\@loopbody\let\next=\@iterate\else\let\next=\relax\fi\next}

\let\TeXend=\end
\def\begin#1{\begingroup\def\@@blockname{#1}\usename{begin#1}}
\def\end#1{\usename{end#1}\def\hack@{#1}%
     \ifx\@@blockname\hack@
          \endgroup
     \else\err@badgroup\hack@\@@blockname
     \fi}
\def\@@blockname{}

\def\defaultoption[#1]#2{%
     \def\hack@{\ifx\hack@ii[\toks@={#2}\else\toks@={#2[#1]}\fi\the\toks@}%
     \futurelet\hack@ii\hack@}

\def\markup#1{\let\@@marksf=\empty
     \ifhmode\edef\@@marksf{\spacefactor=\the\spacefactor\relax}\/\fi
     ${}^{\hbox{\subscriptfonts#1}}$\@@marksf}

%************** Time registers ******************************

\newtoks\shortyear
\newtoks\militaryhour
\newtoks\standardhour
\newtoks\minute
\newtoks\amorpm

\def\settime{\count@=\time\divide\count@ by60
     \militaryhour=\expandafter{\number\count@}%
     {\multiply\count@ by-60 \advance\count@ by\time
          \xdef\hack@{\ifnum\count@<10 0\fi\number\count@}}%
     \minute=\expandafter{\hack@}%
     \ifnum\count@<12
          \amorpm={am}
     \else\amorpm={pm}
          \ifnum\count@>12 \advance\count@ by-12 \fi
     \fi
     \standardhour=\expandafter{\number\count@}%
     \def\hack@19##1##2{\shortyear={##1##2}}%
          \expandafter\hack@\the\year}

\def\monthword#1{%
     \ifcase#1
          $\bullet$\err@badcountervalue{monthword}%
          \or January\or February\or March\or April\or May\or June%
          \or July\or August\or September\or October\or November\or December%
     \else$\bullet$\err@badcountervalue{monthword}%
     \fi}

\def\monthabbr#1{%
     \ifcase#1
          $\bullet$\err@badcountervalue{monthabbr}%
          \or Jan\or Feb\or Mar\or Apr\or May\or Jun%
          \or Jul\or Aug\or Sep\or Oct\or Nov\or Dec%
     \else$\bullet$\err@badcountervalue{monthabbr}%
     \fi}

\def\militarytime{\the\militaryhour:\the\minute}
\def\standardtime{\the\standardhour:\the\minute}

%************** Number styles *******************************

\def\@setnumstyle#1#2{\expandafter\global\expandafter\expandafter
     \expandafter\let\expandafter\expandafter
     \csname @\expandafter\blank\string#1style\endcsname
     \csname#2\endcsname}
\def\numstyle#1{\usename{@\expandafter\blank\string#1style}#1}
\def\ifblank#1\then{\useafter\ifx{@\expandafter\blank\string#1}\blank}

\def\blank#1{}

\def\Roman#1{\expandafter\uppercase\expandafter{\romannumeral#1}}
\def\alphabetic#1{%
     \ifcase#1
          $\bullet$\err@badcountervalue{alphabetic}%
          \or a\or b\or c\or d\or e\or f\or g\or h\or i\or j\or k\or l\or m%
          \or n\or o\or p\or q\or r\or s\or t\or u\or v\or w\or x\or y\or z%
     \else$\bullet$\err@badcountervalue{alphabetic}%
     \fi}
\def\Alphabetic#1{\expandafter\uppercase\expandafter{\alphabetic{#1}}}
\def\symbols#1{%
     \ifcase#1
          $\bullet$\err@badcountervalue{symbols}%
          \or*\or\dag\or\ddag\or\S\or$\|$%
          \or**\or\dag\dag\or\ddag\ddag\or\S\S\or$\|\|$%
     \else$\bullet$\err@badcountervalue{symbols}%
     \fi}

%************** String macros *******************************

\catcode`\^^?=13 \def^^?{\relax}

\def\trimleading#1\to#2{\edef#2{#1}%
     \expandafter\@trimleading\expandafter#2#2^^?^^?}
\def\@trimleading#1#2#3^^?{\ifx#2^^?\def#1{}\else\def#1{#2#3}\fi}

\def\trimtrailing#1\to#2{\edef#2{#1}%
     \expandafter\@trimtrailing\expandafter#2#2^^? ^^?\relax}
\def\@trimtrailing#1#2 ^^?#3{\ifx#3\relax\toks@={}%
     \else\def#1{#2}\toks@={\trimtrailing#1\to#1}\fi
     \the\toks@}

\def\trim#1\to#2{\trimleading#1\to#2\trimtrailing#2\to#2}

\catcode`\^^?=15

%************** List macros *********************************

\long\def\additemL#1\to#2{\toks@={\^^\{#1}}\toks@ii=\expandafter{#2}%
     \xdef#2{\the\toks@\the\toks@ii}}

\long\def\additemR#1\to#2{\toks@={\^^\{#1}}\toks@ii=\expandafter{#2}%
     \xdef#2{\the\toks@ii\the\toks@}}

\def\getitemL#1\to#2{\expandafter\@getitemL#1\hack@#1#2}
\def\@getitemL\^^\#1#2\hack@#3#4{\def#4{#1}\def#3{#2}}

%************************************************************
%*
%*             Font-related macros
%*
%************************************************************
\message{font macros,}

%************** Font set-up *********************************

\newdimen\rp@
\newcount\@@sizeindex \@@sizeindex=0
\newcount\@@factori
\newcount\@@factorii
\newcount\@@factoriii
\newcount\@@factoriv

\countdef\maxfam=18
\newfam\itfam
\newfam\bffam
\newfam\bfsfam
\newfam\bmitfam

\def\@mathfontinit{\count@=4
     \loop\textfont\count@=\nullfont
          \scriptfont\count@=\nullfont
          \scriptscriptfont\count@=\nullfont
          \ifnum\count@<\maxfam\advance\count@ by\@ne
     \repeat}

\def\@fontstyleinit{%
     \def\it{\err@fontnotavailable\it}%
     \def\bf{\err@fontnotavailable\bf}%
     \def\bfs{\err@bfstobf}%
     \def\bmit{\err@fontnotavailable\bmit}%
     \def\sc{\err@fontnotavailable\sc}%
     \def\sl{\err@sltoit}%
     \def\ss{\err@fontnotavailable\ss}%
     \def\tt{\err@fontnotavailable\tt}}

\def\@parameterinit#1{\rm\rp@=.1em \@getscaling{#1}%
     \let\^^\=\@doscaling\scalingskipslist
     \setbox\strutbox=\hbox{\vrule
          height.708\baselineskip depth.292\baselineskip width\z@}}

\def\@getfactor#1#2#3#4{\@@factori=#1 \@@factorii=#2
     \@@factoriii=#3 \@@factoriv=#4}

\def\@getscaling#1{\count@=#1 \advance\count@ by-\@@sizeindex\@@sizeindex=#1
     \ifnum\count@<0
          \let\@mulordiv=\divide
          \let\@divormul=\multiply
          \multiply\count@ by\m@ne
     \else\let\@mulordiv=\multiply
          \let\@divormul=\divide
     \fi
     \edef\@@scratcha{\ifcase\count@                {1}{1}{1}{1}\or
          {1}{7}{23}{3}\or     {2}{5}{3}{1}\or      {9}{89}{13}{1}\or
          {6}{25}{6}{1}\or     {8}{71}{14}{1}\or    {6}{25}{36}{5}\or
          {1}{7}{53}{4}\or     {12}{125}{108}{5}\or {3}{14}{53}{5}\or
          {6}{41}{17}{1}\or    {13}{31}{13}{2}\or   {9}{107}{71}{2}\or
          {11}{139}{124}{3}\or {1}{6}{43}{2}\or     {10}{107}{42}{1}\or
          {1}{5}{43}{2}\or     {5}{69}{65}{1}\or    {11}{97}{91}{2}\fi}%
     \expandafter\@getfactor\@@scratcha}

\def\@doscaling#1{\@mulordiv#1by\@@factori\@divormul#1by\@@factorii
     \@mulordiv#1by\@@factoriii\@divormul#1by\@@factoriv}

%************* Size-changing commands ***********************

\newskip\headskip
\newskip\footskip

\def\typesize=#1pt{\count@=#1 \advance\count@ by-10
     \ifcase\count@
          \@setsizex\or\err@badtypesize\or
          \@setsizexii\or\err@badtypesize\or
          \@setsizexiv
     \else\err@badtypesize
     \fi}

\def\@setsizex{\getixpt
     \def\subsubscriptfonts{\vpt}%
          \def\subsubscriptsize{\vpt\@parameterinit{-8}}%
     \def\subscriptfonts{\viipt}\def\subscriptsize{\viipt\@parameterinit{-4}}%
     \def\footnotefonts{\viiipt}\def\footnotesize{\viiipt\@parameterinit{-2}}%
     \def\smallfonts{\ixpt}\def\smallsize{\ixpt\@parameterinit{-1}}%
     \def\normalfonts{\xpt}\def\normalsize{\xpt\@parameterinit{0}}%
     \def\bigfonts{\xiipt}\def\bigsize{\xiipt\@parameterinit{2}}%
     \def\Bigfonts{\xivpt}\def\Bigsize{\xivpt\@parameterinit{4}}%
     \def\biggfonts{\xviipt}\def\biggsize{\xviipt\@parameterinit{6}}%
     \def\Biggfonts{\xxipt}\def\Biggsize{\xxipt\@parameterinit{8}}%
     \def\tinyfonts{\vpt}\def\tinysize{\vpt\@parameterinit{-8}}%
     \def\HUGEFONTS{\xxvpt}\def\HUGESIZE{\xxvpt\@parameterinit{10}}%
     \normalsize\fixedskipslist}

\def\@setsizexii{\getxipt
     \def\subsubscriptfonts{\vipt}%
          \def\subsubscriptsize{\vipt\@parameterinit{-6}}%
     \def\subscriptfonts{\viiipt}%
          \def\subscriptsize{\viiipt\@parameterinit{-2}}%
     \def\footnotefonts{\xpt}\def\footnotesize{\xpt\@parameterinit{0}}%
     \def\smallfonts{\xipt}\def\smallsize{\xipt\@parameterinit{1}}%
     \def\normalfonts{\xiipt}\def\normalsize{\xiipt\@parameterinit{2}}%
     \def\bigfonts{\xivpt}\def\bigsize{\xivpt\@parameterinit{4}}%
     \def\Bigfonts{\xviipt}\def\Bigsize{\xviipt\@parameterinit{6}}%
     \def\biggfonts{\xxipt}\def\biggsize{\xxipt\@parameterinit{8}}%
     \def\Biggfonts{\xxvpt}\def\Biggsize{\xxvpt\@parameterinit{10}}%
     \def\tinyfonts{\vpt}\def\tinysize{\vpt\@parameterinit{-8}}%
     \def\HUGEFONTS{\xxvpt}\def\HUGESIZE{\xxvpt\@parameterinit{10}}%
     \normalsize\fixedskipslist}

\def\@setsizexiv{\getxiiipt
     \def\subsubscriptfonts{\viipt}%
          \def\subsubscriptsize{\viipt\@parameterinit{-4}}%
     \def\subscriptfonts{\xpt}\def\subscriptsize{\xpt\@parameterinit{0}}%
     \def\footnotefonts{\xiipt}\def\footnotesize{\xiipt\@parameterinit{2}}%
     \def\smallfonts{\xiiipt}\def\smallsize{\xiiipt\@parameterinit{3}}%
     \def\normalfonts{\xivpt}\def\normalsize{\xivpt\@parameterinit{4}}%
     \def\bigfonts{\xviipt}\def\bigsize{\xviipt\@parameterinit{6}}%
     \def\Bigfonts{\xxipt}\def\Bigsize{\xxipt\@parameterinit{8}}%
     \def\biggfonts{\xxvpt}\def\biggsize{\xxvpt\@parameterinit{10}}%
     \def\Biggfonts{\err@sizetoolarge\Biggfonts\HUGEFONTS}%
          \def\Biggsize{\err@sizetoolarge\Biggsize\HUGESIZE}%
     \def\tinyfonts{\vpt}\def\tinysize{\vpt\@parameterinit{-8}}%
     \def\HUGEFONTS{\xxvpt}\def\HUGESIZE{\xxvpt\@parameterinit{10}}%
     \normalsize\fixedskipslist}

\def\subsubscriptfonts{\vpt} \def\subsubscriptsize{\vpt\@parameterinit{-8}}
\def\subscriptfonts{\viipt}  \def\subscriptsize{\viipt\@parameterinit{-4}}
\def\footnotefonts{\viiipt}  \def\footnotesize{\viiipt\@parameterinit{-2}}
\def\smallfonts{\err@sizenotavailable\smallfonts}
                             \def\smallsize{\ixpt\@parameterinit{-1}}
\def\normalfonts{\xpt}       \def\normalsize{\xpt\@parameterinit{0}}
\def\bigfonts{\xiipt}        \def\bigsize{\xiipt\@parameterinit{2}}
\def\Bigfonts{\xivpt}        \def\Bigsize{\xivpt\@parameterinit{4}}
\def\biggfonts{\xviipt}      \def\biggsize{\xviipt\@parameterinit{6}}
\def\Biggfonts{\xxipt}       \def\Biggsize{\xxipt\@parameterinit{8}}
\def\tinyfonts{\vpt}         \def\tinysize{\vpt\@parameterinit{-8}}
\def\HUGEFONTS{\xxvpt}       \def\HUGESIZE{\xxvpt\@parameterinit{10}}

%************************************************************
%*
%*             Document layout
%*
%************************************************************
\message{document layout,}

%************** Page format *********************************

\newtoks\everyoutput \everyoutput={}
\newdimen\depthofpage
\newcount\pagenum \pagenum=0

\newdimen\oddtopmargin  \newdimen\eventopmargin
\newdimen\oddleftmargin \newdimen\evenleftmargin
\newtoks\oddhead        \newtoks\evenhead
\newtoks\oddfoot        \newtoks\evenfoot

\def\topmargin{\afterassignment\@seteventop\oddtopmargin}
\def\leftmargin{\afterassignment\@setevenleft\oddleftmargin}
\def\head{\afterassignment\@setevenhead\oddhead}
\def\foot{\afterassignment\@setevenfoot\oddfoot}

\def\@seteventop{\eventopmargin=\oddtopmargin}
\def\@setevenleft{\evenleftmargin=\oddleftmargin}
\def\@setevenhead{\evenhead=\oddhead}
\def\@setevenfoot{\evenfoot=\oddfoot}

\def\pagenumstyle#1{\@setnumstyle\pagenum{#1}}

\newif\ifdraft
\def\draft{\drafttrue\leftmargin=.5in \overfullrule=5pt }

\def\outputstyle#1{\global\expandafter\let\expandafter
          \@outputstyle\csname#1output\endcsname
     \usename{#1setup}}

\output={\@outputstyle}

\def\normaloutput{\the\everyoutput
     \global\advance\pagenum by\@ne
     \ifodd\pagenum
          \voffset=\oddtopmargin \hoffset=\oddleftmargin
     \else\voffset=\eventopmargin \hoffset=\evenleftmargin
     \fi
     \advance\voffset by-1in  \advance\hoffset by-1in
     \count0=\pagenum
     \expandafter\shipout\pagebox
     \ifnum\outputpenalty>-\@MM\else\dosupereject\fi}

\newdimen\fullhsize
\newbox\leftpage
\newcount\leftpagenum
\newcount\outputpagenum \outputpagenum=0
\let\leftorright=L

\def\twoupoutput{\the\everyoutput
     \global\advance\pagenum by\@ne
     \if L\leftorright
          \global\setbox\leftpage=\leftline{\pagebox}%
          \global\leftpagenum=\pagenum
          \global\let\leftorright=R%
     \else\global\advance\outputpagenum by\@ne
          \ifodd\outputpagenum
               \voffset=\oddtopmargin \hoffset=\oddleftmargin
          \else\voffset=\eventopmargin \hoffset=\evenleftmargin
          \fi
          \advance\voffset by-1in  \advance\hoffset by-1in
          \count0=\leftpagenum \count1=\pagenum
          \shipout\vbox{\hbox to\fullhsize
               {\box\leftpage\hfil\leftline{\pagebox}}}%
          \global\let\leftorright=L%
     \fi
     \ifnum\outputpenalty>-\@MM
     \else\dosupereject
          \if R\leftorright
               \globaldefs=\@ne\head={\hfil}\foot={\hfil}\globaldefs=\z@
               \null\newpage
          \fi
     \fi}

\def\pagebox{\vbox{\makeheadline\pagebody\makefootline}}

\def\makeheadline{%
     \vbox to\z@{\baselinestretch=\@m
          \vskip\topskip\vskip-.708\baselineskip\vskip-\headskip
          \line{\vbox to\ht\strutbox{}%
               \ifodd\pagenum\the\oddhead\else\the\evenhead\fi}%
          \vss}%
     \nointerlineskip}

\def\pagebody{\vbox to\vsize{%
     \boxmaxdepth\maxdepth
     \ifvoid\topins\else\unvbox\topins\fi
     \depthofpage=\dp255
     \unvbox255
     \ifraggedbottom\kern-\depthofpage\vfil\fi
     \ifvoid\footins
     \else\vskip\skip\footins
          \footnoterule
          \unvbox\footins
          \vskip-\footnoteskip
     \fi}}

\def\makefootline{\baselineskip=\footskip
     \line{\ifodd\pagenum\the\oddfoot\else\the\evenfoot\fi}}

%************** Sectioning commands *************************

\newskip\abovechapterskip
\newskip\belowchapterskip
\newskip\abovesectionskip
\newskip\belowsectionskip
\newskip\abovesubsectionskip
\newskip\belowsubsectionskip

\def\chapterstyle#1{\global\expandafter\let\expandafter\@chapterstyle
     \csname#1text\endcsname}
\def\sectionstyle#1{\global\expandafter\let\expandafter\@sectionstyle
     \csname#1text\endcsname}
\def\subsectionstyle#1{\global\expandafter\let\expandafter\@subsectionstyle
     \csname#1text\endcsname}

\def\chapter#1{%
     \ifdim\lastskip=17sp \else\chapterbreak\vskip\abovechapterskip\fi
     \@chapterstyle{\ifblank\chapternumstyle\then
          \else\newchapternum=\next\chapternumformat\ \fi#1}%
     \nobreak\vskip\belowchapterskip\vskip17sp }

\def\section#1{%
     \ifdim\lastskip=17sp \else\sectionbreak\vskip\abovesectionskip\fi
     \@sectionstyle{\ifblank\sectionnumstyle\then
          \else\newsectionnum=\next\sectionnumformat\ \fi#1}%
     \nobreak\vskip\belowsectionskip\vskip17sp }

\def\subsection#1{%
     \ifdim\lastskip=17sp \else\subsectionbreak\vskip\abovesubsectionskip\fi
     \@subsectionstyle{\ifblank\subsectionnumstyle\then
          \else\newsubsectionnum=\next\subsectionnumformat\ \fi#1}%
     \nobreak\vskip\belowsubsectionskip\vskip17sp }

%************** Text formatting commands ********************

\let\TeXunderline=\underline
\let\TeXoverline=\overline
\def\underline#1{\relax\ifmmode\TeXunderline{#1}\else
     $\TeXunderline{\hbox{#1}}$\fi}
\def\overline#1{\relax\ifmmode\TeXoverline{#1}\else
     $\TeXoverline{\hbox{#1}}$\fi}

\def\baselinestretch{\afterassignment\@baselinestretch\count@}
\def\@baselinestretch{\baselineskip=\normalbaselineskip
     \divide\baselineskip by\@m\baselineskip=\count@\baselineskip
     \setbox\strutbox=\hbox{\vrule
          height.708\baselineskip depth.292\baselineskip width\z@}%
     \bigskipamount=\the\baselineskip
          plus.25\baselineskip minus.25\baselineskip
     \medskipamount=.5\baselineskip
          plus.125\baselineskip minus.125\baselineskip
     \smallskipamount=.25\baselineskip
          plus.0625\baselineskip minus.0625\baselineskip}

\def\\{\ifhmode\ifnum\lastpenalty=-\@M\else\hfil\penalty-\@M\fi\fi
     \ignorespaces}
\def\newpage{\vfil\break}

\def\lefttext#1{\par{\@text\leftskip=\z@\rightskip=\centering
     \noindent#1\par}}
\def\righttext#1{\par{\@text\leftskip=\centering\rightskip=\z@
     \noindent#1\par}}
\def\centertext#1{\par{\@text\leftskip=\centering\rightskip=\centering
     \noindent#1\par}}
\def\@text{\parindent=\z@ \parfillskip=\z@ \everypar={}%
     \spaceskip=.3333em \xspaceskip=.5em
     \def\\{\ifhmode\ifnum\lastpenalty=-\@M\else\penalty-\@M\fi\fi
          \ignorespaces}}

\def\beginleft{\par\@text\leftskip=\z@ \rightskip=\centering}
     
\def\beginright{\par\@text\leftskip=\centering\rightskip=\z@ }
     
\def\begincenter{\par\@text\leftskip=\centering\rightskip=\centering}

\def\beginnarrow{\defaultoption[\parindent]\@beginnarrow}
\def\@beginnarrow[#1]{\par\advance\leftskip by#1\advance\rightskip by#1}

\begingroup
\catcode`\[=1 \catcode`\{=11 \gdef\beginignore[\endgroup\bgroup
     \catcode`\e=0 \catcode`\\=12 \catcode`\{=11 \catcode`\f=12 \let\or=\relax
     \let\nd{ignor=\fi \let\}=\egroup
     \iffalse}
\endgroup

\long\def\marginnote#1{\leavevmode
     \edef\@marginsf{\spacefactor=\the\spacefactor\relax}%
     \ifdraft\strut\vadjust{%
          \hbox to\z@{\hskip\hsize\hskip.1in
               \vbox to\z@{\vskip-\dp\strutbox
                    \marginnoteformat
                    \vskip-\ht\strutbox
                    \noindent\strut#1\par
                    \vss}%
               \hss}}%
     \fi
     \@marginsf}

%************** The \bye command ****************************

\newtoks\everybye \everybye={\par\vfil}
\outer\def\bye{\the\everybye
     \footnotecheck
     \prelabelcheck
     \streamcheck
     \supereject
     \TeXend}

%************************************************************
%*
%*             Footnotes
%*
%************************************************************
\message{footnotes,}

\newcount\footnotenum \footnotenum=0
\newskip\footnoteskip
\let\@footnotelist=\empty

\def\footnotenumstyle#1{\@setnumstyle\footnotenum{#1}%
     \useafter\ifx{@footnotenumstyle}\symbols
          \global\let\@footup=\empty
     \else\global\let\@footup=\markup
     \fi}

\def\footnote{\footnotecheck\defaultoption[]\@footnote}
\def\@footnote[#1]{\@footnotemark[#1]\@footnotetext}

\def\footnotemark{\defaultoption[]\@footnotemark}
\def\@footnotemark[#1]{\let\@footsf=\empty
     \ifhmode\edef\@footsf{\spacefactor=\the\spacefactor\relax}\/\fi
     \ifnoarg#1\then
          \global\advance\footnotenum by\@ne
          \@footup{\footnotenumformat}%
          \edef\@@foota{\footnotenum=\the\footnotenum\relax}%
          \expandafter\additemR\expandafter\@footup\expandafter
               {\@@foota\footnotenumformat}\to\@footnotelist
          \global\let\@footnotelist=\@footnotelist
     \else\markup{#1}%
          \additemR\markup{#1}\to\@footnotelist
          \global\let\@footnotelist=\@footnotelist
     \fi
     \@footsf}

\def\footnotetext{%
     \ifx\@footnotelist\empty\err@extrafootnotetext\else\@footnotetext\fi}
\def\@footnotetext{%
     \getitemL\@footnotelist\to\@@foota
     \global\let\@footnotelist=\@footnotelist
     \insert\footins\bgroup
     \footnoteformat
     \splittopskip=\ht\strutbox\splitmaxdepth=\dp\strutbox
     \interlinepenalty=\interfootnotelinepenalty\floatingpenalty=\@MM
     \noindent\llap{\@@foota}\strut
     \bgroup\aftergroup\@footnoteend
     \let\@@scratcha=}
\def\@footnoteend{\strut\par\vskip\footnoteskip\egroup}

\def\footnoterule{\normalfonts
     \kern-.3em \hrule width2in height.04em \kern .26em }

\def\footnotecheck{%
     \ifx\@footnotelist\empty
     \else\err@extrafootnotemark
          \global\let\@footnotelist=\empty
     \fi}

%************************************************************
%*
%*             Labelling macros
%*
%************************************************************
\message{labels,}

\let\@@labeldef=\xdef
\newif\if@labelfile
\newwrite\@labelfile
\let\@prelabellist=\empty

\def\label#1#2{\trim#1\to\@@labarg\edef\@@labtext{#2}%
     \edef\@@labname{lab@\@@labarg}%
     \useafter\ifundefined\@@labname\then\else\@yeslab\fi
     \useafter\@@labeldef\@@labname{#2}%
     \ifstreaming
          \expandafter\toks@\expandafter\expandafter\expandafter
               {\csname\@@labname\endcsname}%
          \immediate\write\streamout{\noexpand\label{\@@labarg}{\the\toks@}}%
     \fi}
\def\@yeslab{%
     \useafter\ifundefined{if\@@labname}\then
          \err@labelredef\@@labarg
     \else\useif{if\@@labname}\then
               \err@labelredef\@@labarg
          \else\global\usename{\@@labname true}%
               \useafter\ifundefined{pre\@@labname}\then
               \else\useafter\ifx{pre\@@labname}\@@labtext
                    \else\err@badlabelmatch\@@labarg
                    \fi
               \fi
               \if@labelfile
               \else\global\@labelfiletrue
                    \immediate\write\sixt@@n{--> Creating file \jobname.lab}%
                    \immediate\openout\@labelfile=\jobname.lab
               \fi
               \immediate\write\@labelfile
                    {\noexpand\prelabel{\@@labarg}{\@@labtext}}%
          \fi
     \fi}

\def\putlab#1{\trim#1\to\@@labarg\edef\@@labname{lab@\@@labarg}%
     \useafter\ifundefined\@@labname\then\@nolab\else\usename\@@labname\fi}
\def\@nolab{%
     \useafter\ifundefined{pre\@@labname}\then
          \undefinedlabelformat
          \err@needlabel\@@labarg
          \useafter\xdef\@@labname{\undefinedlabelformat}%
     \else\usename{pre\@@labname}%
          \useafter\xdef\@@labname{\usename{pre\@@labname}}%
     \fi
     \useafter\newif{if\@@labname}%
     \expandafter\additemR\@@labarg\to\@prelabellist}

\def\prelabel#1{\useafter\gdef{prelab@#1}}

\def\ifundefinedlabel#1\then{%
     \expandafter\ifx\csname lab@#1\endcsname\relax}
\def\useiflab#1\then{\csname iflab@#1\endcsname}

\def\prelabelcheck{{%
     \def\^^\##1{\useiflab{##1}\then\else\err@undefinedlabel{##1}\fi}%
     \@prelabellist}}

%************************************************************
%*
%*             Equation numbering
%*
%************************************************************
\message{equation numbering,}

\newcount\chapternum
\newcount\sectionnum
\newcount\subsectionnum
\newcount\equationnum
\newcount\subequationnum
\newcount\figurenum
\newcount\subfigurenum
\newcount\tablenum
\newcount\subtablenum

\newif\if@subeqncount
\newif\if@subfigcount
\newif\if@subtblcount

\def\newchapternum{\newsectionnum=\z@\@resetnum\chapternum}
\def\newsectionnum{\newsubsectionnum=\z@\@resetnum\sectionnum}
\def\newsubsectionnum{\newequationnum=\z@\newfigurenum=\z@\newtablenum=\z@
     \@resetnum\subsectionnum}
\def\newequationnum{\newsubequationnum=\z@\@resetnum\equationnum}
\def\newsubequationnum{\@resetnum\subequationnum}
\def\newfigurenum{\newsubfigurenum=\z@\@resetnum\figurenum}
\def\newsubfigurenum{\@resetnum\subfigurenum}
\def\newtablenum{\newsubtablenum=\z@\@resetnum\tablenum}
\def\newsubtablenum{\@resetnum\subtablenum}

\def\@resetnum#1{\global\advance#1by1 \edef\next{\the#1\relax}\global#1}

\newchapternum=0

\def\chapternumstyle#1{\@setnumstyle\chapternum{#1}}
\def\sectionnumstyle#1{\@setnumstyle\sectionnum{#1}}
\def\subsectionnumstyle#1{\@setnumstyle\subsectionnum{#1}}
\def\equationnumstyle#1{\@setnumstyle\equationnum{#1}}
\def\subequationnumstyle#1{\@setnumstyle\subequationnum{#1}%
     \ifblank\subequationnumstyle\then\global\@subeqncountfalse\fi
     \ignorespaces}
\def\figurenumstyle#1{\@setnumstyle\figurenum{#1}}
\def\subfigurenumstyle#1{\@setnumstyle\subfigurenum{#1}%
     \ifblank\subfigurenumstyle\then\global\@subfigcountfalse\fi
     \ignorespaces}
\def\tablenumstyle#1{\@setnumstyle\tablenum{#1}}
\def\subtablenumstyle#1{\@setnumstyle\subtablenum{#1}%
     \ifblank\subtablenumstyle\then\global\@subtblcountfalse\fi
     \ignorespaces}

\def\eqnlabel#1{%
     \if@subeqncount
          \newsubequationnum=\next
     \else\newequationnum=\next
          \ifblank\subequationnumstyle\then
          \else\global\@subeqncounttrue
               \newsubequationnum=\@ne
          \fi
     \fi
     \label{#1}{\puteqnformat}(\puteqn{#1})%
     \ifdraft\rlap{\hskip.1in{\tt#1}}\fi}

\let\puteqn=\putlab

\def\equation#1#2{\useafter\gdef{eqn@#1}{#2\eqno\eqnlabel{#1}}}
\def\Equation#1{\useafter\gdef{eqn@#1}}

\def\putequation#1{\useafter\ifundefined{eqn@#1}\then
     \err@undefinedeqn{#1}\else\usename{eqn@#1}\fi}

\def\eqnseriesstyle#1{\gdef\@eqnseriesstyle{#1}}
\def\begineqnseries{\subequationnumstyle{\@eqnseriesstyle}%
     \defaultoption[]\@begineqnseries}
\def\@begineqnseries[#1]{\edef\@@eqnname{#1}}
\def\endeqnseries{\subequationnumstyle{blank}%
     \expandafter\ifnoarg\@@eqnname\then
     \else\label\@@eqnname{\puteqnformat}%
     \fi
     \aftergroup\ignorespaces}

\def\figlabel#1{%
     \if@subfigcount
          \newsubfigurenum=\next
     \else\newfigurenum=\next
          \ifblank\subfigurenumstyle\then
          \else\global\@subfigcounttrue
               \newsubfigurenum=\@ne
          \fi
     \fi
     \label{#1}{\putfigformat}\putfig{#1}%
     {\def\marginnoteformat{\tt}\marginnote{#1}}}

\let\putfig=\putlab

\def\figseriesstyle#1{\gdef\@figseriesstyle{#1}}
\def\beginfigseries{\subfigurenumstyle{\@figseriesstyle}%
     \defaultoption[]\@beginfigseries}
\def\@beginfigseries[#1]{\edef\@@figname{#1}}
\def\endfigseries{\subfigurenumstyle{blank}%
     \expandafter\ifnoarg\@@figname\then
     \else\label\@@figname{\putfigformat}%
     \fi
     \aftergroup\ignorespaces}

\def\tbllabel#1{%
     \if@subtblcount
          \newsubtablenum=\next
     \else\newtablenum=\next
          \ifblank\subtablenumstyle\then
          \else\global\@subtblcounttrue
               \newsubtablenum=\@ne
          \fi
     \fi
     \label{#1}{\puttblformat}\puttbl{#1}%
     {\def\marginnoteformat{\tt}\marginnote{#1}}}

\let\puttbl=\putlab

\def\tblseriesstyle#1{\gdef\@tblseriesstyle{#1}}
\def\begintblseries{\subtablenumstyle{\@tblseriesstyle}%
     \defaultoption[]\@begintblseries}
\def\@begintblseries[#1]{\edef\@@tblname{#1}}
\def\endtblseries{\subtablenumstyle{blank}%
     \expandafter\ifnoarg\@@tblname\then
     \else\label\@@tblname{\puttblformat}%
     \fi
     \aftergroup\ignorespaces}

%************************************************************
%*
%*             Reference numbering
%*
%************************************************************
\message{reference numbering,}

\newcount\referencenum \referencenum=0
\newcount\@@prerefcount \@@prerefcount=0
\newcount\@@thisref
\newcount\@@lastref
\newcount\@@loopref
\newcount\@@refseq
\newdimen\refnumindent
\let\@undefreflist=\empty

\def\referencenumstyle#1{\@setnumstyle\referencenum{#1}}

\def\referencestyle#1{\usename{@ref#1}}

\def\@refsequential{%
     \gdef\@refpredef##1{\global\advance\referencenum by\@ne
          \let\^^\=0\label{##1}{\^^\{\the\referencenum}}%
          \useafter\gdef{ref@\the\referencenum}{{##1}{\undefinedlabelformat}}}%
     \gdef\@reference##1##2{%
          \ifundefinedlabel##1\then
          \else\def\^^\####1{\global\@@thisref=####1\relax}\putlab{##1}%
               \useafter\gdef{ref@\the\@@thisref}{{##1}{##2}}%
          \fi}%
     \gdef\endputreferences{%
          \loop\ifnum\@@loopref<\referencenum
                    \advance\@@loopref by\@ne
                    \expandafter\expandafter\expandafter\@printreference
                         \csname ref@\the\@@loopref\endcsname
          \repeat
          \par}}

\def\@refpreordered{%
     \gdef\@refpredef##1{\global\advance\referencenum by\@ne
          \additemR##1\to\@undefreflist}%
     \gdef\@reference##1##2{%
          \ifundefinedlabel##1\then
          \else\global\advance\@@loopref by\@ne
               {\let\^^\=0\label{##1}{\^^\{\the\@@loopref}}}%
               \@printreference{##1}{##2}%
          \fi}
     \gdef\endputreferences{%
          \def\^^\####1{\useiflab{####1}\then
               \else\reference{####1}{\undefinedlabelformat}\fi}%
          \@undefreflist
          \par}}

\def\beginprereferences{\par
     \def\reference##1##2{\global\advance\referencenum by1\@ne
          \let\^^\=0\label{##1}{\^^\{\the\referencenum}}%
          \useafter\gdef{ref@\the\referencenum}{{##1}{##2}}}}
\def\endprereferences{\global\@@prerefcount=\the\referencenum\par}

\def\beginputreferences{\par
     \refnumindent=\z@\@@loopref=\z@
     \loop\ifnum\@@loopref<\referencenum
               \advance\@@loopref by\@ne
               \setbox\z@=\hbox{\referencenum=\@@loopref
                    \referencenumformat\enskip}%
               \ifdim\wd\z@>\refnumindent\refnumindent=\wd\z@\fi
     \repeat
     \putreferenceformat
     \@@loopref=\z@
     \loop\ifnum\@@loopref<\@@prerefcount
               \advance\@@loopref by\@ne
               \expandafter\expandafter\expandafter\@printreference
                    \csname ref@\the\@@loopref\endcsname
     \repeat
     \let\reference=\@reference}

\def\@printreference#1#2{\ifx#2\undefinedlabelformat\err@undefinedref{#1}\fi
     \noindent\ifdraft\rlap{\hskip\hsize\hskip.1in \tt#1}\fi
     \llap{\referencenum=\@@loopref\referencenumformat\enskip}#2\par}

\def\reference#1#2{{\par\refnumindent=\z@\putreferenceformat\noindent#2\par}}

\def\putref#1{\trim#1\to\@@refarg
     \expandafter\ifnoarg\@@refarg\then
          \toks@={\relax}%
     \else\@@lastref=-\@m\def\@@refsep{}\def\@more{\@nextref}%
          \toks@={\@nextref#1,,}%
     \fi\the\toks@}
\def\@nextref#1,{\trim#1\to\@@refarg
     \expandafter\ifnoarg\@@refarg\then
          \let\@more=\relax
     \else\ifundefinedlabel\@@refarg\then
               \expandafter\@refpredef\expandafter{\@@refarg}%
          \fi
          \def\^^\##1{\global\@@thisref=##1\relax}%
          \global\@@thisref=\m@ne
          \setbox\z@=\hbox{\putlab\@@refarg}%
     \fi
     \advance\@@lastref by\@ne
     \ifnum\@@lastref=\@@thisref\advance\@@refseq by\@ne\else\@@refseq=\@ne\fi
     \ifnum\@@lastref<\z@
     \else\ifnum\@@refseq<\thr@@
               \@@refsep\def\@@refsep{,}%
               \ifnum\@@lastref>\z@
                    \advance\@@lastref by\m@ne
                    {\referencenum=\@@lastref\putrefformat}%
               \else\undefinedlabelformat
               \fi
          \else\def\@@refsep{--}%
          \fi
     \fi
     \@@lastref=\@@thisref
     \@more}

%************************************************************
%*
%*             Job streaming
%*
%************************************************************
\message{streaming,}

\newif\ifstreaming

\def\streamto{\defaultoption[\jobname]\@streamto}
\def\@streamto[#1]{\global\streamingtrue
     \immediate\write\sixt@@n{--> Streaming to #1.str}%
     \newwrite\streamout\immediate\openout\streamout=#1.str }

\def\streamfrom{\defaultoption[\jobname]\@streamfrom}
\def\@streamfrom[#1]{\newread\streamin\openin\streamin=#1.str
     \ifeof\streamin
          \expandafter\err@nostream\expandafter{#1.str}%
     \else\immediate\write\sixt@@n{--> Streaming from #1.str}%
          \let\@@labeldef=\gdef
          \ifstreaming
               \edef\@elc{\endlinechar=\the\endlinechar}%
               \endlinechar=\m@ne
               \loop\read\streamin to\@@scratcha
                    \ifeof\streamin
                         \streamingfalse
                    \else\toks@=\expandafter{\@@scratcha}%
                         \immediate\write\streamout{\the\toks@}%
                    \fi
                    \ifstreaming
               \repeat
               \@elc
               \input #1.str
               \streamingtrue
          \else\input #1.str
          \fi
          \let\@@labeldef=\xdef
     \fi}

\def\streamcheck{\ifstreaming
     \immediate\write\streamout{\pagenum=\the\pagenum}%
     \immediate\write\streamout{\footnotenum=\the\footnotenum}%
     \immediate\write\streamout{\referencenum=\the\referencenum}%
     \immediate\write\streamout{\chapternum=\the\chapternum}%
     \immediate\write\streamout{\sectionnum=\the\sectionnum}%
     \immediate\write\streamout{\subsectionnum=\the\subsectionnum}%
     \immediate\write\streamout{\equationnum=\the\equationnum}%
     \immediate\write\streamout{\subequationnum=\the\subequationnum}%
     \immediate\write\streamout{\figurenum=\the\figurenum}%
     \immediate\write\streamout{\subfigurenum=\the\subfigurenum}%
     \immediate\write\streamout{\tablenum=\the\tablenum}%
     \immediate\write\streamout{\subtablenum=\the\subtablenum}%
     \immediate\closeout\streamout
     \fi}

%************************************************************
%*
%*             Error messages
%*
%************************************************************

\def\err@badtypesize{%
     \errhelp={The limited availability of certain fonts requires^^J%
          that the base type size be 10pt, 12pt, or 14pt.^^J}%
     \errmessage{--> Illegal base type size}}

\def\err@badsizechange{\immediate\write\sixt@@n
     {--> Size change not allowed in math mode, ignored}}

\def\err@sizetoolarge#1{\immediate\write\sixt@@n
     {--> \noexpand#1 too big, substituting HUGE}}

\def\err@sizenotavailable#1{\immediate\write\sixt@@n
     {--> Size not available, \noexpand#1 ignored}}

\def\err@fontnotavailable#1{\immediate\write\sixt@@n
     {--> Font not available, \noexpand#1 ignored}}

\def\err@sltoit{\immediate\write\sixt@@n
     {--> Style \noexpand\sl not available, substituting \noexpand\it}%
     \it}

\def\err@bfstobf{\immediate\write\sixt@@n
     {--> Style \noexpand\bfs not available, substituting \noexpand\bf}%
     \bf}

\def\err@badgroup#1#2{%
     \errhelp={The block you have just tried to close was not the one^^J%
          most recently opened.^^J}%
     \errmessage{--> \noexpand\end{#1} doesn't match \noexpand\begin{#2}}}

\def\err@badcountervalue#1{\immediate\write\sixt@@n
     {--> Counter (#1) out of bounds}}

\def\err@extrafootnotemark{\immediate\write\sixt@@n
     {--> \noexpand\footnotemark command
          has no corresponding \noexpand\footnotetext}}

\def\err@extrafootnotetext{%
     \errhelp{You have given a \noexpand\footnotetext command without first
          specifying^^Ja \noexpand\footnotemark.^^J}%
     \errmessage{--> \noexpand\footnotetext command has no corresponding
          \noexpand\footnotemark}}

\def\err@labelredef#1{\immediate\write\sixt@@n
     {--> Label "#1" redefined}}

\def\err@badlabelmatch#1{\immediate\write\sixt@@n
     {--> Definition of label "#1" doesn't match value in \jobname.lab}}

\def\err@needlabel#1{\immediate\write\sixt@@n
     {--> Label "#1" cited before its definition}}

\def\err@undefinedlabel#1{\immediate\write\sixt@@n
     {--> Label "#1" cited but never defined}}

\def\err@undefinedeqn#1{\immediate\write\sixt@@n
     {--> Equation "#1" not defined}}

\def\err@undefinedref#1{\immediate\write\sixt@@n
     {--> Reference "#1" not defined}}

\def\err@nostream#1{%
     \errhelp={You have tried to input a stream file that doesn't exist.^^J}%
     \errmessage{--> Stream file #1 not found}}

%************************************************************
%*
%*             Initialization
%*
%************************************************************
\message{jyTeX initialization}

\everyjob{\immediate\write16{--> jyTeX version \fmtversion}%
     \edef\@@jobname{\jobname}%
%     \openin0=\inputpath jysupp
%     \ifeof0
%     \else\closein0
%          \immediate\write16{--> Additional macros loaded from jysupp.tex}%
%          \jyinput jysupp
%     \fi
%     \openin0=\inputpath jylocal
%     \ifeof0
%     \else\closein0
%          \immediate\write16{--> Additional macros loaded from jylocal.tex}%
%          \jyinput jylocal
%     \fi
     \edef\jobname{\@@jobname}%
     \settime
     \openin0=\jobname.lab
     \ifeof0
     \else\closein0
          \immediate\write16{--> Getting labels from file \jobname.lab}%
          \input\jobname.lab
     \fi}

%************** Spacing *************************************

\def\fixedskipslist{%
     \^^\{\topskip}%
     \^^\{\splittopskip}%
     \^^\{\maxdepth}%
     \^^\{\skip\topins}%
     \^^\{\skip\footins}%
     \^^\{\headskip}%
     \^^\{\footskip}}

\def\scalingskipslist{%
     \^^\{\p@renwd}%
     \^^\{\delimitershortfall}%
     \^^\{\nulldelimiterspace}%
     \^^\{\scriptspace}%
     \^^\{\jot}%
     \^^\{\normalbaselineskip}%
     \^^\{\normallineskip}%
     \^^\{\normallineskiplimit}%
     \^^\{\baselineskip}%
     \^^\{\lineskip}%
     \^^\{\lineskiplimit}%
     \^^\{\bigskipamount}%
     \^^\{\medskipamount}%
     \^^\{\smallskipamount}%
     \^^\{\parskip}%
     \^^\{\parindent}%
     \^^\{\abovedisplayskip}%
     \^^\{\belowdisplayskip}%
     \^^\{\abovedisplayshortskip}%
     \^^\{\belowdisplayshortskip}%
     \^^\{\abovechapterskip}%
     \^^\{\belowchapterskip}%
     \^^\{\abovesectionskip}%
     \^^\{\belowsectionskip}%
     \^^\{\abovesubsectionskip}%
     \^^\{\belowsubsectionskip}}

%************** Document layout *****************************

\def\twoupsetup{%                                % setup for twoup style
     \topmargin=.75in
     \leftmargin=.5in
     \vsize=6.9in
     \hsize=4.75in
     \fullhsize=10in
     \let\draft=\relax}

\outputstyle{normal}                             % page style

\def\marginnoteformat{\subscriptsize             % paragraphing of margin notes
     \hsize=1in \baselinestretch=1000 \everypar={}%
     \tolerance=5000 \hbadness=5000 \parskip=0pt \parindent=0pt
     \leftskip=0pt \rightskip=0pt \raggedright}

\head={\ifdraft\normalfonts\it\hfil DRAFT\hfil   % format of headline
     \llap{\number\day\ \monthword\month\ \militarytime}\else\hfil\fi}
\foot={\hfil\normalfonts\numstyle\pagenum\hfil}  % format of footline

\normalbaselineskip=12pt                         % usual \baselineskip
\normallineskip=0pt                              % usual \lineskip
\normallineskiplimit=0pt                         % usual \lineskiplimit
\normalbaselines                                 % set \baselineskip

\topskip=.85\baselineskip \splittopskip=\topskip \headskip=2\baselineskip
\footskip=\headskip

\pagenumstyle{arabic}                            % counter style

\parskip=0pt                                     % no skip between paragraphs
\parindent=20pt                                  % usual \parindent

\baselinestretch=1000                            % set \big-, \med-, \smallskip

%************** Sectioning **********************************

\chapterstyle{left}                              % position of heading
\chapternumstyle{blank}                          % counter style
\def\chapterbreak{\newpage}                      % break before heading
\abovechapterskip=0pt                            % space before heading
\belowchapterskip=1.5\baselineskip               % space after heading
     plus.38\baselineskip minus.38\baselineskip
\def\chapternumformat{\numstyle\chapternum.}     % format of heading counter

\sectionstyle{left}                              % position of heading
\sectionnumstyle{blank}                          % counter style
\def\sectionbreak{\vskip0pt plus4\baselineskip\penalty-100
     \vskip0pt plus-4\baselineskip}              % break before heading
\abovesectionskip=1.5\baselineskip               % space before heading
     plus.38\baselineskip minus.38\baselineskip
\belowsectionskip=\the\baselineskip              % space after heading
     plus.25\baselineskip minus.25\baselineskip
\def\sectionnumformat{%                          % format of heading counter
     \ifblank\chapternumstyle\then\else\numstyle\chapternum.\fi
     \numstyle\sectionnum.}

\subsectionstyle{left}                           % position of heading
\subsectionnumstyle{blank}                       % counter style
\def\subsectionbreak{\vskip0pt plus4\baselineskip\penalty-100
     \vskip0pt plus-4\baselineskip}              % break before heading
\abovesubsectionskip=\the\baselineskip           % space before heading
     plus.25\baselineskip minus.25\baselineskip
\belowsubsectionskip=.75\baselineskip            % space after heading
     plus.19\baselineskip minus.19\baselineskip
\def\subsectionnumformat{%                       % format of heading counter
     \ifblank\chapternumstyle\then\else\numstyle\chapternum.\fi
     \ifblank\sectionnumstyle\then\else\numstyle\sectionnum.\fi
     \numstyle\subsectionnum.}

%************** Footnotes ***********************************

\footnotenumstyle{symbols}                       % counter style
\footnoteskip=0pt                                % jyTeX spacing parameter
\def\footnotenumformat{\numstyle\footnotenum}    % \footnotemark format
\def\footnoteformat{\footnotesize                % paragraphing of text
     \everypar={}\parskip=0pt \parfillskip=0pt plus1fil
     \leftskip=1em \rightskip=0pt
     \spaceskip=0pt \xspaceskip=0pt
     \def\\{\ifhmode\ifnum\lastpenalty=-10000
          \else\hfil\penalty-10000 \fi\fi\ignorespaces}}

%************** Labels **************************************

\def\undefinedlabelformat{$\bullet$}             % mark for undefined label

%************** Equation numbering **************************

\equationnumstyle{arabic}                        % counter style
\subequationnumstyle{blank}                      % counter style
\figurenumstyle{arabic}                          % counter style
\subfigurenumstyle{blank}                        % counter style
\tablenumstyle{arabic}                           % counter style
\subtablenumstyle{blank}                         % counter style

\eqnseriesstyle{alphabetic}                      % sub-counter style for series
\figseriesstyle{alphabetic}                      % sub-counter style for series
\tblseriesstyle{alphabetic}                      % sub-counter style for series

\def\puteqnformat{\hbox{%                        % equation number format
     \ifblank\chapternumstyle\then\else\numstyle\chapternum.\fi
     \ifblank\sectionnumstyle\then\else\numstyle\sectionnum.\fi
     \ifblank\subsectionnumstyle\then\else\numstyle\subsectionnum.\fi
     \numstyle\equationnum
     \numstyle\subequationnum}}
\def\putfigformat{\hbox{%                        % figure number format
     \ifblank\chapternumstyle\then\else\numstyle\chapternum.\fi
     \ifblank\sectionnumstyle\then\else\numstyle\sectionnum.\fi
     \ifblank\subsectionnumstyle\then\else\numstyle\subsectionnum.\fi
     \numstyle\figurenum
     \numstyle\subfigurenum}}
\def\puttblformat{\hbox{%                        % table number format
     \ifblank\chapternumstyle\then\else\numstyle\chapternum.\fi
     \ifblank\sectionnumstyle\then\else\numstyle\sectionnum.\fi
     \ifblank\subsectionnumstyle\then\else\numstyle\subsectionnum.\fi
     \numstyle\tablenum
     \numstyle\subtablenum}}

%************** Reference numbering *************************

\referencestyle{sequential}                      % referencing method
\referencenumstyle{arabic}                       % counter style
\def\putrefformat{\numstyle\referencenum}        % format of reference citation
\def\referencenumformat{\numstyle\referencenum.} % format of number in list
\def\putreferenceformat{%                        % paragraphing of list
     \everypar={\hangindent=1em \hangafter=1 }%
     \def\\{\hfil\break\null\hskip-1em \ignorespaces}%
     \leftskip=\refnumindent\parindent=0pt \interlinepenalty=1000 }

%************** Font initialization *************************

\normalsize

%*****************************************************************************

\def\fmtversion{2.6M (June 1992)}

\catcode`\@=12
% ------------------ End of jytex.tex -----------------

%\input jytex.tex   % available from hep-th
\typesize=10pt \magnification=1200 \baselineskip17truept
%\baselineskip25truept
\footnotenumstyle{arabic} \hsize=6truein\vsize=8.5truein
%\input castess.lab
%\draft
%\leftmargin=1.25in
%\oddleftmargin=.5in
%\evenleftmargin=1.5in
\sectionnumstyle{blank}
\chapternumstyle{blank}
\chapternum=1
\sectionnum=1
\pagenum=0
%\referencestyle{preordered}
% title style follows

\def\begintitle{\pagenumstyle{blank}\parindent=0pt
\begin{narrow}[0.4in]}
\def\endtitle{\end{narrow}\newpage\pagenumstyle{arabic}}

% exercise style follows

\def\beginexercise{\vskip 20truept\parindent=0pt\begin{narrow}[10
truept]}
\def\endexercise{\vskip 10truept\end{narrow}}

% **************    my jyTeX abbreviations   *****************

\def\eql#1{\eqno\eqnlabel{#1}}
\def\ref{\reference}
\def\peq{\puteqn}
\def\pref{\putref}

\def\mgn{\marginnote}
\def\bex{\begin{exercise}}
\def\eex{\end{exercise}}

% *********************** My definitions ************************

\font\open=msbm10 %scaled\magstep1 % For VAX. Borde p195.

 %scaled\magstep1 % For VAX. Borde p195.
%\font\open=msym10 %scaled\magstep1 % For Arbortxt on PC
%\font\opens=msym8 %scaled\magstep1 % For Arbortxt on PC
  % For Arbortxt on PC, and VAX. Borde p199

%\font\smsb=cmss8
\def\StretchRtArr#1{{\count255=0\loop\relbar\joinrel\advance\count255 by1
\ifnum\count255<#1\repeat\rightarrow}}
\def\StretchLtArr#1{\,{\leftarrow\!\!\count255=0\loop\relbar
\joinrel\advance\count255 by1\ifnum\count255<#1\repeat}}

\def\StretchLRtArr#1{\,{\leftarrow\!\!\count255=0\loop\relbar\joinrel\advance
\count255 by1\ifnum\count255<#1\repeat\rightarrow\,\,}}
\def\OverArrow#1#2{\;{\buildrel #1\over{\StretchRtArr#2}}\;}

\def\mbox#1{{\leavevmode\hbox{#1}}}

\def\hspace#1{{\phantom{\mbox#1}}}
\def\oR{\mbox{\open\char82}}

\def\oZ{\mbox{\open\char90}}

\def\al{\alpha}
 %in jyTeX
 %in jyTeX
 %in jyTeX
 %in jyTeX
 %in jyTeX
 %in jyTeX
 %in jyTeX
 %in jyTeX
 %in jyTeX
% in jyTeX
% in jyTeX
\def\bom{{\bmit\omega}}% in jyTeX
\def\be{\beta}
\def\ga{\gamma}
\def\de{\delta}
\def\Ga{\Gamma}

\def\la{\lambda}

\def\om{\omega}

\def\si{\sigma}
\def\Si{\Sigma}
\def\th{\theta}

\def\ze{\zeta}

\def\caL{{\cal L}}

\def\caH{{\cal H}}
\def\det{{\rm det\,}}

\def\Real{{\rm Re\,}}

\def\sc{{\rm sc }}

\def\zf{$\zeta$--function}
\def\zfs{$\zeta$--functions}

     % Newline

\def\frac#1/#2{\leavevmode\kern.1em
\raise.5ex\hbox{\the\scriptfont0 #1}\kern-.1em/\kern-.15em
\lower.25ex\hbox{\the\scriptfont0 #2}}
\def\sfrac#1/#2{\leavevmode\kern.1em
\raise.5ex\hbox{\the\scriptscriptfont0 #1}\kern-.1em/\kern-.15em
\lower.25ex\hbox{\the\scriptscriptfont0 #2}}

\def\gtorder{\mathrel{\raise.3ex\hbox{$>$}\mkern-14mu
             \lower0.6ex\hbox{$\sim$}}}
\def\ltorder{\mathrel{\raise.3ex\hbox{$<$}\mkern-14mu
             \lower0.6ex\hbox{$\sim$}}}

\def\semidirprod{\rlap{\ss C}\raise1pt\hbox{$\mkern.75mu\times$}}
\def\for{\lower6pt\hbox{$\Big|$}}
\def\fish{\kern-.25em{\phantom{abcde}\over \phantom{abcde}}\kern-.25em}

 %triple
%dot
 %double
%dot
 %double dot
%for small #1

\def\boxit#1{\vbox{\hrule\hbox{\vrule\kern3pt
        \vbox{\kern3pt#1\kern3pt}\kern3pt\vrule}\hrule}}
\def\dalemb#1#2{{\vbox{\hrule height .#2pt
        \hbox{\vrule width.#2pt height#1pt \kern#1pt \vrule
                width.#2pt} \hrule height.#2pt}}}

\def\ol{\overline}
        %double stroke
\def\frac#1#2{{{#1}\over{#2}}}
 %lower covariant deriv.
 %upper covariant deriv.
 %lower covariant deriv semicolon.
    %lower ordinary  deriv.
    %lower ordinary  deriv comma.

\def\noin{\noindent}

      %Connection
    %Connection'
\def\comb#1#2{{\left(#1\atop#2\right)}}

\def\cosec{{\rm cosec\,}}

\def\eg{{\it e.g.}}
\def\ie{{\it i.e. }}
\def\cf{{\it cf }}
\def\pa{\partial}

 %gives average <#1>
 %gives thermal average <<#1>>
   %gives bracket <#1|#2>
   %gives comma bracket <#1,#2>
 %gives round bracket (#1,#2)
 %gives round bracket (#1,|#2)
 %gives big bracket <#1|#2>
  %gives
%matrix element <#1|#2|#3>
  %gives reduced matrix element
%<#1||#2||#3>

\def\curl{{\rm curl\,}}
\def\div{{\rm div\,}}
\def\grad{{\rm grad\,}}

\def\sumdasht#1#2{{\mathop{{\sum}'}_{#1}^{#2}}}

\def\3j#1#2#3#4#5#6{\left\lgroup\matrix{#1&#2&#3\cr#4&#5&#6\cr}
\right\rgroup}

\def\man{{\cal M}}

\def\m?{\mgn{?}}
% KK's defs

\def\pa{\partial}

\def\beq{\begin{eqnarray}}
\def\eeq{\end{eqnarray}}

%  *******************  Journal refs **********************

\def\aop#1#2#3{{\it Ann. Phys.} {\bf {#1}} ({#2}) #3}

\def\cmp#1#2#3{{\it Comm. Math. Phys.} {\bf {#1}} ({#2}) #3}
\def\cqg#1#2#3{{\it Class. Quant. Grav.} {\bf {#1}} ({#2}) #3}

\def\jmp#1#2#3{{\it J. Math. Phys.} {\bf {#1}} ({#2}) #3}
\def\jpa#1#2#3{{\it J. Phys.} {\bf A{#1}} ({#2}) #3}

\def\np#1#2#3{{\it Nucl. Phys.} {\bf B{#1}} ({#2}) #3}
\def\pl#1#2#3{{\it Phys. Lett.} {\bf {#1}} ({#2}) #3}

\def\prp#1#2#3{{\it Phys. Rep.} {\bf {#1}} ({#2}) #3}
\def\pr#1#2#3{{\it Phys. Rev.} {\bf {#1}} ({#2}) #3}
\def\prA#1#2#3{{\it Phys. Rev.} {\bf A{#1}} ({#2}) #3}

\def\prD#1#2#3{{\it Phys. Rev.} {\bf D{#1}} ({#2}) #3}

\def\rmp#1#2#3{{\it Rev. Mod. Phys.} {\bf {#1}} ({#2}) #3}

\def\zfp#1#2#3{{\it Z. f. Phys.} {\bf {#1}} ({#2}) #3}

\def\cras#1#2#3{{\it Comptes Rend. Acad. Sci. (Paris)} {\bf{#1}} (#2) #3}
\def\prs#1#2#3{{\it Proc. Roy. Soc.} {\bf A{#1}} ({#2}) #3}
\def\pcps#1#2#3{{\it Proc. Camb. Phil. Soc.} {\bf{#1}} ({#2}) #3}

\def\amsh#1#2#3{{\it Abh. Math. Sem. Ham.} {\bf {#1}} ({#2}) #3}
\def\am#1#2#3{{\it Acta Mathematica} {\bf {#1}} ({#2}) #3}
\def\aim#1#2#3{{\it Adv. in Math.} {\bf {#1}} ({#2}) #3}
\def\ajm#1#2#3{{\it Am. J. Math.} {\bf {#1}} ({#2}) #3}

\def\aom#1#2#3{{\it Ann. of Math.} {\bf {#1}} ({#2}) #3}
\def\cjm#1#2#3{{\it Can. J. Math.} {\bf {#1}} ({#2}) #3}
\def\bams#1#2#3{{\it Bull.Am.Math.Soc.} {\bf {#1}} ({#2}) #3}

\def\cmh#1#2#3{{\it Comm. Math. Helv.} {\bf {#1}} ({#2}) #3}

\def\dmj#1#2#3{{\it Duke Math. J.} {\bf {#1}} ({#2}) #3}
\def\invm#1#2#3{{\it Invent. Math.} {\bf {#1}} ({#2}) #3}

\def\jdg#1#2#3{{\it J. Diff. Geom.} {\bf {#1}} ({#2}) #3}

\def\joa#1#2#3{{\it J. of Algebra} {\bf {#1}} ({#2}) #3}
\def\jram#1#2#3{{\it J. f. reine u. Angew. Math.} {\bf {#1}} ({#2}) #3}
\def\jims#1#2#3{{\it J. Indian. Math. Soc.} {\bf {#1}} ({#2}) #3}
\def\jlms#1#2#3{{\it J. Lond. Math. Soc.} {\bf {#1}} ({#2}) #3}
\def\jmpa#1#2#3{{\it J. Math. Pures. Appl.} {\bf {#1}} ({#2}) #3}
\def\ma#1#2#3{{\it Math. Ann.} {\bf {#1}} ({#2}) #3}

\def\mz#1#2#3{{\it Math. Zeit.} {\bf {#1}} ({#2}) #3}
\def\ojm#1#2#3{{\it Osaka J.Math.} {\bf {#1}} ({#2}) #3}

\def\plb#1#2#3{{\it Phys. Letts.} {\bf {B#1}} ({#2}) #3}
\def\plms#1#2#3{{\it Proc. Lond. Math. Soc.} {\bf {#1}} ({#2}) #3}
\def\pgma#1#2#3{{\it Proc. Glasgow Math. Ass.} {\bf {#1}} ({#2}) #3}
\def\qjm#1#2#3{{\it Quart. J. Math.} {\bf {#1}} ({#2}) #3}

\def\rmjm#1#2#3{{\it Rocky Mountain J. Math.} {\bf {#1}} ({#2}) #3}

\def\tams#1#2#3{{\it Trans.Am.Math.Soc.} {\bf {#1}} ({#2}) #3}

% *******************   Main text *********************
\begin{title}
\vglue 1truein
%\righttext {MUTP/96/23}
%\righttext{hep-th/96}
\vskip15truept
%\leftline{\today}
%\vskip 30truept
\centertext {\Bigfonts \bf $p$--form spectra and Casimir energies}
\vskip10truept \centertext{\Bigfonts \bf on  spherical tesselations} \vskip
20truept \centertext{J.S.Dowker\footnote{dowker@a35.ph.man.ac.uk}} \vskip
7truept \centertext{\it Theory Group,} \centertext{\it School of Physics and
Astronomy,} \centertext{\it The University of Manchester,} \centertext{\it
Manchester, England} \vskip40truept
\begin{narrow}
Casimir energies on space--times having the fundamental domains of
semi--regular spherical tesselations of the three--sphere as their spatial
sections are computed for scalar and Maxwell fields. The spectral theory of
$p$--forms on the fundamental domains is also developed and degeneracy
generating functions computed. Absolute and relative boundary conditions are
encountered naturally. Some aspects of the heat--kernel expansion are
explored. The expansion is shown to terminate with the constant term which is
computed to be $1/2$ on all tesselations for a coexact 1--form and shown to
be so by topological arguments. Some practical points concerning generalised
Bernoulli numbers are given.
\end{narrow}
\vskip 5truept
%\righttext {August 1996}
\vskip 60truept
%\righttext{Typeset in \jyTeX}
\vfil
\end{title}
\pagenum=0
\newpage

\section{\bf 1. Introduction.}

A recent paper [\pref{BHS}], see also, [\pref{BMO}], [\pref{BandO}], on some
field theory effects on half an Einstein Universe has prompted me to exhume
and extend some previous work on space--times that include this case. In fact,
some aspects of (free) scalar field theories on a hemisphere had been early
discussed, [\pref{Kenn,KandU,Dow30}], using both modes and images. Similar
calculations were undertaken later in [\pref{BandO1}].

However it is not this path I wish to take here. I would rather consider the
hemisphere as a (simple) example of a fundamental domain of a regular
spherical tesselation and continue with the calculations begun in
[{\pref{ChandD}]. This reference deals with scalar fields, but Chang's thesis,
[\pref{Chang}], contains higher spin results and I will be incorporating these
since the works [{\pref{BHS,BandO}] are concerned specifically with Maxwell
fields. I will consider the spectral theory of $p$--forms in spherical
tesselations and use it in the relevant, $p=1$, case.

Apart from its general interest, half an Einstein Universe is conformally
related to anti de Sitter space and is of relevance to supersymmetry.

I will concentrate on tesselations of the 3--sphere, although some results
hold for the general sphere. The calculation is presented partly as an
exercise in spectral geometry on bounded domains.

I will be concerned to evaluate global, \ie integrated, quantities such as the
total Casimir energy, so the explicit form of the modes is not as vital as the
spectrum.

\section{\bf 2. The geometry}

Space--time is static of the form $\oR\times$ S$^d/\Gamma$ where $\Ga$ is a
finite group of isometries of the sphere, in particular the complete symmetry
group of a $(d+1)$--dimensional regular polytope. The projection of this onto
its circumscribing hypersphere, S$^d$, yields a spherical tesselation, or
honeycomb, the cells of which are the projections of the $d$-dimensional
faces of the polytope.  $\Gamma$ is generated, for example, by reflections in
$d+1$ concurrent hyperplanes. These hyperplanes ($d$--flats) intersect the
circumscribing S$^d$ in a set of reflecting great $(d-1)$--spheres and divide
up the honeycomb into a finer tesselation by forming the boundaries of
$|\Gamma|$ spherical $d$--simplices that are transitively permuted by $\Ga$.
One chosen such simplex can be taken as the fundamental domain for this
action. For $d=3$ the fundamental domain is a spherical tetrahedron. In
particular situations the tetrahedron can degenerate. An example is the
3--hemisphere when $\Ga$ has a single reflective action.

Although my main interest is with analysis on the fundamental domain, it is
sometimes helpful to think of this as being the intersection of the
hypersphere with that portion of the ambient $\oR^{d+1}$ cut out by the
reflecting hypersurfaces. It is, therefore, embedded in this ambient infinite
alcove, hyper--kaleidoscope or M\"obius corner, as it is variously termed,
\eg\ [\pref{BandB}], [\pref{DandCh}], [\pref{Dowl}] . In the following, I
sometimes use $n=d+1$.

The elements of $\Ga$ separate into two, equal sized sets distinguished by
whether an element contains an even or an odd number of reflections. The even
set is a (finite) subgroup of SO$(d+1)$ and this rotational part of $\Ga$ is
sometimes referred to as the polytope group and its elements as `direct'
rotations. The whole of $\Ga$ is the extended polytope group, as it can be
obtained from the rotational part by adjoining a reflection. This split means
that the above tesselation actually consists of $|\Ga|/2$ copies of the
fundamental domain and an equal number of copies of the reflected fundamental
domain. It is, therefore, technically not a regular tesselation. I term it a
semi--regular tesselation. If one restricts to just the rotational part of
$\Ga,$ then one does get a regular tesselation whose fundamental domain is
obtained by sticking together a $d$--simplex and its reflection.

The 3--sphere is special because of the isomorphism SO$(4)\sim$ SU$(2)\times$
SU$(2)/\oZ_2$ and the rotational action of $\Ga$ on S$^3$ can be neatly
effected by left and right group actions on SU$(2)$ since this is isomorphic
to S$^3$. This structure means that the semi--regular tesselations of the
3--sphere are more numerous than for higher dimensions, \eg\ [\pref{SeandT,
Warner}]. This is, of course the same question as the number of regular
polytopes, which goes a long way back to Schl\"afli.

\section{\bf 3. The degeneracy problem}

In previous work, [\pref{DandB, Dow11, Dow12}], I have considered quantum
field theory on quotients of the 3--sphere. Since the fixed--point free action
of $\Ga$ employed there plays no role in the preliminary mode analysis and in
the basic construction of the \zfs, I can, for rapidity, use the formulae in
these references, with suitable modifications.

Even with fixed points, the eigenvalues of the relevant operators are a subset
of those on the full sphere. There are two main approaches to the calculation
of the required degeneracies. One uses the L$\times$R structure of $\Ga$ and
writes the degeneracy as a function of the left and right `rotation' angles
which arise through the isomorphism, SO$(3)\sim $ SU$(2)/\oZ_2$, applied to
each group element $\ga,=(\ga_L,\ga_R)\in\Ga$. Since quantities, like the
heat--kernel or Casimir energy, evaluated on the factored sphere, \ie on the
fundamental domain, involve a group average over $\Ga$ (a preimage sum), they
become just sums over the rotation angles and have to be evaluated term by
term using explicit values for these angles. A snag is that this method only
works nicely for the rotational part of $\Ga$. Fortunately, it can be shown
that the reflective part does not contribute to the Casimir energy for odd
spheres.

The other approach computes a generating function for the degeneracies as
a closed form involving the {\it degrees} associated with the polytope
group. This method is generally to be preferred but numerical agreement
between the two routes provides a comforting check, \cf\ [\pref{Dow11}].

To illustrate the techniques I look first at the scalar field and repeat some
material from [\pref{ChandD}], which uses degrees, and then outline the angle
sum approach which can be applied to the spin--1 case. Later, I will describe
the degree approach for the Maxwell field after dealing with $p$--forms.

\section{\bf 4. Scalar Casimir energy by degrees}

Using the degree method allows factors of the $d$--sphere to be discussed
without many problems and I will do this as far as possible easily.

All I say in this paper will be for the simplest case of a conformally
coupled scalar field. Then, on the unit $d$--sphere, the eigenvalues of
the Helmholtz--like operator are well known to be perfect squares,
  $$
    \la_n=\bigg({n+d-2\over2}\bigg)^2,\quad n=1,3,\ldots
  \eql{eigvals}
  $$
The generating function for the degeneracies, $d(l)$, of the eigenvalues
is defined to be
  $$
  h(\si)=\sum_{l=0}^\infty d(l)\,\si^l\,,
  \eql{genfun}
  $$
and of course depends only on the labelling, $l=0,1,\ldots$, not on the
specific form of the eigenvalues.

I introduce a new parameter $\tau$ by $\si=e^{-\tau}$ so that $h(\si)$,
which I also write as $h(\tau)$, is seen to be related to the cylinder
kernel, $T(\tau)$, \ie the kernel for the square root of the
Helmholtz--like operator, by
  $$
  T(\tau)=e^{-(d-1)\tau/2}\,h(\tau)\,.
  \eql{cylkern}
  $$
$T(\tau)$ has a thermal significance.

The idea now is to construct $h(\tau)$ by independent reasoning. The
argument depends on the old fact that the eigenfunctions on the
$d$--sphere arise from harmonic polynomials in an ambient Euclidean
$\oR^{d+1}$. The number of these is determined by Molien's theorem. The
harmonic generating function is, [\pref{Meyer}], the group average,
  $$
  h_\chi(\si)={1-\si^2\over |\Ga|}\sum_A{\chi^*(A)\over \det(1-\si A)}\,.
  \eql{genfun2}
  $$
The notation is that $A$ are the $(d+1)\times (d+1)$ matrix representatives of
$\Ga$ considered as a finite subgroup of O$(d+1)$. For flexibility, a
twisting, $\chi(A)$, has been included, corresponding to the equivariant
Molien theorem. $\chi(A)\equiv\chi(\ga)$ is the character in some
representation. I will use only the simplest cases so need not enlarge on its
general significance, see [\pref{Stanley1}].

The usual harmonic generating function, $h_N(\tau)$, is when $\chi(A)$ is
the trivial character, $=1,\,\forall A$. Its explicit evaluation depends
on Invariant Theory with the result that
  $$
  h_N(\si)=(1-\si^2)\prod_{i=1}^{d+1}{1\over(1-\si^{d_i})}\,,
  $$
where the degrees $d_i$, $i=1,2,\ldots,d+1$ are the degrees of the
linearly independent generating members (basis) of the invariant
polynomial ring associated with the action of $\Ga$ on the
$(d+1)$--dimensional ambient vector space, $\oR^{d+1}$. Since there is
always the basic invariant $x^2+y^2+z^2+\ldots$, of degree 2 = $d_{d+1}$,
say, the last factor on the denominator cancels the `harmonic factor'
$(1-\si^2)$ leaving just the degrees $(d_1,d_2,\ldots,d_d)\equiv {\bf d}$
and
  $$
  h_N(\si)=\prod_{i=1}^{d}{1\over(1-\si^{d_i})}\,.
  \eql{genfunn}
  $$

The subscript on $h_N$ stands for `Neumann' for the reason that all elements
of $\Ga$, including the odd ones, enter with the same sign in (\peq{genfun2})
which is the combination of images that gives Neumann conditions on the
boundary of the physical region, here a fundamental domain. To obtain the {\it
Dirichlet generating function}, $h_D(\si)$, the odd elements have to come in
negatively which can be achieved by setting $\chi(A)$ equal to the sign
character, $\det A$, to get, (\cf [\pref{BandB}]),
  $$\eqalign{
   h_D(\si)=&{1-\si^2\over |\Ga|}\sum_A{\det(A)\over \det(1-\si A)}\cr
   =&\,\si^{d_0}\prod_{i=1}^{d}{1\over(1-\si^{d_i})}\,,
   }
   \eql{genfund}
  $$
where $d_0\equiv\sum_{i=1}^{d+1}(d_i-1)$ is the degree of the Jacobian
and equals the number of reflecting hyperplanes. I note for future
reference that $h_N(\si)$ is the generating function for a $0$--form and
$h_D(\si)$ that for a $(d+1)$--form, on $\oR^{d+1}$, proportional to the
volume form.

Adding Neumann and Dirichlet gives the generating function when the
sphere is factored by just the rotational subgroup of $\Ga$. In this case
the boundary conditions are periodic ones on the doubled fundamental
domain.

A very simple, but instructive, example is the relation between the full
(periodic!) sphere and the hemisphere. The generating functions for the former
are trivially obtained by making $\Ga$ consist of just the identity\footnote{
In this case all the degrees are one because every polynomial is invariant and
the basis invariants can be taken simply to be the cartesian coordinates,
$x_1,\ldots,x_n$, on $\oR^n$.} so that, in Molien's theorem, $A$ is restricted
to be the unit $n\times n$ matrix. Then (\peq{genfun2}) gives
  $$
  h_{sphere}(\si)={1+\si\over(1-\si)^d}={1\over(1-\si)^d}
  +{\si\over(1-\si)^d}\,.
  $$
The two parts on the right--hand side are the Neumann and Dirichlet
generating functions for the hemisphere for which the degrees, ${\bf d}$,
are all one, $(1,1,\ldots)$ (see below) and $d_0=1$. To make this plain,
apply (\peq{genfun2}) to the two element group $\Ga=\{id,R\}$ where $R$
is a reflection represented by the diagonal matrix
  $$
  R=\left(\matrix{1&0&0&\ldots&0\cr
            0&1&0&\ldots&0\cr
            \vdots&\vdots&\vdots&\vdots&\vdots\cr
            0&0&0&\ldots&-1\cr}\right)\,.
  $$
A simple calculation yields the Neumann hemisphere expression if the $id$
and $R$ contributions are added and the Dirichlet one if they are
subtracted.

All the required spectral information is now in place so I just state the
expression for the total Casimir energy, $E$, in terms of the \zf,
$\ze(s)$, constructed from the eigenvalues (\peq{eigvals}) and
degeneracies, $d(l)$,
  $$
  E={1\over2}\,\ze(-1/2)\,.
  \eql{casen1}
  $$
This relation is reasonable if there are no divergences, as here.

There is no need to extract the degeneracies from the generating function
$h(\tau)$ since the relation (\peq{cylkern}) enables the \zf\ to be written
down immediately,
  $$
  \ze(s)={i\Ga(1-2s)\over2\pi}\int_{C_0}d\tau\,(-\tau)^{2s-1}
  e^{-(d-1)\tau/2}\,h(\tau)\,,
  \eql{zet0}
  $$
where $C_0$ is the Hankel contour.

Setting $s$ equal to $-1/2$ in (\peq{zet0}), the evaluation of $E$ reduces to
residues and the Taylor expansion of the integrand. Before giving the answer,
I note that the \zfs\ are examples of Barnes \zfs, [\pref{Barnesa,Barnesb}],
the general definition of which is,
  $$\eqalign{ \zeta_d(s,a|{\bom})=&{i\Gamma(1-s)\over2\pi}\int_{C_0}
  d\tau {\exp(-a\tau)
  (-\tau)^{s-1}\over\prod_{i=1}^d\big(1-\exp(-\om_i\tau)\big)}\cr
  =&\sum_{{\bf {m}}={\bf 0}}^\infty{1\over(a+{\bf {m.}\,{\bom}})^s},\qquad
  \Real\, s>d\,.}
  \eql{barn}
  $$
For simplicity, the $\om_i$ are taken positive. If $a$ is zero, the origin
${\bf m=0}$ is to be excluded. One has the specific relations,
  $$\eqalign{
  \ze_N(s)=\,&\ze_d(2s,(d-1)/2|{\bf d})\cr
  \ze_D(s)=\,&\ze_d(2s,\Si d_i-(d-1)/2\,|\,{\bf d})\,,
  }
  \eql{sczet}
  $$
and the calculation of the Casimir energy (\peq{casen1}) gives generalised
Bernoulli polynomials,
  $$
  E_0=-(\mp)^{d+1}{1\over (d+1)!\, 2\prod d_i }
  B^{(d)}_{d+1}\big((d-1)/2\,|\,{\bf d}\big)\,,
  \eql{casen2}
  $$
where the upper sign is for Neumann and the lower for Dirichlet conditions.
For odd spheres these are equal which can be seen earlier, and more generally.
From (\peq{zet0}), the residue gives,
  $$
  \ze(-1/2)=-{\pa\over\pa\tau}\, T(\tau)\bigg|_0\,,
  $$
and the only contribution comes from the part of $T(\tau)$ odd under
$\tau\to-\tau$. Using the variable $\si$ it easy to show that the odd
part is,
  $$
  T(\si)-T(1/\si)={\si^{(d-1)/2}(1-\si^2)
  \over|\Ga|}\sum_A{\chi^*(A)\big(1-
  (-1)^d\det(A)\big)\over\det(1-\si A)}\,.
  \eql{teeodd}
  $$

Two conclusions can be drawn from this equation. For even spheres, the
sum reduces to one over the odd, reflective part of $\Ga$. Hence the
Casimir energy is zero on even spheres quotiented by just the rotation
subgroup of $\Ga$. (I only need this for $\chi=1$ but it holds for all
twistings.) By contrast, for odd spheres, the sum runs over only the
rotational part of $\Ga$ which means that the Neumann and Dirichlet
Casimir energies are equal (see below). This implies that on the
hemisphere both of these are half the full sphere value since adding
Neumann and Dirichlet gives the full sphere quantities. The result of
Kennedy and Unwin, [\pref{KandU}], agrees with this but goes further
because it says that the energy {\it densities} are equal.

It is worth noting that another way of writing (\peq{teeodd}) is the
inversion relation,
  $$
   T_N(1/\si)=(-1)^d\,T_D(\si)\,.
   \eql{invers}
  $$
Such relations are used by Stanley [\pref{Stanley}]. In fact
(\peq{invers}) is equivalent to Theorem 8.1 in [\pref{Stanley}] and, in
commutative algebra parlance, implies that the invariant polynomial ring
is Gorenstein.

For $d=3$, (\peq{casen2}) gives,
  $$
  E_0=E_N=E_D={1\over4!\,|\Ga|}\,B^{(3)}_4(1\,|\,d_1,d_2,d_3)\,,
  \eql{casen3}
  $$
where I have used one of the basic properties of the degrees, $|\Ga|=2\prod
d_i$.

The Bernoulli polynomial, $B^{(3)}_4(a| d_1,d_2,d_3)$, and some properties, is
given in Appendix A and leads to the following numbers for the four {\it
extended} polytope groups,
  $$\eqalign{
  E(3,3,3)&=-{601\over28800}\,,\quad
  E(3,3,4)=-{3557\over46080}\cr
  \noalign{\vskip5truept}
  E(3,4,3)&=-{69391\over414720}\,,\quad
  E(3,3,5)=-{3178447\over5184000}\,.\cr
  }
  \eql{casnum}
  $$
It is straightforward to compute $E$ for higher (odd) dimensions.

The hemisphere is an example, ($q=1$), of the lune, the fundamental
domain of the group whose degrees are $(q,1,1,\ldots)$, [\pref{Dow3}].
(\peq{casen3}) gives for the lune of dihedral angle, $\pi/q$,
  $$
  E_q={q^4+5q^2-3\over1440q}\,,
  \eql{lunecas0}
  $$
and so, for the hemisphere, $E_{hs}={1/480}$, half the full sphere value,
as expected.

\section{\bf 5. Scalar and Maxwell Casimir energies by angles}

This approach has been used by ourselves in several previous works
[\pref{DandB,Dow11,Dow12}], [\pref{DandJ,Dow13}] and so I feel I can proceed
rapidly and just write down the \zf\ derived in these papers from the
character expression for the degeneracies after some minor manipulation,
  $$
  \ze(s)={h(j)\over2|\Ga|}
  \sum_{\al,\be}{2\over\cos\be-\cos\al}
  \sumdasht{n=j}{\infty} {1\over n^{2s}}\,\big(\cos n\be\cos j\al-
  \cos n\al\cos j\be\big)\,,
  \eql{zeta1}
  $$
where $h(0)=2$ and $h(1)=1$. The angles $\al$ and $\be$ are defined by,
  $$
  \al=\th_R+\th_L\,,\quad \be=\th_R-\th_L\,,
  $$
where $\th_L$ and $\th_R$ are the rotation angles corresponding to the left
and right SU(2) factors in the rotational part of $\Ga$. As explained
previously, it is sufficient in odd sphere dimensions, to sum over just the
rotation part of $\Ga$. The sum over $\al$ and $\be$ is the sum over this
part.

The identity term $\al=0,\,\be=0$ formally diverges at $s=-1/2$ and has
to be treated separately.
  For the two spins, the identity \zfs\ are,
  $$
  \ze^{id}(s)={h(j)\over|\Ga|}\,\big(\ze_R(2s-2)-j\,\ze_R(2s)
  \big)\,,\quad j=0,1\,,
  \eql{idzet}
  $$
which differs from the full sphere expression only by the $1/|\Ga|$ volume
factor. The non--free action of $\Ga$ means that other group elements have
also to be treated separately. It turns out that it is sufficient to extract
those elements with $\al=0,\be\ne0$ and those with $\th_L=0,\th_R\ne0$. The
first condition means that the element fixes a rotation plane ($2$--flat), the
four--dimensional analogue of a rotation axis in three dimensions. The element
belongs to the isotropy subgroup (the adjoint group) of $\Ga=$ SU(2)$\times$
SU(2), defined by the action $g\to \xi g \xi^{-1}$ on $g\in$ SU(2) which group
leaves fixed only the poles, $+1$ and $-1$, the intersections of all the fixed
S${^1}$'s on S$^3$.

For computational completeness, the contributions of these special group
elements to the Casimir energy are written out,
  $$\eqalign{
  E_{j=0}^{(0,\be)}&={1\over
  4|\Ga|}\bigg({1\over4}\cosec^4\be/2-{1\over12}\cosec^2\be/2\bigg)\cr
  E_{j=1}^{(0,\be)}&=2E_{j=0}^{(0,\be)}+{1\over12|\Ga|}\cr
  E_j^{(\be,\be)}&={h(j)\over2}\bigg(-{1\over4}\cosec^4\be/2
  +\de_{j1}\cosec^2\be/2\bigg)\,.
  }
  $$

The next step is to find, for each polytope group, the conjugacy classes with
their orders, sizes and the values of $\al$ and $\be$. The group is then
decomposed into these classes and the group sum over $\al$ and $\be$ performed
class by class. The possible classes have been determined by Hurley,
[\pref{Hurley}], in his computation of the crystal classes in four dimensions.
Using his notation the class decompositions of the rotation groups are,
[\pref{Chang}],
  $$\eqalign{
  \{3,3,3\}\,=\,&I\oplus15E\oplus20K\oplus(2\times 12)L'\cr
  \{3,3,4\}\,=\,&I\oplus
  I'\oplus(2\times24)A\oplus(2\times6)D\oplus(6+12+24)E\cr
  &\oplus32(K\oplus  K')\oplus12(R\oplus R')\cr
  \{3,4,3\}\,=\,&I\oplus I'\oplus(2\times
  72)A\oplus(2\times48)C\oplus(2\times6)D\oplus(18+72)E\cr
  &\oplus(2\times32)(K\oplus K')
  \oplus36(R\oplus R')\oplus(2\times8)(S\oplus
  S')\cr
  \{3,3,5\}\,=\,&I\oplus I'\oplus(2\times
  600)C\oplus(2\times30)D\oplus450E\oplus 400(K\oplus K')\cr
  &\oplus(2\times12)(W\oplus W'\oplus X\oplus X')
  \oplus(2\times 240)(Y\oplus
  Y'\oplus Z\oplus Z')\cr
  &\oplus(2\times360)(a\oplus b)\,.
  }
  \eql{classdecom}
  $$
Most of this information can be found in Hurley, [\pref{Hurley}], except that
for $\{3,3,5\}$, which is not crystallographic and so the new classes
$U,V,W,X,Y,Z$ and their dashed counterparts with opposite trace have been
introduced, as well as classes $a,b$.

The classes occurring in the above decompositions are given below, the
notation being that $C(\chi,\si,\det, \al/\pi,\be/\pi)$ refers to class `$C$'
with invariants defined by,
  $$
  \det(\la 1-A)=\la^4-\chi(A)\la^3+\si(A)\la^2
  -\chi(A)\la+\det A\,,\quad A\in
  C\subset\Ga\,,
  $$
and the angles $\al$ and $\be$ have already been defined. Hurley does not
give the values of $\al$ and $\be$, see [\pref{Chang}].
  $$\eqalign{
  &I(4,6,1,0,0),\,I'(4,6,-1,1,1),\cr
  &A(0,0,1,1/4,3/4),\,C(0,-1,1,1/6,5/6),\,
  D(0,2,1,1/2,1/2),\,E(0,-2,1,0,1),\cr
  &K(1,0,1,0,2/3)\,,K'(-1,0,1,1,1/3)\,,L(1,1,1,1/5,3/5)
  ,\,L'(-1,1,1,4/5,2/5),\cr
  &R(2,2,1,0,1/2),\,R'(-2,2,1,1,1/2),
  \,S(2,3,1,1/3,1/3),\,S'(-2,3,1,2/3,2/3)\,.
  }
  $$

Hurley also gives the decompositions of the extended polytope groups. For
comparison I just exhibit that for $\{3^3\}$, of order 120, from his
Table 2.b,
  $$
  \{3,3,3\}=1\oplus15E\oplus24L'\oplus30F\oplus20N\oplus10T'\,.
  $$
The last three classes form the odd part of the group and, perforce,
contribute nothing to the scalar Casimir energy, a fact not obvious in
this approach.

The calculation of the scalar Casimir energies  via these angle forms
yields the values obtained before, (\peq{casnum}), as should be.

It turns out that the odd group part goes out for spin--one as it does
for spin-zero. Essentially this is owing to a cancellation between ${\bf
E}$ and ${\bf H}$ and will be detailed later. Then the Maxwell Casimir
values computed again via (\peq{casen1}), (\peq{zeta1}) and (\peq{idzet})
are,
  $$\eqalign{
  E_1(3,3,3)&={2639\over14400}\,,\quad
  E_1(3,3,4)={791\over2880}\cr
  \noalign{\vskip5truept}
  E_1(3,4,3)&={10453\over25920}\,,\quad
  E_1(3,3,5)={309421\over324000}\,.\cr
  }
  \eql{casnum2}
  $$

Incidentally, it is possible to give a simple check of the numbers in the
decomposition (\peq{classdecom}) which I exemplify in the $\{3^3\}$ case.
Define the Shephard--Todd numbers by saying that $b_r$ elements of $\Ga$
fix a $(d+1-r)$--flat but no flat of higher dimensions. They are related
to the degrees by Solomon's theorem,
  $$
   \prod_{i=1}^{d+1}\big(1+(d_i-1)t\big)=\sum_{r=0}^{d+1}b_r\,t^r\,.
  $$
The degrees are algebraic while the Shephard--Todd numbers are geometric.

For $\{3^3\}$, $b_0=1,b_1=10,b_2=35,b_3=50,b_4=24$, $d_1=3,d_2=4,d_3=5$
and these values agree with the class decomposition since the elements
with $\al=0,\be\ne0$ total $15+20=35$ and those with $\al\ne0,\be\ne0$
total 24 which is the number of those elements that fix a point and
nothing else. Only even $b$'s are relevant for the rotation group.

I just record the $(b_1,b_2,b_3,b_4)$ for the other cases:
$(16,86,176,105)$ for $\{3,3,4\}$; $(24,190,552,385)$ for $\{3,4,3\}$;
$(60,1138,7140,6061)$ for $\{3,3,5\}$.

\section{\bf 6. Invariant $p$--form theory}

As a prelude to Maxwell theory, I will look at $p$--forms, the eigenproblem
for which on spheres is quite standard. One way of extending this to
tessellations is to adapt the technique of Gallot and Meyer, [\pref{GandM}],
which is the form analogue of the classic $\oR^n$ embedding harmonic
polynomial method with the coefficients of the ambient forms (in a local
cartesian basis) being homogeneous harmonic polynomials. Finite O(4) subgroup
invariant theory can then be applied to these coefficients. If the full $\Ga$
is employed, the fundamental domain has a boundary and the form is
conventionally chosen to satisfy absolute or relative boundary conditions, as
will be clarified later.

The extension of Molien's theorem to forms is used in the basic paper by
Solomon, [\pref{Solomon}]. Flatto, [\pref{Flatto}], gives a summary. Although
I am most concerned with $d=3$, in order to render the discussion a little
more general, I will choose at the appropriate moment, for odd $d$, the form
order $p=(d-1)/2$, so that the eigenvalues of the de Rham Laplacian for {\it
coexact} $p$--forms on the sphere are again perfect squares, \cf\
[\pref{Dow8}], [\pref{CandH}],
 $$\mu(p,l)= (l+p+1)^2\,,\quad l=0,1,\ldots\,\,,
 \eql{ceig1}
 $$
the general expression being,
  $$
  \la^{CE}(p,l)=\big(l+(d+1)/2\big)^2-\big((d-1)/2-p\big)^2\,,
  \quad l=0,1,\ldots\,\,.
  \eql{ceig2}
  $$
The eigenvalues on the factored sphere are the same, only the
degeneracies change.

A $p$--form,
  $$
  \al=\sum_{\mu_1<\mu_2\ldots\atop1}^n
   \overline a_{\mu_1\ldots\mu_p}(x)\,dx^{\mu_1}\wedge\ldots\wedge
  dx^{\mu_p}\,,
  \eql{pform1}
  $$
is said to be invariant under $\Ga\subset$ O($n$), if for $\ga\in\Ga$
acting on $x^1,\ldots,x^n$ (written $\ga x$),
  $$
  \al=\ga\al\equiv\sum \overline a_{\mu_1\ldots\mu_p}(\ga x)\,
  dx^{\mu_1}(\ga^{-1}x)\wedge\ldots\wedge dx^{\mu_p}(\ga^{-1}x)\,,\quad
  \forall\ga\,,
  $$
or
  $$
  \overline a_{\mu_1\ldots\mu_p}(\ga x)
  =\overline a_{\nu_1\ldots\nu_p}(x)\,
  A^{\nu_1}_{\mu_1}(\ga)\ldots A^{\nu_p}_{\mu_p}(\ga)\,,
  $$
where $A(\ga)$ is the $n\times n$ fundamental (vector) representation of
$\ga$.

A twisting, corresponding to (\peq{genfun2}), could be introduced by
requiring $\al=\chi(\ga)\,\ga\al$. I will be concerned only with the
trivial and the sign twistings, $\chi(\ga)=1$ and $\chi(\ga)=\det\ga$. In
the latter case the forms are sometimes referred to as anti-invariant.

When $\Ga$ is just $\{$id$\}$ every function is an invariant function. In
particular {\it every} homogeneous polynomial is invariant. This is the
situation discussed by Gallot and Meyer, [\pref{GandM}], appropriate for
the full sphere embedded in $\oR^n$. In the following I will sometimes
refer to this as the `full sphere' or `trivial' case.

For a general reflective group, the role of the cartesian coordinates is
played by the polynomials, $(I^1,\ldots,I^n)$, of the invariant basis and it
is a theorem that any invariant $p$--form, (\peq{pform1}), can be written as,
  $$
  \al=\sum_{\{\mu\}} a_{\mu_1\ldots\mu_p}\,dI^{\mu_1}\wedge\ldots\wedge
  dI^{\mu_p}\,,
  \eql{pform2}
  $$
where the coefficients are polynomials in $I^1,\ldots,I^n$, written
$a_*\in k[I^1,\ldots,I^n].$ which stands for the ring of polynomials
generated by the $I^i$ over the algebraic field $k$ (here just the real
or complex numbers).

In the standard method, [\pref{GandM}], it is shown that the form $\al$ is
harmonic if the cartesian coefficients, $\overline a_{\{\mu\}}$, are harmonic
scalar functions. This is because $\oR^n$ is flat and the differentials
$dx^\mu$ are parallel transported. The scaling properties of the form are then
used to determine the radial derivatives which arise when the difference
between the ambient and intrinsic Laplacians is computed. The label $`l$' in
(\peq{ceig2}) is the degree of the polynomial coefficients in (\peq{pform1}).
The form also picks up a scaling factor of $`p$' from the differentials in
(\peq{pform1}) (see [\pref{GandM}] \S11.9).

The important fact is that exactly the same holds on the factored sphere,
leading again to (\peq{ceig2}). The only difference is that (\peq{pform1}) has
to take the invariant structure (\peq{pform2}).

Because of this theorem, the differentials,
  $$
  (I^1)^{k_1}\ldots (I^n)^{k_n}\,dI^{i_1}
  \wedge\ldots\wedge dI^{i_p}\,,\quad
  k_i\in\oZ\,,
  \eql{ipbas}
  $$
form a basis for the space of {\it invariant} homogeneous $p$--forms of degree
$l$. One has to remember that it is the cartesian $\overline a_*$ which are
homogeneous degree $l$, not the $a_*$. These coefficients are connected by the
usual tensor relation under change of coordinates and so the degree, $l$, of
the $\overline a_*$ corresponding to the basis (\peq{ipbas}) is easily seen to
be,
  $$
   l = k_1 d_1+\ldots+k_n d_n+ m_{i_1}+\ldots+m_{i_p}\,,\quad m_i=d_i-1\,.
   \eql{deg1}
  $$

A basic ambient, $\oR^n$, result concerns the dimension, $d(p,l)$, of the
vector space of (unrestricted) $p$--forms whose coefficients are homogeneous,
polynomials of degree $l$ invariant under $\Ga$. The generating function is,
[\pref{Flatto}] Theorem 3.16, [\pref{Solomon}], following from (\peq{deg1}),
  $$\eqalign{
  d_a(p,\si)\equiv\sum_{l=0}^\infty d_a(p,l)\,\si^l
  &={e_p\big(\si^{m_1},\ldots,
  \si^{m_n}\big)\over(1-\si^{d_1})\ldots(1-\si^{d_n})}\cr
  &={1\over|\Ga|}\sum_A{e_p\big(\la_1(A),\ldots,\la_n(A)\big)\over
  \det(1-\si A)}\,,
  }
  \eql{pagen1}
  $$
where $e_p(x_1,\ldots,x_n)$ is the $p$-th elementary symmetric function
in $x_1,\ldots,x_n$ and the $\la_i(A)$ are the eigenvalues of $A$. The
$m_i=d_i-1$ are the exponents of the group $\Ga$.

In view of (\peq{invers}), I investigate the behaviour of $T_a(p,\si)\equiv
\si^{n/2}\,d_a(p,\si)$ under $\si\to1/\si$. A simple calculation reveals that,
  $$
  T_a(p,1/\si)=(-1)^n\,T_a(n-p,\si)\,,
  \eql{pdual1}
  $$
which is seen to be an ambient duality statement and can be re--expressed
by defining the relative generating function of forms anti--invariant
under $\Ga$,
  $$
 d_r(p,\si)\equiv{1\over|\Ga|}\sum_A{e_p
  \big(\la_1(A),\ldots,\la_n(A)\big)\,\det A\over\det(1-\si A)}
  =\sum_{l=0}^\infty d_r(p,l)\,\si^l\,,
  \eql{prgen1}
  $$
with also
  $$
  T_r(p,1/\si)=(-1)^n\,T_r(n-p,\si)\,.
  \eql{pdual2}
  $$
Equations (\peq{pdual1}) and (\peq{pdual2}) are true generally. For reflective
groups, more particularly, since $\la_i=1/\la_i$, the invariant and
anti--invariant quantities can be related,
  $$
  T_r(p,\si)=(-1)^nT_a(p,1/\si)=T_a(n-p,\si)\,.
  \eql{pdual3}
  $$
In the trivial case I note the trivial fact that $T_r=T_a$ and the subscript
can be omitted.

As a special case I can check an earlier remark concerning the Dirichlet
function by setting $p$ equal to $n$ in (\peq{pagen1}) to obtain,
  $$
  d_a(n,\si)=\si^{ m_1+m_2+\ldots+m_n}\, d_a(0,\si)\,,
  \eql{drel}
  $$
which is just (\peq{genfund}) since $\sum_i m_i=d_0$.

The conclusion is that an {\it invariant} $n$--form corresponds to an {\it
anti--invariant} $0$--form \ie\ a pseudoscalar, as is well known.

One can extend this, again to a well known statement, that, by duality an
invariant $p$--form corresponds to an anti-invariant $n-p$--form. The boundary
conditions also switch. This will be encountered later. It will turn out that
invariant and anti--invariant forms correspond to forms obeying absolute and
relative boundary conditions, respectively. The notation used above reflects
this.
\section {\bf 7. $p$--form degeneracies}
The degeneracies on the sphere have been computed by Ikeda and Taniguchi,
[\pref{IandT}], and Iwasaki and Katase, [\pref{IandK}] and I will start with
this `trivial' case. Another interesting discussion is provided by Weck and
Witsch, [\pref{WandWi}] esp. \S3 and appendix, using ordinary tensor
analysis, who derive corresponding results and compute the various
degeneracies. The work of Paquet, [\pref{Paquet}] is a useful elaboration and
part correction of [\pref{GandM}]. Some other relevant discussions of
$p$--forms on spheres and on particular manifolds with boundary, such as the
ball, are in Vassilevich ,[\pref{Vass}], [\pref{Vass2}], Elizalde {\it et
al}, [\pref{ELV}],[\pref{ELV2}], Copeland and Toms, [\pref{CandT}], Camporesi
and Higuchi [\pref{CandH}], Cappelli and d'Appollionio, [\pref{CandA}], from
quantum field theory angles. See also Rubin and Ordonez, [\pref{RandO}].

By restricting $A$ to be just the unit matrix, one gets the trivial
expression,
  $$
  d(p,\si)=\comb{d+1}p{1\over(1-\si)^{d+1}}\,,
  \eql{genun}
  $$
and so (or otherwise),
   $$
  d(p,l)\equiv\dim P^p_l=\comb {d+l}l\comb{d+1}p\,.
  \eql{fdeg}
  $$

In this section I evaluate the various mode degeneracies, like (\peq{fdeg}),
on the spherical factors. In the following section, I  give the equivalent
generating functions, like (\peq{genun}), which are more useful to me. The
reason for this doubling of effort is transparency, to make contact with other
calculations and to act as a check.

I define $\caH^p_l$, the set of harmonic, homogeneous $p$--forms of degree
$l$, and also $H^p_l$, the similar module of {\it coclosed}, harmonic forms,
  $$
  H^p_l=\caH^p_l\cap\ker\overline \de\,,
  $$
in terms of which the required closed degeneracy on the sphere is,
[\pref{IandK}], [\pref{IandT}],
  $$
   \dim\big(H^p_l\cap\ker\overline d\big)\,,
  $$
as brought up later. Here, $\ol d$ and $\ol \de$ are the ambient derivatives.

It can be shown, in the present polynomial case on $\oR^n$, that closed
implies exact, unless $l+p=0$, and coclosed coexact, unless $n-p+l=0$ , and so
there exists the exact sequence generated by $\overline d$,
  $$
  0\longrightarrow P^0_{l+p}\OverArrow {\ol d} 2\ldots
  \OverArrow {\ol d} 2P^{p-1}_{l+1}\OverArrow {\ol d} 2
  P^p_l\OverArrow {\ol d} 2
  P^{p+1}_{l-1}\OverArrow {\ol d} 2\ldots
  \OverArrow {\ol d} 2 P^{p+l}_0\longrightarrow0\,.
 \eql{pexseq}$$

Restricting to the harmonic sub--modules in (\peq{pexseq}) gives the
exact sequences,
 $$
  0\longrightarrow \caH^0_{l+p}\OverArrow {\ol d} 2\ldots
  \OverArrow {\ol d} 2\caH^{p-1}_{l+1}\OverArrow {\ol d} 2
  \caH^p_l\OverArrow {\ol d} 2
  \caH^{p+1}_{l-1}\OverArrow {\ol d} 2\ldots
  \OverArrow {\ol d} 2 \caH^{p+l}_0\longrightarrow0
 \eql{dexseq}
 $$
and
  $$
  0\longrightarrow H^0_{l+p}\OverArrow {\ol d} 2\ldots
  \OverArrow {\ol d} 2H^{p-1}_{l+1}\OverArrow {\ol d} 2
  H^p_l\OverArrow {\ol d} 2
  H^{p+1}_{l-1}\OverArrow {\ol d} 2\ldots
  \OverArrow {\ol d} 2 H^{p+l}_0\longrightarrow0\,.
 \eql{dexseq2}
 $$

I treat these sequences in turn. All give recursions of the same form.
Sequence (\peq{pexseq}) yields the relations for the unrestricted
degeneracies,
  $$
  \dim P^p_l=\dim\big(P^p_l\cap\ker\ol d\big)+
  \dim\big(P^{p+1}_{l-1}\cap\ker\ol d\big)\,,
  $$
written,
  $$
  d_b(p,l)=d_b^C(p,l)+d_b^C(p+1,l-1)\,.
  \eql{recurs3}
  $$
The recursion can be solved by either increasing or decreasing $p$. The
first route gives,
  $$
  d_b^C(p,l)=\sum_{j=1}^p(-1)^{j-1}\,d_b(p-j,l+j)
  +\de_{ba}\de_{p0}\, \de_{l0}\,,
  \eql{dclo}
  $$
where the final term arises from the $p=0$ zero mode (a $0$--form is
automatically coclosed and, if closed, it is harmonic and constant,
$l=0$, $\grad\phi=0$). It has to be added in by hand since the exact
sequence doesn't cover this case. It exists only for invariant functions.
In boundary condition language, invariant functions give Neumann
conditions on the fundamental domain and anti--invariant ones, Dirichlet
and there is no zero mode in the latter case.

I have included a subscript, $b$, to indicate the invariance, $b=a$, or
anti--invariance, $b=r$, of the forms. Being duals, I will set $*a=r$ and
$*r=a$. This extension of the trivial situation is allowed because the exact
sequences are graded by the periodicity type. This follows from the invariant
form, (\peq{ipbas}), and its anti--invariant partner. A simple direct
demonstration is given in section 8.

The `higher' part of the exact sequence leads to the equivalent formula,
  $$
  d_b^C(p,l)=\sum_{j=0}^{n-p}(-1)^j\,d_b(p+j,l-j)\,,
  \eql{dclo1}
  $$
and thence the identity, (either equate (\peq{dclo}) and (\peq{dclo1}) or
set $p=0$ in these equations),
   $$
  \sum_{j=0}^n(-1)^j\,d_b(j,k-j)=\de_{ba}\de_{k0}\,.
  \eql{dclo4}
  $$
This can be checked for the trivial case (\peq{fdeg}),
  $$
  \sum_{j=0}^n (-1)^j\comb{n+k-j-1}{k-j}\comb{n}{j}=\comb{k-1}k\,,
  \eql{tr1}
  $$
which is zero if $k>0$. However, when $k=0$ the summand on the left--hand
side is zero unless $j=0$ when it is, unambiguously,
  $$
  \comb{n-1}0 \comb{n}0=1\,.
  $$
This defines the right--hand side. I have used the identity,
  $$
  \comb{D-n} k =\sum_{j=0}^n (-1)^j\comb nj\comb{D-j}{k-j}=
  \sum_{j=0}^n (-1)^j\comb nj \comb{D-j}{D-k}\,,
  $$
with $D=n+k-1$ and which follows either by recursion, or from the
equation,
  $$
  (1+x)^{D-n}=(1+x)^D\bigg(1-{x\over1+x}\bigg)^n\,.
  $$

We have the end values from (\peq{dclo}) and (\peq{dclo1}),
  $$
  d_b^C(0,l)=\de_{ba}\de_{l0},\,\quad d_b^C(n,l)=d_b(n,l)\,.
  \eql{endv}
  $$

A general check of the identity, (\peq{dclo4}), not just for the trivial case
is given later using the generating function version of the above which is
somewhat neater.

For the next sequence, (\peq{dexseq}), the dimensions are related by,
  $$
  \dim\caH^p_l=\dim\big(\caH^p_l\cap \ker\ol d\big) +
  \dim\big(\caH^{p+1}_{l-1}\cap \ker\ol d\big)\,,
  \eql{recurs}
  $$
which I write,
  $$
  h_b(p,l)=h_b^C(p,l)+h_b^C(p+1,l-1)\,.
  \eql{recurs1}
  $$

As before, this can be solved by either increasing or decreasing $p$. The
first way gives,
  $$
  h_b^C(p,l)=\sum_{j=1}^p(-1)^{j-1}\,h_b(p-j,l+j)
  +\de_{ba}\de_{p0}\de_{l0}\,,
  \eql{hclo}
  $$
where the last term again arises from the effect of the $p=0$ mode.

The other route gives,
  $$
  h_b^C(p,l)=\sum_{j=0}^{n-p}(-1)^j\,h_b(p+j,l-j)+
  \de_{ba}\de_{p0}\,\de_{l2}\,.
  \eql{hclo4}
  $$
The last term is added to give the required value \footnote{ If this term is
not added, one obtains the spurious value $h^C_a(0,2)=-1$. Indeed, calculation
of \eg\ $\dim H^2_2$ on $\oR^2$ according to the formula at the bottom of
p.143 of [\pref{IandK}] gives $-1$. In my notation this corresponds to
$h^{CC}_r(2,2)$ (see later) {\it without} the addition, which equals the
uncorrected $h^C_a(0,2)$ by duality.},

  $$
  h_b^C(0,l)=\de_{l0}\,.
  $$

Equating these two expressions, or setting $p=0$, gives the identity,
  $$
   \sum_{j=0}^n(-1)^j h_b(j,l-j)=\de_{ba}(\de_{l0}-\de_{l2})\,.
   \eql{hid}
  $$

$h^C$ is not what I want however. In the trivial case, the degeneracy,
$h^C(p,l)$, for closed harmonic $p$--forms on $\oR^n$, is not the degeneracy
on the sphere. For an ambient $\oR^n$ form, $\om$, there is the collar split
around the sphere,
  $$
  \om=\om_1+dr\wedge\om_2\,,
  $$
where $\om_1$ and $\om_2$ are ambient forms of orders $p$ and $p-1$
respectively.

If $\om$ is ambiently closed, the `tangential' part $\om_1$ is closed on
the sphere, as is well known and easily checked. Projecting $\om$ to the
sphere by the inclusion, $i^*$, leaves just $\om_1$ and so the degrees of
freedom in the $\om_2$ have to be removed from $h^C$, which counts all the
closed harmonic $\om$ on $\oR^n$. The easiest way of doing this is,
[\pref{IandT}], to use the harmonic coclosed module, $H^p_l$.

Define $V^p_\la$ as the subspace of the space of differential forms {\it on
the sphere} S$^d$ consisting of eigenforms associated with each eigenvalue
$\la$ of the Laplacian on the sphere. Then there exists the isomorphism
between $\oR^{d+1}$ and S$^d$,
  $$
  H^p_l\cap\ker\overline d \longleftrightarrow V^p_\la\cap\ker d\,,
  \eql{isom1}
  $$
where $\la=(l+p)(d-p+l+1)$. The dimension of the right--hand side is the
dimension of the left and is the required closed degeneracy on the sphere.

Turning last to the sequence, (\peq{dexseq2}), the dimensions are related by,
  $$
  \dim H^p_l=\dim\big(H^p_l\cap \ker\ol d\big) +
  \dim\big(H^{p+1}_{l-1}\cap \ker\ol d\big)\,,
  \eql{recursc}
  $$
or,
  $$
  h_b^{CC}(p,l)=h_b^{CCC}(p,l)+h_b^{CCC}(p+1,l-1)\,.
  \eql{recursc1}
  $$
$h_b^{CCC}$ is the final quantity I want and the recursion can be solved
as before. The low road gives
  $$
  h_b^{CCC}(p,l)=\sum_{j=1}^p(-1)^{j-1}\,h_b^{CC}(p-j,l+j)
  +\de_{ba}\de_{p0}\de_{l0}+\de_{br}\de_{pn}\de_{l0}\,,
  \eql{hcloc}
  $$
and the high road,
  $$
  h_b^{CCC}(p,l)=\sum_{j=0}^{n-p}(-1)^j\,h_b^{CC}(p+j,l-j)\,.
  \eql{hcloc2}
  $$
Duality on $\oR^n$, implies
 $$
  h_b^{CC}(p,l)=h_{*b}^C(n-p,l)\,,
  \eql{ndual}
  $$
and so, from (\peq{hclo}),
  $$\eqalign{
  h_b^{CC}(p,l)&=\sum_{j=1}^{n-p}(-1)^{j-1}\,h_{*b}(n-p-j,l+j)
  +\de_{br}\de_{n-p,0}\,\de_{l0}\cr
  &=\sum_{j=1}^{n-p}(-1)^{j-1}\,h_b(p+j,l+j)
  +\de_{br}\de_{n-p,0}\,\de_{l0}\,.
  }
  \eql{hclod}
  $$
Equivalently, again using duality on $\oR^n$, from (\peq{hclo4}),
  $$\eqalign{
  h_b^{CC}(p,l)&=\sum_{j=0}^{p}(-1)^j\,h_{*b}(n-p+j,l-j)
  +\de_{br}\de_{n-p,0}\cr
  &=\sum_{j=0}^{p}(-1)^j\,h_{b}(p-j,l-j)
  +\de_{br}\de_{n-p,0}\,.
  }
  \eql{hclod3}
  $$

As usual, setting $p=0$ in (\peq{hcloc}) and (\peq{hcloc2}) gives an
identity
  $$
  \sum_{j=0}^{n}(-1)^j\,h_b^{CC}(j,l-j)=\de_{l0}\,\de_{ba}\,.
  \eql{hcloc3}
  $$

Thus the problem of finding $h^{CCC}$ is reduced to finding the dimension,
$h(p,l)$, of the {\it harmonic} $p$--form space on $\oR^n$. This follows
easily in the trivial case since each form coefficient is a scalar so that the
total number of harmonic forms is just the dimension of the $p$--form
multiplied by the number of harmonic homogeneous scalar polynomials which has
already been found. Therefore, in the trivial, sphere case,
  $$\eqalign{
  h(p,l)&=\comb{d+1}p\bigg[\comb{d+l}l-\comb{d+l-2}{l-2}\bigg]\cr
  \noalign {\vskip 5truept}
  &=d(p,l)-d(p,l-2)\,,
  }
  \eql{hun1}
  $$
and this extends to,
  $$\eqalign{
  h_b(p,l)&= d_b(p,l)-d_b(p,l-2)\cr
        &=h_{*b}(n-p,l)\,,
  }
  \eql{hun2}
  $$
in general. This is confirmed in Appendix B. The $p=0$ case is discussed
by B\'erard and Besson, [\pref{BandB}].

The solutions (\peq{hclo}) and (\peq{hclo4}) become,
  $$
  h_b^C(p,l)=\sum_{j=1}^p(-1)^{j-1}\big(d_b(p-j,l+j)
  -d_b(p-j,l+j-2)\big)
  +\de_{ba}\de_{p0}\,\de_{l0}\,,
  \eql{hclot1}
  $$
and
  $$
  h_b^C(p,l)=\sum_{j=0}^{n-p}(-1)^j\big(d_b(p+j,l-j)-d_b(p+j,l-j-2)\big)
  +\de_{ba}\de_{p0}\,\de_{l2}\,.
  \eql{hclot2}
  $$

Everything is now in place to compute $h^{CCC}$. I am not interested in
producing the standard expression for the degeneracy on the sphere,
[\pref{IandK}], only in numerical evaluation, which shows agreement in this
trivial case using (\peq{hun1}). For practical purposes I prefer to use
generating functions.
\section{\bf 7. p-form generating functions}

I now construct the generating functions and start with the unrestricted
degeneracies, $d(p,l)$. Firstly the recursion ({\peq{recurs3}), which
holds for the $d$'s, becomes
  $$
  d_b(p,\si)=d_b^C(p,\si)+\si\,d_b^C(p+1,\si)\,.
  \eql{drecurs}
  $$

We can now implement the recursion from this and compare with what the
solutions (\peq{dclo}) and (\peq{dclo1}) give. From these two equations I
find
  $$\eqalign{
  d_b^C(p,\si)&=\sum_{j=1}^p(-1)^{j-1}\sum_{l=0}^\infty d_b(p-j,l+j)\si^l
  +\de_{ba}\sum_{l=0}^\infty\de_{p+l,0}\si^l\cr
  &=\sum_{j=1}^p(-1)^{j-1}\si^{-j}\sum_{l=j}^\infty d_b(p-j,l)\si^l
  +\de_{ba}\de_{p0}\cr
  &=\sum_{j=1}^p(-1)^{j-1}\si^{-j}\bigg(d_b(p-j,\si)-\sum_{m=0}^{j-1}
  {\si^m\over m!}\,d_b^{(m)}(p-j,\si)\bigg|_0\bigg)
  +\de_{ba}\de_{p0}\,,\cr
  }
  \eql{dclo2}
  $$
and
  $$\eqalign{
  d_b^C(p,\si)&=\sum_{j=0}^{n-p}\sum_{l=0}^\infty
  (-1)^j\,d_b(p+j,l-j)\si^l\cr
  &=\sum_{j=0}^{n-p}
  (-1)^j\si^j\sum_{l=0}^\infty\,d_b(p+j,l-j)\si^{l-j}\cr
  &=\sum_{j=0}^{n-p}
  (-1)^j\si^j\sum_{l=0}^\infty\,d_b(p+j,l)\si^{l}\cr
  &=\sum_{j=0}^{n-p}
  (-1)^j\si^j\,d_b(p+j,\si)\,,\cr
  }
  \eql{dclo7}
  $$
where, in the step from the second to third line, the fact that
$d_b(p,l)=0$ for $l<0$ has been used. These equations follow, of course,
by direct solution of the generating function recursion, (\peq{drecurs}).

The identity (\peq{dclo4}) becomes, or set $p=0$ in (\peq{dclo7}),
  $$
  \sum_{j=0}^n(-1)^j\si^j\,d_b(j,\si)=\de_{ba}\,,
  \eql{dclo5}
  $$
which again can be checked directly. The general invariant situation is,
  $$
{1\over (1-\si^{d_1})\ldots(1-\si^{d_{n}})}
\sum_{j=0}^{n}(-1)^j\si^j\,e_j\big(\si^{m_1},\ldots,
  \si^{m_{n}}\big)=1\,,
  \eql{invs}
  $$
by a basic identity. The general anti--invariant case is,
  $$\eqalign{
  &{1\over (1-\si^{d_1})\ldots(1-\si^{d_{n}})}
  \sum_{j=0}^{n}(-1)^j\si^j\,e_{n-j}\big(\si^{m_1},\ldots,
  \si^{m_{n}}\big)\cr
  &=(-1)^n{(\si-\si^{m_1})\ldots(\si-\si^{m_n})
  \over (1-\si^{d_1})\ldots(1-\si^{d_{n}})}\,,\cr
  &=0\,,
  }
  \eql{ainv}
  $$
since, apart from the trivial case, $m_n=1$. These results confirm
(\peq{dclo5}) in general.

The same equations hold for the various harmonic modules. For the purely
harmonic one, $\caH^p_l$, from (\peq{hclo}),
  $$\eqalign{
  h_b^C(p,\si)
  =\sum_{j=1}^p(-1)^{j-1}\si^{-j}\bigg(h_b(p-j,\si)-\sum_{m=0}^{j-1}&
  {\si^m\over m!}\,h_b^{(m)}(p-j,\si)\bigg|_0\bigg)\cr
  &+\de_{ba}\de_{p0}\,,\cr
  }
  \eql{hclo9}
  $$
(the final term reflects the zero mode) and
  $$\eqalign{
  h_b^C(p,\si)&=\sum_{j=0}^{n-p}
  (-1)^j\si^j\,h_b(p+j,\si)\cr
  }
  \eql{hclo10}
  $$
where, as before, the fact that $h(p,l)=0$ for $l<0$ has been used.

For the harmonic coclosed module, $H^p_l$,
  $$\eqalign{
  h_b^{CCC}(p,\si)
  &=\sum_{j=1}^p(-1)^{j-1}\si^{-j}\bigg(h_b^{CC}(p-j,\si)-\sum_{m=0}^{j-1}
  {\si^m\over m!}\,h_b^{CC(m)}(p-j,\si)\bigg|_0\bigg)\cr
  &\hspace{***************************}+\de_{ba}\de_{p0}+\de_{br}\de_{pn}\,,
  }
  \eql{hcloc5}
  $$
(the final term reflects the zero modes) and
  $$\eqalign{
  h_b^{CCC}(p,\si)
  &=\sum_{j=0}^{n-p}
  (-1)^j\si^j\,h_b^{CC}(p+j,\si)\,.\cr
  }
  \eql{hcloc4}
  $$
The relation (\peq{hclod}) between $h^{CC}$ and $h$ is,
  $$\eqalign{
  h_b^{CC}(p,\si)=\sum_{j=1}^{n-p}(-1)^{j-1}
  \si^{-j}\bigg(h_{*b}(n-p-j,\si)-\sum_{m=0}^{j-1}
  {\si^m\over m!}&\,h_{*b}^{(m)}(n-p-j,\si)\bigg|_0\bigg)\cr
  &+\de_{br}\de_{n-p,0}\cr
   =\sum_{j=1}^{n-p}(-1)^{j-1}
  \si^{-j}\bigg(h_{b}(p+j,\si)-\sum_{m=0}^{j-1}
  {\si^m\over m!}&\,h_{b}^{(m)}(p+j,\si)\bigg|_0\bigg)
  \cr
  &+\de_{br}\de_{n-p,0}\,,
  }
  \eql{hclod2}
  $$
or, equivalently, from (\peq{hclod3}),
  $$\eqalign{
  h_b^{CC}(p,\si)&=\sum_{j=0}^{p}(-1)^j\si^j\,h_{*b}(n-p+j,\si)
  +\si^2\de_{br}\de_{n-p,0}\cr
  &=\sum_{j=0}^{p}(-1)^j\si^j\,h_{b}(p-j,\si)
  +\si^2\de_{br}\de_{n-p,0}\,.
  }
  \eql{hclod4}
  $$

Finally, from (\peq{hun1}), the harmonic generating function is
  $$
  h_b(p,\si)\equiv \sum_{l=0}^\infty h_b(p,l)\,\si^l
  =(1-\si^2)\,d_b(p,\si)\,.
  \eql{harmgen3}
  $$
The identity (\peq{hid}) becomes
  $$
  \sum_{j=0}^n(-1)^j\si^j\,h_b(j,\si)=\de_{ba}(1-\si^2)\,.
  \eql{hid2}
  $$
which, in view of (\peq{harmgen3}) is the same as (\peq{dclo5}). For
(\peq{harmgen3}), one has therefore, in general,
  $$
  h_a(p,\si)
  ={1-\si^2\over(1-\si^{d_1})\ldots(1-\si^{d_n})}
  \,\,e_{p}\big(\si^{m_1},\ldots,
  \si^{m_n}\big)\,,
  \eql{harmgen2}
  $$
and
  $$
  h_r(p,\si)
  ={1-\si^2\over(1-\si^{d_1})\ldots(1-\si^{d_n})}
  \,\,e_{n-p}\big(\si^{m_1},\ldots,
  \si^{m_n}\big)\,.
  \eql{harmgen2a}
  $$
Except in the trivial case, the numerator will cancel against the last
factor in the denominator. Technically, it is to use these expressions
that the formalism has been written in terms of generating functions.

In particular for the `hemisphere', a useful special case, the generating
functions are, from (\peq{pagen1}) and (\peq{prgen1}),
  $$\eqalign{
  h_a^{hemisphere}(p,\si)&={1\over(1-\si)^d}
  \bigg[\si\comb{d}{p-1}+\comb  dp\bigg]\cr
  h^{hemisphere}_r(p,\si)&={1\over(1-\si)^d}
    \bigg[\comb{d}{p-1}+\si\comb dp\bigg]\,.\cr
  }
  \eql{hsdeg}
  $$
As a check, adding these gives (\peq{genun}).

\section{\bf 8. Boundary conditions}
Before proceeding with the calculation, the question of boundary conditions
arises more acutely. This is related to the conditions imposed on the form
under the action of $\Ga$. I have considered both invariant and
anti--invariant forms \ie\ those twisted by $\det\ga$. As has been mentioned,
for functions ($0$--forms), these two types correspond to Neumann and
Dirichlet boundary conditions. In the $p$--form case they give {\it absolute}
and {\it relative} conditions, [\pref{Gilkey}], [\pref{Gilkey2}]. This can be
seen as follows. At a frontier hypersurface of a fundamental domain choose a
boundary adapted coordinate system either in $\oR^n$ or in S$^d$, it's not too
important which. Let's select $\oR^n$. It is easily seen that, under the
reflection in the frontier, for an invariant form the tangential components
don't change sign, while the normal ones do, and conversely for the
anti--invariant case. At the boundary therefore, either the normal components
vanish or the tangential ones do. This is the definition of absolute and
relative and I can say that absolute $=$ invariant, relative =
anti--invariant.

In the specific physical case of the Maxwell field, ${\bf E}$ is absolute and
${\bf H}$ is relative. The two contributions to spectral quantities have to be
combined. Looking back to the relevant generating functions, (\peq{pagen1})
and (\peq{prgen1}), this removes the indirect elements of $\Ga$, leaving, as
advertised, just the rotation part. Another way of deducing this is via
duality on $\oR^4$ as discussed in the next section.

Incidentally, and somewhat surprisingly, anti--invariant $p$--form theory is
not so straightforward as the invariant one and seems to be of somewhat
specialist interest. The problem is to find the basis corresponding to the
invariant one, (\peq{ipbas}). For functions, the anti--invariant polynomial
algebra is obtained from the invariant one simply by multiplying by the
Jacobian skew polynomial, $J$, of degree $d_0$ leading to (\peq{genfund}). For
$p$--forms there exists, [\pref{Solomon2}], a construction of basis
anti--invariant $1$--forms, a summary of which, for a more general situation,
can be found in Solomon and Terao [\pref{SandT}] \S1, 6, and extensive
geometric information in Orlik and Terao, [\pref{OandT}]. See also
[\pref{Shepler}], \S4. However, since I am interested only in the generating
functions, I can use, and have done, the isomorphism between the space of
invariant $p$--forms and the space of anti--invariant $(n-p)$--forms (see
[\pref{SandT}] (6.6) for $p=1$).

This seems the appropriate place to confirm that the boundary conditions
filter through the exact sequences (\peq{pexseq}), (\peq{dexseq}) and
(\peq{dexseq2}).

It is sufficient to concentrate on one reflective hyperplane which, for
convenience, I choose to be $x_n=0$. Consider the $p$--form $\al$,
  $$
  \al=\sum_{\mu_1<\mu_2\ldots\atop1}^n
   \al_{\mu_1\ldots\mu_p}(x)\,dx^{\mu_1}\wedge\ldots\wedge
  dx^{\mu_p}\,.
  \eql{pform3}
  $$
Separate the coordinates\footnote{ It is easy to transcribe this discussion
into the more elegant boundary adapted collar split, but I don't need such
generality.} according to the index split, $\mu=(i,n)$ where $i$ runs from $1$
to $n-1$. Then $\al$ is invariant (upper sign) or anti--invariant (lower sign)
according to
  $$\eqalign{
  \al_{i_1\ldots i_p}(x^i,-x^n)&=\pm \al_{i_1\ldots i_p}(x^i,x^n)\cr
  \al_{i_1\ldots i_{p-1}n}(x^i,-x^n)&=\mp \al_{i_1\ldots
  i_{p-1}n}(x^i,x^n)\,.
  }
  \eql{rule}
  $$
Exhibit $\ol d\al$ as,
  $$\eqalign{
  \ol d\al&=\sum_{\mu_0<\mu_1<\mu_2\ldots\atop1}^n
   \pa_{\mu_0}\al_{\mu_1\ldots\mu_p}(x)\,dx^{\mu_0}\wedge
   dx^{\mu_1}\wedge\ldots\wedge dx^{\mu_p}\cr
   &\equiv\sum_{\mu_0<\mu_1<\mu_2\ldots\atop1}^n
    \be_{\mu_0\mu_1\ldots\mu_p}(x)\,dx^{\mu_0}\wedge
   dx^{\mu_1}\wedge\ldots\wedge dx^{\mu_p}\,.
   }
  \eql{dpform}
  $$
It is easy to show that $\be$ satisfies the relations (\peq{rule}) with the
same signs as $\al$. To spell things out: if $n$ occurs in the set $\mu_i$,
($i=0$ to $p$) the coefficients are of the types $\pa_n\al_{i\ldots j}$ and
$\pa_i\al_{n\ldots j}$. Using both parts of (\peq{rule}) shows that each of
these types satisfies the bottom equation of (\peq{rule}) with the same signs
that $\al$ obeys, while, if the $\mu_i$ does not contain $n$, the coefficients
are typically $\pa_i\al_{j\ldots k}$ and obey the top equation of (\peq{rule})
with the same sign as $\al$.

It is generally true that the exterior derivative preserves absolute boundary
conditions and its dual relative ones, \eg\ [\pref{Gilkey}]. The present
special geometry goes beyond this in that $\ol d$ and $\ol \de$ preserve both
sorts.

\section{\bf 9. Casimir energy for Maxwell by degrees}

In this section I construct specific generating functions and derive some
field theoretic consequences in the form of Casimir energies. In the
quantization of massless anti--symmetric tensor fields ($p$--forms) the
demands of gauge invariance can be met by the well known ghosts--for--ghosts
procedure, [\pref{Obukhov}], [\pref{CandT}]. On a static space--time,
$T\times\man$, the Obukhov alternating combination of forms reduces just to
the coexact $p$--form on $\man$ plus a harmonic combination, essentially by
t\'elescopage, [\pref{Dow8}] \footnote{ The collapse of an alternating sum
frequently happens. Early instances in the $p$--form setting occur in
[\pref{MandS}] and [\pref{Patodi}] and termed `t\'elescopage' in [\pref{BGM}].
See also [\pref{GandS}]. Being topological in essence, it appears, for
example, in computations of analytic torsion, [\pref{Ray}].}. There are some
question marks concerning the role of the zero modes which come into play at
finite temperature and in the computation of the effective action. The zero
temperature vacuum energy is unambiguous and it is this I shall look at here.

The relation of the coexact and closed quantities is,
  $$
  \la^C(p,l)=\la^{CE}(p-1,l)\,,\quad g^C(p,l)=g^{CE}(p-1,l)\,,
  $$
on the sphere, which hold even in the presence of a boundary.

As the main example I look at coexact $1$--forms, $p=1$, on the
3--sphere, $d=3$, since this is equivalent to Maxwell theory, and seek to
find $g_b^{CE}(1,\si)\equiv h_b^{CCC}(2,\si)$. It is sufficient to
compute for just one of the symmetry types because, by duality on
$\oR^4$,
  $$
  g_{*b}^{CE}(1,\si)=h_{*b}^{CCC}(2,\si)=h_b^{CCC}
  (2,\si)=g_b^{CE}(1,\si)\,.
  \eql{cedual}
  $$

Although the computational equations can be amalgamated, they are easily
subjected to symbolic manipulation, so I leave them as they are. From
(\peq{hcloc5}), $h_b^{CCC}(2,\si)$ is seen to depend on $h_b^{CC}(1,\si)$ and
$h_b^{CC}(0,\si)$ which I evaluate via (\peq{hclod4}) giving,
  $$
  h_b^{CC}(0,\si)=h_b(0,\si)\,,\quad h_b^{CC}(1,\si)=h_b(1,\si)-\si
  h_b(0,\si)\,,
  $$
where the $h_b$ are given by (\peq{harmgen2}) and (\peq{harmgen2a}). Just to
check, I look at the full sphere. In this case,
  $$
  h_a(0,\si)={1+\si\over(1-\si)^3}\,,\quad
   h_a(1,\si)={(1+\si)(4-\si)\over(1-\si)^3}\,,
  $$
and, from (\peq{hcloc5}),
  $$
  g_{b}^{CE}(1,\si)=h_{b}^{CCC}(2,\si)={2(3-\si)\over(1-\si)^3}\,,
  \eql{gsph}
  $$
which is equivalent to the standard degeneracies,
$g_{b}^{CE}(1,l)=2(l+1)(l+3)$.

In the general case combining the ingredients yields,
  $$
  g_b^{CE}(1,\si)={\si^{d_1+d_2}
  +\si^{d_2+d_3}+\si^{d_1+d_3}-
  \si^{d_1+d_2+d_3}\over\si^2(1-\si^{d_1})(1-\si^{d_2})(1-\si^{d_3})}\,.
  \eql{fdgfun}
  $$
This expression is due to Chang, [{\pref{Chang}], who derived it by averaging
the spin--one character over the group. It is one of our basic results.

An important special  case is the lune with degrees $(q,1,1)$. I find,
  $$\eqalign{
  h_a(0,\si)&={1\over(1-\si)^2(1-\si^q)}\,,\quad
  h_a(1,\si)={2+\si^{q-1}+\si\over(1-\si)^2(1-\si^q)}\cr
  h_r(0,\si)&={\si^q\over(1-\si)^2(1-\si^q)}\,,\quad
  h_r(1,\si)={2\si^q+\si^{q-1}+\si\over(1-\si)^2(1-\si^q)}
  }
  $$
and
  $$\eqalign{
  h_a^{CC}(0,\si)&={1\over(1-\si)^2(1-\si^q)}\,,\quad
  h_a^{CC}(1,\si)={2+\si^{q-1}\over(1-\si)^2(1-\si^q)}\cr
  h_r^{CC}(0,\si)&={\si^q\over(1-\si)^2(1-\si^q)}\,,\quad
  h_r^{CC}(1,\si)={2\si^q+\si^{q-1}+\si
  -\si^{q+1}\over(1-\si)^2(1-\si^q)}\,.
  }
  $$
Calculation produces,
  $$
  g_b^{CE}(1,\si)={1+\si^{q-1}(2-\si)\over(1-\si)^2(1-\si^q)}\,,
  $$
as also follows from (\peq{fdgfun}), as a check.

The hemisphere case is $q=1$,
  $$
  g^{CE}_{b,hemisphere}(1,\si)={3-\si\over(1-\si)^3}\,,
  \eql{hemis}
  $$
half the full sphere value, (\peq{gsph}).

The coexact eigenvalues are given by (\peq{ceig1}) as $\mu(1,l)=(l+2)^2$ with
$l=0,1,\ldots$ and so the cylinder kernel is given by,
  $$
  T^{CE}_b(1,\si)=\si^2\,g_b^{CE}(1,\si)\,.
  $$
To repeat, $T$ is the kernel for the positive square root of the de Rham
Laplacian,
  $$
  T(\tau)=e^{-\tau\sqrt(d\de+\de d)}\,.
  \eql{ck1}
  $$

Using (\peq{fdgfun}), the \zf\ can be obtained in terms of Barnes functions,
(\peq{barn}), as the basic equation, [\pref{Chang}],
  $$
  \ze^{CE}(1,s)=\sum_{i=1}^3\ze_3\big(2s,\Si d-d_i|{\bf d}\big)-
  \ze_3\big(2s,\Si d|{\bf d}\big)\,,
  \eql{zetace}
  $$
which holds, as said, for both symmetry types, \ie boundary conditions.

The Casimir energy from (\peq{casen1}) follows as (\cf (\peq{casen2}) for
scalars), [\pref{Chang}],
  $$
  E_1=-{1\over2\,.\,4!\prod d_j}\bigg(\sum_{i=1}^3 B^{(3)}_4(d_i|{\bf d})-
  B^{(3)}_4(0|{\bf d})\bigg)\,.
  \eql{en1}
  $$

The numbers so computed agree with those found earlier by the more clumsy
sum--over--angles technique, (\peq{casnum2}). The lune of dihedral angle
$\pi/q$, gives, (\cf (\peq{lunecas0})),
  $$
  E_1={q^4+5q^2+30q-3\over720 q}\,,
  \eql{lunecas1}
  $$
and for the hemisphere $E_1=11/240$ half the full sphere value, as
anticipated.
\section{\bf 10. Higher spatial dimensions}

The simplifying fact in the preceding that allows the \zf\ to be readily found
is really that the eigenvalues are perfect squares. This occurs for a coexact
$p$--form in the `middle' dimension $p=(d-1)/2$, (\peq{ceig2}). I refer to
this as `middle' because the corresponding closed form has order $p+1=n/2$,
assuming $n$ is even. One might conveniently term these forms `self dual' but
this has other meanings.

As an example of higher dimensions, I briefly treat a middle form in a
lune fundamental domain.

The harmonic generating function (\peq{harmgen2}) is easily shown to be,
  $$
  h_a(p,\si)={e_p(\si^{q-1},1,1,\ldots,1,\si)\over(1-\si)^{n-2}(1-\si^q)}\,,
  $$
where
  $$
  e_p(\si^{q-1},1,1,\ldots,1,\si)=\comb{n-2}p+(\si
  +\si^{q-1})\,\comb{n-2}{p-1}+\si^q\comb{n-2}{p-2}\,,
  $$
and putting this into the mill yields finally the middle dimension
$g_r^{CE}(p,\si)$ in the form,
  $$
  g_r^{CE}(p,\si)={P_p(\si)+\si^{q-1}Q_p(\si)\over (1-\si)^{2p}(1-\si^q)}\,,
  $$
where $P_p$ and $Q_p$ are polynomials in $\si$. For example,
  $$\eqalign{
  P_2(\si)&=\si^2-4\si+6,\quad Q_2(\si)=-(\si-4)\cr
  P_3(\si)&=\si^3-6\si^2+15\si-20,\quad Q_3(\si)=\si^2-6\si+15\cr
  P_4(\si)&=\si^4-8\si^3+28\si^2-56\si+70,\quad Q_4(\si)=-(\si^3-8\si^2
  +28\si-56)\,.
  }
  $$
The relation is
  $$
  Q_p(\si)={(-1)^{p/2}\over\si}\big(P_p(\si)-P_p(0))\,.
  $$

A sample result is the Casimir energy for a $2$--form (of either
symmetry) on a 5--dimensional lune of dihedral angle $\pi/q$, obtained by
straightforward residues as,
  $$
  E_2(q)={5q^6+48q^5+165q^4+220q^3+75q^2+138q-136\over360q}\,.
  $$
The hemisphere value is $E_2(1)=103/72$.
\section{\bf 11. The heat--kernel expansion}

Having the \zf\ on S$^3/\Ga$, it is a standard matter to compute the
short--time heat--kernel expansion coefficients which often have an
independent significance. In the present context, aspects of polytope and
fundamental domain geometry have been related to the scalar ($p=0$)
coefficients, [\pref{Dow20}]. In this section I compute a few of these
important quantities and explore some features.

To begin, I state my present convention for the traced heat--kernel expansion
which is
  $$
  K(t)\sim {1\over t^{3/2}}\sum_{n=0} C_{n/2}\, t^{n/2}\,,
  \eql{hk1}
  $$
with labels for the specific quantity under investigation.

The first non--trivial example I consider is $C_{1/2}$ which is an integral
over the boundary, $\pa\man$, of the fundamental domain, $\man$. The general
formula is given by Bla\v{z}i\'c {\it et al}, (`BBG'), [\pref{BBG}], for a
smooth manifold, but the result extends unchanged to a singular one by
dimensions. I will obtain it for a one--form on S$^3/\Ga$ from the \zf,
(\peq{zetace}), using the generic relation on a $d$--dimensional manifold,
  $$
  \ze(s)\sim {1\over\Ga\big((d-m)/2\big)}\,{C_{m/2}\over s-(d-m)/2}\,,
  \quad s\to(d-m)/2\,,
  \eql{resid}
  $$
for $m=0,1,\ldots,d-1,d+1,d+3,\ldots$. The known residues of the Barnes \zf\
give,
  $$
  C^{CE}_{1/2}({\bf d})={2\over|\Ga|}\,\bigg(B^{(3)}_1(0\,|\,{\bf d})-\sum_i
  B^{(3)}_1(d_i\,|\,{\bf d})\bigg)\,,
  \eql{c12}
  $$
which computes to zero. Some technicalities are given in Appendix A. This
vanishing says that there is as much Dirichlet as Neumann content in the {\it
coexact} 1--form. The fact that the surface term is zero for the Maxwell
field is an old result, [\pref{BaandD,BaandH,DandCa,Kennedy2}], and follows
from a cancellation between the $E$ and $H$ modes, where these exist.

My next example is the constant term, $C_{d/2}$, in the expansion which plays
an important role, \eg\ [\pref{MandP}], and takes contributions from
singularities in the manifold. For example, in three dimensions, $C_{3/2}$,
being a pure boundary quantity, involves an integration over the
2--dimensional frontier, $\pa\man$, of the fundamental domain, plus an
integration over the 1--dimensional intersections (`edges') of the boundary
pieces plus the effect of the 0--dimensional vertices of $\man$ (which can be
pictured as a spherical tetrahedron). Because all parts of $\pa\man$, of the
various codimensions, are geodesically embedded in S$^3$, all extrinsic
curvatures vanish and one is left with an integration of scalar curvatures
plus the vertex parts. In [\pref{Dow21}] the latter quantities were evaluated
for spin zero. Since the general expression for $C_{3/2}$ has been evaluated
for a general $p$--form (on a smooth manifold) in [\pref{BBG}], it is now
possible to repeat this calculation in the form setting. I will again pursue
the calculation only for $1$--forms on S$^3/\Ga$ to give a flavour of the
method.

The total $p$--form \zf\ is a sum of coexact $p$-- and $(p-1)$--forms,
  $$
  \ze_b^{tot}(p,s)=\ze_b^{CE}(p,s)+\ze_b^{CE}(p-1,s)\,,
  \eql{zetot}
  $$
which holds on any manifold with or without boundary \footnote{ The theory of
forms on bounded manifolds is discussed by Duff, [\pref{Duff}], and Duff and
Spenser, [\pref{DandS}]. Probably the most complete analysis is given by
Conner, [\pref{Conner}], who introduces spaces of relative and absolute forms
denoted by ${\caL}$ and ${\man}$ respectively. He gives a careful treatment of
the Hodge decomposition in the presence of a boundary. The situation is not so
straightforward. There are (at least) two orthogonal decompositions of an {\it
arbitrary} form, one `adapted' to absolute conditions and one to relative. The
upshot, however, is that one can proceed as in the compact case, and
(\peq{zetot}) is justified in general.}.

The BBG expressions are for the general $p$--form but I find  that, rather
than use the combination (\peq{zetot}), it is better to extract the coexact
part, which is done by inverting (\peq{zetot}) in the familiar t\'elescopage
fashion. Decreasing $p$ gives,
  $$
  \ze_b^{CE}(p,s)=\sum_{q=0}^p(-1)^{p-q}\,\ze_b^{tot}(q,s)\,,
  \eql{zeinv}
  $$
or the dual form,
  $$
  \ze_{*b}^{CE}(p,s)=(-1)^d\sum_{q=d-p}^d(-1)^{p-q}\,\ze_b^{tot}(q,s)\,,
  \eql{zeinvd}
  $$
while increasing $p$ produces,
  $$
 \ze_b^{CE}(p,s)=-\sum_{q=p+1}^d(-1)^{p-q}\,\ze_b^{tot}(q,s)\,.
 \eql{zeinv2}
  $$

In the context of Maxwell theory in a bounded domain of flat 3--space, I note
that this has recently been used, essentially, by Bernasconi {\it et al},
[\pref{BGH}].

Equating (\peq{zeinv}) and (\peq{zeinv2}) gives,
  $$
  \sum_{q=0}^d(-1)^q\,\ze_b^{tot}(q,s)=0\,,
  \eql{tid}
  $$
which is the archetypal t\'elescopage identity. The earlier discussion of
degeneracies and generating functions exhibits similar identities.

The next standard relation I require is, in a $d$--dimensional manifold,
  $$
  \ze(p,0)=C_{d/2}(p)-n_0(p)\,,
  \eql{cezet0}
  $$
where $n_0$ is the number of zero modes. Labels can be attached as required.
For example, setting $s$ to zero in 9\peq{tid}) yields the classic Betti
number relation,
  $$\eqalign{
  \sum_{q=0}^d(-1)^q C^b_{d/2}&=\sum_{q=0}^d n_0^b(q)\cr
&=\chi_b(\man)\,,
  }
  $$
where $\chi_a(\man)$ is the Euler number, $\chi(\man)$, of the manifold and
$\chi_r(\man)$ is the relative Euler number, $\chi(\man,\pa\man)$. For the
factored sphere, $\man=S^d/\Ga$,
  $$
  \chi\big(\man\big)=1\,,\quad \chi\big(\man,\pa\man)=-1\,.
  \eql{chis}
  $$

Simple substitution into (\peq{zetace}) gives, for both symmetries,
  $$\eqalign{
  \ze^{CE}(1,0)&=\sum_{i=1}^3\ze_3\big(0,\Si d-d_i|{\bf d}\big)-
  \ze_3\big(0,\Si d|{\bf d}\big)\cr
  &={1\over3|\Ga|}\bigg(\sum_i B^{(3)}_3(d_i\,|\,{\bf d})-
  B^{(3)}_3(0\,|\,{\bf d})\bigg)\,,
  }
  \eql{c32}
  $$
which evaluates identically to $1/2$ in all cases. Some computational details
are given in Appendix A. However, there is no need for an explicit
computation. except as a check, because there is a topological explanation as
outlined below.

Adding the (equal) values for the two symmetries gives 1 for the fundamental
domain of the purely rotational part of the polytope group. In this paper I
simply present this as a mathematical fact. In particular it holds for the
full sphere which I have discussed, [\pref{Dow8}], in connection with finite
temperature effects.  In fact the value of 1 {\it must} hold for all the
doubled fundamental domains by a topological argument given, for the sphere,
in [\pref{ELV}]. For such a domain, there are no boundary heat--kernel
coefficients for a general $p$--form by general arguments, see
[\pref{Gilkey}] for the smooth case. Hence we have, for odd $d$,
  $$
  C^b_{d/2}(q)+C_{d/2}^{*b}(q)=0\,,\quad\forall q\,.
  \eql{ceeid}
  $$
I will confirm later that fixed points do not spoil this identity.

Adding (\peq{zeinvd}) and (\peq{zeinv2}), and now setting $p=(d-1)/2$ for the
middle form,
  $$
  \ze^{CE}_b(p,s)+\ze^{CE}_{*b}(p,s)=-\sum_{q=p+1}^{2p+1}(-1)^{p-q}
  \big(\ze_b^{tot}(q,s)+\ze_b^{tot}(q,s)\big)\,.
  $$
The two \zfs\ on the left are equal and yield, when $s=0$, the result, for
any $\man$,
  $$
  2\big(C_{d/2}^{CE}(p)-n_0^{CE}(p)\big)=
  -\sum_{q=p+1}^{2p+1}(-1)^{p-q}
  \big(C^b_{d/2}(q)+C^{*b}_{d/2}(q)-n_0^b(q)-n_0^{*b}(q)\big)\,.
  $$

Applying this to a fundamental domain, the known cohomology  means that only
one of the Betti numbers contributes (at $q=2p+1=d$) and then (\peq{ceeid})
produces the final answer,
  $$
  C^{CE}_{d/2}(p)={1\over2}\,(-1)^{p+1}\,.
  \eql{cee32}
  $$

Curiously, the value of $1/2$ accords with the conjecture in Baltes and Hilf,
[\pref{BaandH}], that the constant term in the heat--kernel expansion is
shape independent for the Maxwell field, and equals $1/2$ (see also Baltes,
[\pref{Baltes}]). It is calculated to be $1/2$ for a cylinder of polygonal
cross section. Again, the cause is a cancellation between the $E$-- and
$H$--type modes. For comparison, in the 3--ball, the value is $5/8$.

Turning now to the alternative expression for $C_{3/2}$, I assume, in
general, that it can be written as a smooth part plus the contributions of
the edges and vertices of the manifold, [\pref{DandA}], [\pref{Dow20}],
[\pref{Dow21}]. For a fundamental domain, all extrinsic curvatures vanish
which eliminates the edge effects leaving,
   $$
  C_{3/2}(p)=\overline C_{3/2}(p)+V(p)\,,
  \eql{c32f}
  $$
where the smooth part, $\overline C$, has been given in [\pref{BBG}] Theorem
1.2 for the {\it general} $p$--form.

I now set $s$ to zero in (\peq{zeinv}) and (\peq{zeinv2}) for $p=1$, $d=3$ and
use (\peq{cezet0}). I will also drop the $3/2$ label for ease, then
  $$\eqalign{
  C^{CE}(1)&=C_a(1)-C_a(0)+1\cr
          &=\overline C_a(1)-\overline C_a(0)+1+V_a(1)-V_a(0)
  }
  \eql{ceinv3}
  $$
and the equivalent,
  $$\eqalign{
  C^{CE}(1)&=C_a(2)-C_a(3)\cr
          &=\overline C_a(2)-\overline C_a(3)+V_a(2)-V_a (3)\,.
  }
  \eql{ceinv4}
  $$
I have selected invariant forms, \ie absolute boundary conditions\footnote{
Relative conditions could have been chosen.}. This means that there is a
constant zero mode for $p=0$ which is reflected in the 1 in (\peq{ceinv3})
coming from the $n_0$ in (\peq{cezet0}). The formulae in [\pref{BBG}] allow
the smooth part to be evaluated giving, from (\peq{ceinv3}) and
(\peq{ceinv4}),
  $$\eqalign{
  C^{CE}(1)&=-{1\over8\pi}\,|\pa\man|+1+V_a(1)-V_a(0)\cr
           &= {1\over8\pi}\,|\pa\man|+V_a(2)-V_a(3)\,.
           }
  \eql{c32g}
  $$
Taking into account the computed value, $C^{CE}(1)=1/2$, adding and
subtracting the two equations in (\peq{c32g}) gives,
  $$
  V_a(1)-V_a(0)=V_a(3)-V_a(2)
  \eql{vees}
  $$
and
  $$
  \sum_{q=0}^3(-1)^q\,V_a(q)={1\over4\pi}\,|\pa\man|+1\,.
  $$
The $0$--form vertex contribution, $V_a(0)=V_N(0)$, has been evaluated in
[\pref{Dow21}], [\pref{DandA2}], and equals $-V_D(0)=-V_r(0)$ which, by
duality, equals $-V_a(3)$. Hence from (\peq{vees}) $V_a(1)=-V_a(2)$ and one
can deduce that the vertex numbers change sign with the boundary conditions,
  $$
  V_a(p)+V_r(p)=0\,.
  $$
Adding the invariant and anti--invariant quantities corresponds to using just
the rotational part of the complete polytope group. The cancellations justify
our previous arguments leading to (\peq{cee32}).

I simplify (\peq{c32g}) by giving the size of the boundary,
   $$
   |\pa\man|=2b_1{|S^2|\over|\Ga|}\,\,,
   $$
which is a geometrical statement since $b_1$ is the number of reflecting
hyperplanes. It can be obtained by a consideration of the coefficient
$C_{1/2}$ for spin zero, [\pref{Dow20}].

For the tesselations, combining the above results, I find,
  $$
   V_a(1)-V_a(0)={1\over2}\bigg({d_1+d_2+d_3-2\over d_1 d_1d_3}-1\bigg)\,.
  \eql{vert1}
  $$
This vertex contribution vanishes for the hemisphere, $d_1=d_2=d_3=1$, which
is correct.

I make the assumption that the total vertex term is the sum of the individual
vertex contributions. As in [\pref{Dow21}], in the tesselation fundamental
domain there are four trihedral vertices, each having three dihedral angles,
$(\pi/\al_1,\pi/\al_2,\pi/\al_3)$, where $(\al_1,\al_2,\al_3)=
(q,r,2),\,(p,q,2),\,(p,2,2),\,(r,2,2)$ in terms of the four tesselation types,
$(p,q,r)=(3,3,3),(3,3,4),(3,4,3),(3,3,5)$, in standard notation. So,
  $$
  V_a(1)-V_a(0)=I(q,r,2)+I(p,q,2)+I(p,2,2)+I(r,2,2)\,,
  \eql{vert2}
  $$
where the $I$'s are the individual vertex contributions. They are symmetrical functions of the three angles.

Equation (\peq{vert2}) yields only three independent equations for the four
$I$'s in (\peq{vert2}). However one can apply the shape independent value of
$1/2$ for $C^{CE}_{3/2}$  in the case of an ordinary polyhedron in flat space
to extract information. For example the polygonal cylinder (\cf
[\pref{BaandH}]), yields the $I(p,2,2)$ value as,
  $$
  I(p,2,2)={1\over8}\bigg(1-{1\over p}\bigg)\,,
  $$
if, to repeat, it is assumed that the $1/2$ is distributed amongst the
vertices. This provides the extra information needed to solve for the other
contributions and simple elimination gives (with a check),
  $$\eqalign{
  I(3,3,2)={1\over8}\,,\quad
  I(3,4,2)={5\over32}\,,\quad
  I(3,5,2)={5\over16}\,.
  }
  $$
Together with the known values for the $0$--form vertex contributions, these
numbers allow one to find the corresponding values for the general, \ie
unconstrained, $1$--form.

The Barnes \zf\ has only a finite number of poles which cease at $m= d-1$. The
only remaining, higher coefficients in the expansion (\peq{hk1}) can be found
from,
  $$
  (-1)^k\,k!\,C_{k+d/2}=\ze(-k)\,,\quad k=0,1,\ldots\,,
  $$
which gives, from (\peq{zetace}),
  $$
  C^{CE}_{k+d/2}=
  {(-1)^k2(2k)!\over|\Ga|(3+2k)!k!}\,
  \bigg(\sum_iB^{(3)}_{3+2k}(d_i\,|\,{\bf d})-B^{(3)}_{3+2k}(0\,|\,{\bf d})
  \bigg)\,.
  \eql{ce32k}
  $$

It is shown in Appendix A that this vanishes, as must be the case (at least
for the hemisphere). Expression (\peq{ce32k}) holds for {\it both} symmetry
types and adding these gives the value for the periodic situation which is
zero, since it is known that the heat--kernel expansion for coexact middle
forms on odd dimensional spheres terminates with the $C^{CE}_{d/2}$ term. The
termination is thus generalised here to the tesselation fundamental domains.
This has finite temperature consequences which I put aside for now.

Further information on the heat--kernel coefficients on the full sphere can be
found in [\pref{DandKi}], [\pref{CandH}] and [\pref{ELV}].

\section{\bf 12. Conclusion}
In this paper, values have been found for the Casimir energies of scalar and
Maxwell fields (coexact $1$--forms) in orbifold factors of spheres, in
particular the 3--sphere. The manifolds are the fundamental domains of
semi--regular tesselations and the computations were performed both by direct
summation over the group by summing over angles and by using a degeneracy
generating function to construct the \zf\ which is given in terms of Barnes
\zfs. The results were specialised to the hemisphere, which is a case often
looked at on its own.

I also discussed the spectral geometry of $p$--forms on the fundamental
domains. It allows, for example, the degeneracy generating function for any
$p$ to be found on any factor S$^d/\Ga$.

Aspects of the short--time expansion of the heat--kernel were discussed. In
particular it was shown, as expected, that the surface term, $C_{1/2}$,
vanished. The constant term was also computed to be $1/2$ for the coexact
1--form on all factors and shown to be so topologically. The expression was
analysed in some detail, bringing out the contribution of the vertices. It
was also shown that the expansion terminates with the constant term on the
fundamental domains.

Another path to vertex contributions is via the heat--kernel on the
generalised cone, [\pref{BKD}], which is a bounded M\"obius corner, having as
two--dimensional base a fundamental domain of, this time, S$^2/\Ga$, the
familiar tiling of the two--sphere, [\pref{Watsons}]. I defer this to another
time.

Having the \zfs\ and generating functions allows other spectral quantities to
be evaluated, for example the functional determinants and also finite
temperature quantities.

Although I have looked mainly at $1$-forms on factors of the three sphere, it
is possible to analyse forms of the middle order in any dimension with very
similar results. The \zfs\ are again of Barnes type. Even for forms of any
order, with non perfect square eigenvalues, the calculation can be pursued,
with determination, \eg\ [\pref{CandT}], [\pref{RandTo}], [\pref{ELV}],
[\pref{DandKi2}] .

I have only computed integrated quantities. It is possible to extend the same
techniques to the local objects such as the Green function.

\section{\bf Appendix A}
I give some technical niceties connected with the generalised Bernoulli
polynomials in terms of which the field theory quantities have been expressed.
They satisfy several important identities, [\pref{Norlund}], often leading to
simplifications, or at least variety, in their evaluation. One such is the
recursion
 $$
  B^{(n)}_\nu(d_i\,|\,{\bf d})=B^{(n)}_\nu(0\,|\,{\bf d})+
  d_i\,\nu\,B^{(n-1)}_{\nu-1}(0\,|\,\hat{\bf d}_i)\,,
  \eql{brecurs}
 $$
where $\hat{\bf d}_i$ stands for the set of $n$--degrees, omitting $d_i$,
 $$
 \hat{\bf d}_i=(d_1,\ldots,d_{i-1},d_{i+1},\ldots d_n)\,.
 $$

The Bernoulli polynomials of 0 argument, $B^{(n)}_\nu(0\,|\,{\bf d})$, are the
generalised \break Bernoulli numbers, $B^{(n)}_\nu({\bf d})$, and, as an
example, the coexact 1--form vacuum energy (\peq{en1}) can be re--expressed
purely in terms of numbers as
  $$\eqalign{
  E_1=-{1\over48d_1d_2d_3}\bigg(2B^{(3)}_4(d_1,d_2,d_3)
  +4d_1B^{(2)}_3(d_2,d_3)
  +&4d_2B^{(2)}_3(d_1,d_3)\cr
  &+4d_3B^{(2)}_3(d_1,d_2)\bigg)\,.
  }
  $$
This is not strictly necessary as the general expression for
$B^{(n)}_\nu(x\,|\,{\bf d})$ could be found.

In a similar fashion, the heat--kernel coefficient (\peq{c12}) becomes
  $$\eqalign{
  C^{CE}_{1/2}({\bf d})={1\over d_1d_2d_3}\bigg(2B^{(3)}_1(d_1,d_2,d_3)+
  d_1\,B^{(2)}_0(d_2,d_3)+&d_2\,B^{(2)}_0(d_1,d_3)\cr
  &+d_3\,B^{(2)}_0(d_1,d_2)\bigg)\,.
  }
  \eql{c12a}
  $$
and for (\peq{c32}),
  $$\eqalign{
  C^{CE}_{3/2}({\bf d})&={1\over6d_1d_2d_3}\bigg(2B^{(3)}_3(d_1,d_2,d_3)+
  3d_1\,B^{(2)}_2(d_2,d_3)+3d_2\,B^{(2)}_2(d_1,d_3)\cr
  &\hspace{***********************}+3d_3\,B^{(2)}_2(d_1,d_2)\bigg)\cr
  &={1\over2}\,.
  }
  \eql{c32a}
  $$

The expression, (\peq{ce32k}), for the higher coefficients can be
re--expressed,
  $$
  C^{CE}_{k+3/2}=
  {(-1)^k2(2k)!\over|\Ga|(3+2k)!k!}\,
  \bigg((3+2k)\sum_id_iB^{(2)}_{2+2k}(\hat{\bf d}_i)+2B^{(3)}_{3+2k}({\bf d})
  \bigg)\,.
  \eql{ce32l}
  $$

The spin zero vacuum energy (\peq{casen2}) can also be subjected to some
(inessential) manipulation. On factors of the two--sphere, the form was given
in terms of the two degrees, [\pref{ChandD}],
  $$
  E_0=\pm{d_0\over96d_1d_2}(d_0^2-d_1^2-d_2^2)\,,
  $$
which can also be written in terms of the Shephard--Todd numbers, $b_1$ and
$b_2$, using, [\pref{Dow20}],
  $$
  B^{(d)}_3\big((d-1)/2\,|\,{\bf
  d}\big)={1\over4}\,b_1\bigg(b_1-b_2+{d-3\over2}\bigg)
  \eql{stb3}
  $$
as
  $$
  E_0=\pm{1\over24|\Ga|}\,b_1\bigg(b_1-b_2-{1\over2}\bigg)\,,\quad
  |\Ga|=2d_1d_2\,.
  $$

The advantage of using the Shephard--Todd numbers is that, for a given $\nu$,
$B^{(d)}_\nu(x\,|\,{\bf d})$ can be displayed as an explicit function of the
dimension, $d$. One path to this representation is via Todd Polynomials,
$T_k$, as Hirzebruch, [\pref{Hirzebruch}], gives the relation,
  $$
  T_k(c_1,\ldots,c_k)={(-1)^k\over k!}\,B^{(d)}_k(d_1,\ldots,d_d)\,,
  \quad k\le d\,,
  $$
where the $c_s$ are the elementary symmetric functions of the degrees $d_i$
($i=1,\ldots,d)$. The relations needed to turn the $c_s$ into elementary
symmetric functions, the $b_r$, of the exponents, $m_i$ ($i=1,\ldots,d+1$) are
detailed in [\pref{Dow20}]. For calculational rapidity here, I just use the
$c_s$ and note the first three Todd polynomials,
  $$
  T_1={1\over2}\, c_1\,,\quad T_2={1\over12}\,(c_1^2+c_2)\,,\quad
  T_3={1\over24}
  \,c_1c_2\,,
  $$
leading to,
  $$\eqalign{
  B^{(d)}_1(x\,|\,{\bf d})&=x-{1\over2}\Si_i\, d_i\cr
  B^{(2)}_2(d_1,d_2)&={1\over6}\big((d_1+d_2)^2+d_1d_2\big)\cr
  B^{(3)}_3(d_1,d_2,d_3)&=-{1\over
  4}(d_1+d_2+d_3)(d_1d_2+d_2d_3+d_2d_3)\,,
  }
  $$
employed in computing (\peq{c12a}) and (\peq{c32a}).

Alternatively, one can use the general formula in terms of ordinary Bernoulli
numbers,
  $$
  B^{(n)}_\nu({\bf d})=\sum{\nu!\over s_1!\ldots s_n!}\,d_1^{s_1}\ldots
  d_n^{s_n}\,B_{s_1}\ldots B_{s_n}\,,
  \eql{begen}
  $$
where the sum is over all  $s_1,\ldots,s_n$, positive or zero, satisfying
$s_1+s_2+\ldots+s_n=\nu$.

In the expression for the higher coefficients, (\peq{ce32l}), the lower index
on the $B$'s is now bigger than the upper one and (\peq{begen}) is called for.
Writing out $B^{(3)}_\nu$ where $\nu$ is odd and greater than 3, the condition
$s_1+s_2+s_3=\nu$ means that one, and only one, of $s_1,s_2,s_3$ must equal 1
because only $B_1=-1/2$ of the odd Bernoulli numbers is non--zero. Breaking up
the summation by taking this into account effectively reduces $n$ and $\nu$ by
one at the expense of producing three sums. The outstanding factors then lead
to the identity,
  $$
  2B^{(3)}_\nu({\bf d})+\nu\sum_i d_i\,B^{(2)}_{\nu-1}(\hat{\bf
  d}_i)=0\,,\quad \nu\,\, {\rm odd}\,\,>3\,,
  $$
which was used in the text to show the termination of the coexact heat--kernel
expansion from (\peq{ce32l}).

The same arguments can be applied for $\nu=3$. Then there is the additional
possibility that all three $s_1,s_2,s_3$ equal one. This gives the extra term,
$3d_1d_2d_3$, and is another way of deriving (\peq{c32a}). Higher $n$ can be
treated similarly.

Finally I give a polynomial which I have used,
  $$\eqalign{
  B_4^{(3)}(x\,|\,{\bf d})=&-{1\over30}\big[d_1^4+d_2^4+d_3^4-5d_1^2d_2^2-
  5d_2^2d_3^2-5d_1^2d_3^2\cr
  &\hspace{**********}-15d_1^2d_2d_3-15d_1d_2^2d_3-15d_1d_2d_3^2\big]\cr
  &-x\big[d_1^2d_2+d_1d_2^2+d_1^2d_3+d_2^2d_3+d_1d_3^2+d_2d_3^2
  +3d_1d_2d_3\big]\cr
  &+x^2\big[d_1^2+3d_1d_2+d_2^2+3d_1d_3+3d_2d_3+d_3^2\big]\cr
  &-x^3\big[d_1+d_2+d_3\big]+x^4\,.
  }
  $$
\section{\bf Appendix B}

I have found it constructive to check a specific numerical example. I choose
the following parameters $p=1$, $l=2$ in $\oR^3$, \ie $d=2$ ($n=3$). The
general homogeneous polynomial of second degree is the conic,
  $$
  ax^2+2hxy +2gxz +2fyz +by^2 + cz^2\,,
  $$
and is therefore 6 dimensional on the full sphere. The harmonic condition
imposes one constraint, giving 5--dimensions. A basis would be, $xy,xz,yz,
x^2-y^2, y^2-z^2$. One now has to impose the closed condition. In vector
language for a 1--form, $\equiv{\bf A}$, $d A\equiv\curl{\bf A}$ and so we get
3 identities, each of which is linear in $x,y,z$, giving 9 conditions not all
of which are independent in view of $\div\curl=0$, which is one scalar
identity, making actually 8. This leaves, finally, a $15-8$ = 7 dimensional
module.

The standard expression for the closed degeneracy on S$^2$ derived in the
cited references gives the value $7$ for $h^{CCC}(1,2)$.

Applying now, (\peq{hcloc2}), (\peq{hcloc}), (\peq{hclod3}) and
(\peq{hun2}) in detail, I find,
  $$\eqalign{
  h^{CCC}(1,2)&=h^{CC}(1,2)-h^{CC}(2,1)+h^{CC}(3,0)\cr
         &= \big(h(1,2)-h(0,1)\big) -\big(h(2,1)-h(1,0)\big) +
 \big(h(3,0)-h(2,-1)\big) \cr
 &=\big(d(1,2)-d(1,0)-d(0,1)\big)-\big(d(2,1)-d(1,0)\big)+d(3,0)\cr
          &=\comb31\comb42-\comb31\comb30
          -\comb32\comb32+\comb33\comb20\cr
    &=3.6-3.1 -3.3+1.1=7
  }
  $$
and
  $$\eqalign{
  h^{CCC}(1,2)&=h^{CC}(0,3)=h(0,3)\cr
         &= d(0,3)-d(0,1)\cr
          &=\comb30\bigg(\comb53-\comb31\bigg)\cr
    &=10-3=7\,.
  }
  $$

In the coclosed case one has $\de A=\div{\bf A}=0$ which is just one condition
linear in $x,y,z$, equivalent to 3 conditions so that there are $15-3=12$
independent components and this is born out by the calculation of
$h^C(2,2)=12$ in 3 dimensions. This just reflects the duality on $\oR^3$
between a closed $p$--form and a coclosed $(n-p)=(3-p)$ one.

I can also check (\peq{hun2}) for the harmonic degeneracies in the
invariant/anti--invariant cases. For the hemisphere there is just one
reflection, say $z\to -z$. The polynomial basis splits into the 4--dim
invariant set, $x^2,y^2,z^2,xy$, and the 2--dim anti--invariant one, $xz,yz$.

In cartesian coordinates the invariance/anti--invariance of the 1--form
(vector) under the reflection implies,
  $$\eqalign{
  A_x(x,y,-z)&=\pm A_x(x,y,z)\cr
  A_y(x,y,-z)&=\pm A_y(x,y,z)\cr
  A_z(x,y,-z)&=\mp A_z(x,y,z)\,,
  }
  $$
upper sign for invariance. Therefore, for the invariant case, $A_x$ and $A_y$
need the invariant polynomial basis and $A_z$ the anti--invariant one, and
conversely for an anti--invariant form. This makes explicit the nature of
absolute and relative conditions the tangential components on the hemisphere
being linear combinations of $A_x$ and $A_y$. In another language, invariant
vectors are {\it polar} vectors while anti--invariant ones are termed {\it
axial}.

Therefore,
  $$
  d_a(1,2)=4.2+1.2=10\,,\quad d_r(1,2)=2.2+1.4=8\,.
  $$
For the $l=0$ case, the coefficients are constants and clearly
   $$
  d_a(1,0)=1.2=2\,,\quad d_r(1,0)=1.1=1\,,
  $$
giving the combinations,
  $$
  d_a(1,2)-d_a(1,0)=10-2=8\,,\quad d_r(1,2)-d_r(1,0)=8-1=7
  $$
(adding to give 15).

These values agree with the hemisphere Poincar\'e series,
  $$
  \wp_a(\si)=2+5\si+10\si^2+\ldots\,\quad
  \wp_r(\si)=1+4\si+8\si^2+\ldots\,.
  $$

The number of harmonic invariant/anti--invariant forms can also be counted
directly from the 5.3 harmonic basis. Thus, by inspection, the invariant basis
is $xy,x^2-y^2,y^2-z^2$ and the anti--invariant one $xz,yz$, so
  $$
  h_a(1,2)=2.3+1.2=8\,,\quad h_r(1,2)= 2.2+1.3=7
  $$
in agreement with the above.

The dimension can be increased without difficulty to, say, 4, just to
make sure. The full basis is 10--dim,
  $$
  x^2,y^2,z^2,w^2,xy,xz,xw,yz,yw,zw
  $$
which splits into $a$ (7--dim) and $r$ (3--dim) parts,
  $$
   x^2,y^2,z^2,w^2,xy,xw,yw\,\quad{\rm and}\quad
  xz,yz,zw
  $$
hence
  $$\eqalign{
 d_a(1,2)-d_a(1,0)&=(7.3+1.3)-1.3=21\cr
d_r(1,2)-d_r(1,0)&=(3.3+1.7)-1.1=15
 }
  $$
The total harmonic basis is 9--dim,
  $$
  x^2-y^2,y^2-z^2,z^2-w^2,xy,xz,xw,yz,yw,zw
  $$
and this splits as 6+3,
  $$
  x^2-y^2,y^2-z^2,z^2-w^2,xy,xw,yw\,\quad{\rm and}\quad xz,yz,zw
  $$
giving
  $$
  h_a(1,2)=6.3+1.3=21\,,\quad h_r(1,2)= 3.3+1.6=15
  $$
again in agreement.

It is also advisable to check another $p$, say $p=2$. The  form is 6
dimensional (in 4 dimensions) 3 components being symmetric and 3
anti--symmetric under the reflection.

  $$\eqalign{
  d_a(2,2)-d_a(2,0)&=(3.7+3.3)-1.3=27\cr
  h_a(2,2)&=6.3+3.3=27\cr
 d_r(2,2)-d_r(2,0)&=(3.3+3.7)-1.3=27\cr
  h_r(2,2)&=6.3+3.3=27\,.
  }
  $$
This result also exhibits the duality in (\peq{hun2}).

\newpage

\noin{\bf References.} \vskip5truept
\begin{putreferences}
  \ref{DandA}{Dowker,J.S. and Apps, J.S. \cqg{15}{1998}{1121}.}
  \ref{Weil}{Weil,A., {\it Elliptic functions according to Eisenstein
  and Kronecker}, Springer, Berlin, 1976.}
  \ref{Ling}{Ling,C-H. {\it SIAM J.Math.Anal.} {\bf5} (1974) 551.}
  \ref{Ling2}{Ling,C-H. {\it J.Math.Anal.Appl.}(1988).}
 \ref{BMO}{Brevik,I., Milton,K.A. and Odintsov, S.D. \aop{302}{2002}{120}.}
 \ref{KandL}{Kutasov,D. and Larsen,F. {\it JHEP} 0101 (2001) 1.}
 \ref{KPS}{Klemm,D., Petkou,A.C. and Siopsis {\it Entropy
 bounds, monoticity properties and scaling in CFT's}. hep-th/0101076.}
 \ref{DandC}{Dowker,J.S. and Critchley,R. \prD{15}{1976}{1484}.}
 \ref{AandD}{Al'taie, M.B. and Dowker, J.S. \prD{18}{1978}{3557}.}
 \ref{Dow1}{Dowker,J.S. \prD{37}{1988}{558}.}
 \ref{Dow30}{Dowker,J.S. \prD{28}{1983}{3013}.}
 \ref{DandK}{Dowker,J.S. and Kennedy,G. \jpa{}{1978}{}.}
 \ref{Dow2}{Dowker,J.S. \cqg{1}{1984}{359}.}
 \ref{DandKi}{Dowker,J.S. and Kirsten, K. {\it Comm. in Anal. and Geom.
 }{\bf7} (1999) 641.}
 \ref{DandKe}{Dowker,J.S. and Kennedy,G.\jpa{11}{1978}{895}.}
 \ref{Gibbons}{Gibbons,G.W. \pl{60A}{1977}{385}.}
 \ref{Cardy}{Cardy,J.L. \np{366}{1991}{403}.}
 \ref{ChandD}{Chang,P. and Dowker,J.S. \np{395}{1993}{407}.}
 \ref{DandC2}{Dowker,J.S. and Critchley,R. \prD{13}{1976}{224}.}
 \ref{Camporesi}{Camporesi,R. \prp{196}{1990}{1}.}
 \ref{BandM}{Brown,L.S. and Maclay,G.J. \pr{184}{1969}{1272}.}
 \ref{CandD}{Candelas,P. and Dowker,J.S. \prD{19}{1979}{2902}.}
 \ref{Unwin1}{Unwin,S.D. Thesis. University of Manchester. 1979.}
 \ref{Unwin2}{Unwin,S.D. \jpa{13}{1980}{313}.}
 \ref{DandB}{Dowker,J.S.and Banach,R. \jpa{11}{1978}{2255}.}
 \ref{Obhukov}{Obhukov,Yu.N. \pl{109B}{1982}{195}.}
 \ref{Kennedy}{Kennedy,G. \prD{23}{1981}{2884}.}
 \ref{CandT}{Copeland,E. and Toms,D.J. \np {255}{1985}{201}.}
 \ref{ELV}{Elizalde,E., Lygren, M. and Vassilevich,
 D.V. \jmp {37}{1996}{3105}.}
 \ref{ELV2}{Elizalde,E., Lygren, M. and Vassilevich,
 D.V. \cmp {183}{1997}{645}.}
 \ref{Malurkar}{Malurkar,S.L. {\it J.Ind.Math.Soc} {\bf16} (1925/26) 130.}
 \ref{Glaisher}{Glaisher,J.W.L. {\it Messenger of Math.} {\bf18}
(1889) 1.} \ref{Anderson}{Anderson,A. \prD{37}{1988}{536}.}
 \ref{CandA}{Cappelli,A. and D'Appollonio, \pl{487B}{2000}{87}.}
 \ref{Wot}{Wotzasek,C. \jpa{23}{1990}{1627}.}
 \ref{RandT}{Ravndal,F. and Tollesen,D. \prD{40}{1989}{4191}.}
 \ref{SandT}{Santos,F.C. and Tort,A.C. \pl{482B}{2000}{323}.}
 \ref{FandO}{Fukushima,K. and Ohta,K. {\it Physica} {\bf A299} (2001) 455.}
 \ref{GandP}{Gibbons,G.W. and Perry,M. \prs{358}{1978}{467}.}
 \ref{Dow4}{Dowker,J.S..}
  \ref{Rad}{Rademacher,H. {\it Topics in analytic number theory,}
Springer-Verlag,  Berlin,1973.}
  \ref{Halphen}{Halphen,G.-H. {\it Trait\'e des Fonctions Elliptiques},
  Vol 1, Gauthier-Villars, Paris, 1886.}
  \ref{CandW}{Cahn,R.S. and Wolf,J.A. {\it Comm.Mat.Helv.} {\bf 51}
  (1976) 1.}
  \ref{Berndt}{Berndt,B.C. \rmjm{7}{1977}{147}.}
  \ref{Hurwitz}{Hurwitz,A. \ma{18}{1881}{528}.}
  \ref{Hurwitz2}{Hurwitz,A. {\it Mathematische Werke} Vol.I. Basel,
  Birkhauser, 1932.}
  \ref{Berndt2}{Berndt,B.C. \jram{303/304}{1978}{332}.}
  \ref{RandA}{Rao,M.B. and Ayyar,M.V. \jims{15}{1923/24}{150}.}
  \ref{Hardy}{Hardy,G.H. \jlms{3}{1928}{238}.}
  \ref{TandM}{Tannery,J. and Molk,J. {\it Fonctions Elliptiques},
   Gauthier-Villars, Paris, 1893--1902.}
  \ref{schwarz}{Schwarz,H.-A. {\it Formeln und
  Lehrs\"atzen zum Gebrauche..},Springer 1893.(The first edition was 1885.)
  The French translation by Henri Pad\'e is {\it Formules et Propositions
  pour L'Emploi...},Gauthier-Villars, Paris, 1894}
  \ref{Hancock}{Hancock,H. {\it Theory of elliptic functions}, Vol I.
   Wiley, New York 1910.}
  \ref{watson}{Watson,G.N. \jlms{3}{1928}{216}.}
  \ref{MandO}{Magnus,W. and Oberhettinger,F. {\it Formeln und S\"atze},
  Springer-Verlag, Berlin 1948.}
  \ref{Klein}{Klein,F. {\it Lectures on the Icosohedron}
  (Methuen, London, 1913).}
  \ref{AandL}{Appell,P. and Lacour,E. {\it Fonctions Elliptiques},
  Gauthier-Villars,
  Paris, 1897.}
  \ref{HandC}{Hurwitz,A. and Courant,C. {\it Allgemeine Funktionentheorie},
  Springer,
  Berlin, 1922.}
  \ref{WandW}{Whittaker,E.T. and Watson,G.N. {\it Modern analysis},
  Cambridge 1927.}
  \ref{SandC}{Selberg,A. and Chowla,S. \jram{227}{1967}{86}. }
  \ref{zucker}{Zucker,I.J. {\it Math.Proc.Camb.Phil.Soc} {\bf 82 }(1977)
  111.}
  \ref{glasser}{Glasser,M.L. {\it Maths.of Comp.} {\bf 25} (1971) 533.}
  \ref{GandW}{Glasser, M.L. and Wood,V.E. {\it Maths of Comp.} {\bf 25}
  (1971)
  535.}
  \ref{greenhill}{Greenhill,A,G. {\it The Applications of Elliptic
  Functions}, MacMillan, London, 1892.}
  \ref{Weierstrass}{Weierstrass,K. {\it J.f.Mathematik (Crelle)}
{\bf 52} (1856) 346.}
  \ref{Weierstrass2}{Weierstrass,K. {\it Mathematische Werke} Vol.I,p.1,
  Mayer u. M\"uller, Berlin, 1894.}
  \ref{Fricke}{Fricke,R. {\it Die Elliptische Funktionen und Ihre Anwendungen},
    Teubner, Leipzig. 1915, 1922.}
  \ref{Konig}{K\"onigsberger,L. {\it Vorlesungen \"uber die Theorie der
 Elliptischen Funktionen},  \break Teubner, Leipzig, 1874.}
  \ref{Milne}{Milne,S.C. {\it The Ramanujan Journal} {\bf 6} (2002) 7-149.}
  \ref{Schlomilch}{Schl\"omilch,O. {\it Ber. Verh. K. Sachs. Gesell. Wiss.
  Leipzig}  {\bf 29} (1877) 101-105; {\it Compendium der h\"oheren
  Analysis}, Bd.II, 3rd Edn, Vieweg, Brunswick, 1878.}
  \ref{BandB}{Briot,C. and Bouquet,C. {\it Th\`eorie des Fonctions
  Elliptiques}, Gauthier-Villars, Paris, 1875.}
  \ref{Dumont}{Dumont,D. \aim {41}{1981}{1}.}
  \ref{Andre}{Andr\'e,D. {\it Ann.\'Ecole Normale Superior} {\bf 6} (1877)
  265;
  {\it J.Math.Pures et Appl.} {\bf 5} (1878) 31.}
  \ref{Raman}{Ramanujan,S. {\it Trans.Camb.Phil.Soc.} {\bf 22} (1916) 159;
 {\it Collected Papers}, Cambridge, 1927}
  \ref{Weber}{Weber,H.M. {\it Lehrbuch der Algebra} Bd.III, Vieweg,
  Brunswick 190  3.}
  \ref{Weber2}{Weber,H.M. {\it Elliptische Funktionen und algebraische
  Zahlen},
  Vieweg, Brunswick 1891.}
  \ref{ZandR}{Zucker,I.J. and Robertson,M.M.
  {\it Math.Proc.Camb.Phil.Soc} {\bf 95 }(1984) 5.}
  \ref{JandZ1}{Joyce,G.S. and Zucker,I.J.
  {\it Math.Proc.Camb.Phil.Soc} {\bf 109 }(1991) 257.}
  \ref{JandZ2}{Zucker,I.J. and Joyce.G.S.
  {\it Math.Proc.Camb.Phil.Soc} {\bf 131 }(2001) 309.}
  \ref{zucker2}{Zucker,I.J. {\it SIAM J.Math.Anal.} {\bf 10} (1979) 192,}
  \ref{BandZ}{Borwein,J.M. and Zucker,I.J. {\it IMA J.Math.Anal.} {\bf 12}
  (1992) 519.}
  \ref{Cox}{Cox,D.A. {\it Primes of the form $x^2+n\,y^2$}, Wiley,
  New York, 1989.}
  \ref{BandCh}{Berndt,B.C. and Chan,H.H. {\it Mathematika} {\bf42} (1995)
  278.}
  \ref{EandT}{Elizalde,R. and Tort.hep-th/}
  \ref{KandS}{Kiyek,K. and Schmidt,H. {\it Arch.Math.} {\bf 18} (1967) 438.}
  \ref{Oshima}{Oshima,K. \prD{46}{1992}{4765}.}
  \ref{greenhill2}{Greenhill,A.G. \plms{19} {1888} {301}.}
  \ref{Russell}{Russell,R. \plms{19} {1888} {91}.}
  \ref{BandB}{Borwein,J.M. and Borwein,P.B. {\it Pi and the AGM}, Wiley,
  New York, 1998.}
  \ref{Resnikoff}{Resnikoff,H.L. \tams{124}{1966}{334}.}
  \ref{vandp}{Van der Pol, B. {\it Indag.Math.} {\bf18} (1951) 261,272.}
  \ref{Rankin}{Rankin,R.A. {\it Modular forms} CUP}
  \ref{Rankin2}{Rankin,R.A. {\it Proc. Roy.Soc. Edin.} {\bf76 A} (1976) 107.}
  \ref{Skoruppa}{Skoruppa,N-P. {\it J.of Number Th.} {\bf43} (1993) 68 .}
  \ref{Down}{Dowker.J.S. \np {104}{2002}{153}.}
  \ref{Eichler}{Eichler,M. \mz {67}{1957}{267}.}
  \ref{Zagier}{Zagier,D. \invm{104}{1991}{449}.}
  \ref{Lang}{Lang,S. {\it Modular Forms}, Springer, Berlin, 1976.}
  \ref{Kosh}{Koshliakov,N.S. {\it Mess.of Math.} {\bf 58} (1928) 1.}
  \ref{BandH}{Bodendiek, R. and Halbritter,U. \amsh{38}{1972}{147}.}
  \ref{Smart}{Smart,L.R., \pgma{14}{1973}{1}.}
  \ref{Grosswald}{Grosswald,E. {\it Acta. Arith.} {\bf 21} (1972) 25.}
  \ref{Kata}{Katayama,K. {\it Acta Arith.} {\bf 22} (1973) 149.}
  \ref{Ogg}{Ogg,A. {\it Modular forms and Dirichlet series} (Benjamin,
  New York,
   1969).}
  \ref{Bol}{Bol,G. \amsh{16}{1949}{1}.}
  \ref{Epstein}{Epstein,P. \ma{56}{1903}{615}.}
  \ref{Petersson}{Petersson.}
  \ref{Serre}{Serre,J-P. {\it A Course in Arithmetic}, Springer,
  New York, 1973.}
  \ref{Schoenberg}{Schoenberg,B., {\it Elliptic Modular Functions},
  Springer, Berlin, 1974.}
  \ref{Apostol}{Apostol,T.M. \dmj {17}{1950}{147}.}
  \ref{Ogg2}{Ogg,A. {\it Lecture Notes in Math.} {\bf 320} (1973) 1.}
  \ref{Knopp}{Knopp,M.I. \dmj {45}{1978}{47}.}
  \ref{Knopp2}{Knopp,M.I. \invm {}{1994}{361}.}
  \ref{LandZ}{Lewis,J. and Zagier,D. \aom{153}{2001}{191}.}
  \ref{DandK1}{Dowker,J.S. and Kirsten,K. {\it Elliptic functions and
  temperature inversion symmetry on spheres} hep-th/.}
  \ref{HandK}{Husseini and Knopp.}
  \ref{Kober}{Kober,H. \mz{39}{1934-5}{609}.}
  \ref{HandL}{Hardy,G.H. and Littlewood, \am{41}{1917}{119}.}
  \ref{Watson}{Watson,G.N. \qjm{2}{1931}{300}.}
  \ref{SandC2}{Chowla,S. and Selberg,A. {\it Proc.Nat.Acad.} {\bf 35}
  (1949) 371.}
  \ref{Landau}{Landau, E. {\it Lehre von der Verteilung der Primzahlen},
  (Teubner, Leipzig, 1909).}
  \ref{Berndt4}{Berndt,B.C. \tams {146}{1969}{323}.}
  \ref{Berndt3}{Berndt,B.C. \tams {}{}{}.}
  \ref{Bochner}{Bochner,S. \aom{53}{1951}{332}.}
  \ref{Weil2}{Weil,A.\ma{168}{1967}{}.}
  \ref{CandN}{Chandrasekharan,K. and Narasimhan,R. \aom{74}{1961}{1}.}
  \ref{Rankin3}{Rankin,R.A. {} {} ().}
  \ref{Berndt6}{Berndt,B.C. {\it Trans.Edin.Math.Soc}.}
  \ref{Elizalde}{Elizalde,E. {\it Ten Physical Applications of Spectral
  Zeta Function Theory}, \break (Springer, Berlin, 1995).}
  \ref{Allen}{Allen,B., Folacci,A. and Gibbons,G.W. \pl{189}{1987}{304}.}
  \ref{Krazer}{Krazer}
  \ref{Elizalde3}{Elizalde,E. {\it J.Comp.and Appl. Math.} {\bf 118}
  (2000) 125.}
  \ref{Elizalde2}{Elizalde,E., Odintsov.S.D, Romeo, A. and Bytsenko,
  A.A and
  Zerbini,S.
  {\it Zeta function regularisation}, (World Scientific, Singapore,
  1994).}
  \ref{Eisenstein}{Eisenstein}
  \ref{Hecke}{Hecke,E. \ma{112}{1936}{664}.}
  \ref{Terras}{Terras,A. {\it Harmonic analysis on Symmetric Spaces} (Springer,
  New York, 1985).}
  \ref{BandG}{Bateman,P.T. and Grosswald,E. {\it Acta Arith.} {\bf 9}
  (1964) 365.}
  \ref{Deuring}{Deuring,M. \aom{38}{1937}{585}.}
  \ref{Guinand}{Guinand.}
  \ref{Guinand2}{Guinand.}
  \ref{Minak}{Minakshisundaram.}
  \ref{Mordell}{Mordell,J. \prs{}{}{}.}
  \ref{GandZ}{Glasser,M.L. and Zucker, {}.}
  \ref{Landau2}{Landau,E. \jram{}{1903}{64}.}
  \ref{Kirsten1}{Kirsten,K. \jmp{35}{1994}{459}.}
  \ref{Sommer}{Sommer,J. {\it Vorlesungen \"uber Zahlentheorie}
  (1907,Teubner,Leipzig).
  French edition 1913 .}
  \ref{Reid}{Reid,L.W. {\it Theory of Algebraic Numbers},
  (1910,MacMillan,New York).}
  \ref{Milnor}{Milnor, J. {\it Is the Universe simply--connected?},
  IAS, Princeton, 1978.}
  \ref{Milnor2}{Milnor, J. \ajm{79}{1957}{623}.}
  \ref{Opechowski}{Opechowski,W. {\it Physica} {\bf 7} (1940) 552.}
  \ref{Bethe}{Bethe, H.A. \zfp{3}{1929}{133}.}
  \ref{LandL}{Landau, L.D. and Lishitz, E.M. {\it Quantum
  Mechanics} (Pergamon Press, London, 1958).}
  \ref{GPR}{Gibbons, G.W., Pope, C. and R\"omer, H., \np{157}{1979}{377}.}
  \ref{Jadhav}{Jadhav,S.P. PhD Thesis, University of Manchester 1990.}
  \ref{DandJ}{Dowker,J.S. and Jadhav, S. \prD{39}{1989}{1196}.}
  \ref{CandM}{Coxeter, H.S.M. and Moser, W.O.J. {\it Generators and
  relations of finite groups} Springer. Berlin. 1957.}
  \ref{Coxeter2}{Coxeter, H.S.M. {\it Regular Complex Polytopes},
   (Cambridge University Press,
  Cambridge, 1975).}
  \ref{Coxeter}{Coxeter, H.S.M. {\it Regular Polytopes}.}
  \ref{Stiefel}{Stiefel, E., J.Research NBS {\bf 48} (1952) 424.}
  \ref{BandS}{Brink and Satchler {\it Angular momentum theory}.}
  %\ref{Racah1}
  \ref{Rose}{Rose}
  \ref{Schwinger}{Schwinger,J.}
  \ref{Bromwich}{Bromwich, T.J.I'A. {\it Infinite Series},
  (Macmillan, 1947).}
  \ref{Ray}{Ray,D.B. \aim{4}{1970}{109}.}
  \ref{Ikeda}{Ikeda,A. {\it Kodai Math.J.} {\bf 18} (1995) 57.}
  \ref{Kennedy}{Kennedy,G. \prD{23}{1981}{2884}.}
  \ref{Ellis}{Ellis,G.F.R. {\it General Relativity} {\bf2} (1971) 7.}
  \ref{Dow8}{Dowker,J.S. \cqg{20}{2003}{L105}.}
  \ref{IandY}{Ikeda, A and Yamamoto, Y. \ojm {16}{1979}{447}.}
  \ref{BandI}{Bander,M. and Itzykson,C. \rmp{18}{1966}{2}.}
  \ref{Schulman}{Schulman, L.S. \pr{176}{1968}{1558}.}
  \ref{Bar1}{B\"ar,C. {\it Arch.d.Math.}{\bf 59} (1992) 65.}
  \ref{Bar2}{B\"ar,C. {\it Geom. and Func. Anal.} {\bf 6} (1996) 899.}
  \ref{Vilenkin}{Vilenkin, N.J. {\it Special functions},
  (Am.Math.Soc., Providence, 1968).}
  \ref{Talman}{Talman, J.D. {\it Special functions} (Benjamin,N.Y.,1968).}
  \ref{Miller}{Miller,W. {\it Symmetry groups and their applications}
  (Wiley, N.Y., 1972).}
  \ref{Dow3}{Dowker,J.S. \cmp{162}{1994}{633}.}
  \ref{Cheeger}{Cheeger, J. \jdg {18}{1983}{575}.}
  \ref{Dow6}{Dowker,J.S. \jmp{30}{1989}{770}.}
  \ref{Dow20}{Dowker,J.S. \jmp{35}{1994}{6076}.}
  \ref{Dow21}{Dowker,J.S. {\it Heat kernels and polytopes} in {\it
   Heat Kernel Techniques and Quantum Gravity}, ed. by S.A.Fulling,
   Discourses in Mathematics and its Applications, No.4, Dept.
   Maths., Texas A\&M University, College Station, Texas, 1995.}
  \ref{Dow9}{Dowker,J.S. \jmp{42}{2001}{1501}.}
  \ref{Dow7}{Dowker,J.S. \jpa{25}{1992}{2641}.}
  \ref{Warner}{Warner.N.P. \prs{383}{1982}{379}.}
  \ref{Wolf}{Wolf, J.A. {\it Spaces of constant curvature},
  (McGraw--Hill,N.Y., 1967).}
  \ref{Meyer}{Meyer,B. \cjm{6}{1954}{135}.}
  \ref{BandB}{B\'erard,P. and Besson,G. {\it Ann. Inst. Four.} {\bf 30}
  (1980) 237.}
  \ref{PandM}{Polya,G. and Meyer,B. \cras{228}{1948}{28}.}
  \ref{Springer}{Springer, T.A. Lecture Notes in Math. vol 585 (Springer,
  Berlin,1977).}
  \ref{SeandT}{Threlfall, H. and Seifert, W. \ma{104}{1930}{1}.}
  \ref{Hopf}{Hopf,H. \ma{95}{1925}{313}. }
  \ref{Dow}{Dowker,J.S. \jpa{5}{1972}{936}.}
  \ref{LLL}{Lehoucq,R., Lachi\'eze-Rey,M. and Luminet, J.--P. {\it
  Astron.Astrophys.} {\bf 313} (1996) 339.}
  \ref{LaandL}{Lachi\'eze-Rey,M. and Luminet, J.--P.
  \prp{254}{1995}{135}.}
  \ref{Schwarzschild}{Schwarzschild, K., {\it Vierteljahrschrift der
  Ast.Ges.} {\bf 35} (1900) 337.}
  \ref{Starkman}{Starkman,G.D. \cqg{15}{1998}{2529}.}
  \ref{LWUGL}{Lehoucq,R., Weeks,J.R., Uzan,J.P., Gausman, E. and
  Luminet, J.--P. \cqg{19}{2002}{4683}.}
  \ref{Dow10}{Dowker,J.S. \prD{28}{1983}{3013}.}
  \ref{BandD}{Banach, R. and Dowker, J.S. \jpa{12}{1979}{2527}.}
  \ref{Jadhav2}{Jadhav,S. \prD{43}{1991}{2656}.}
  \ref{Gilkey}{Gilkey,P.B. {\it Invariance theory,the heat equation and
  the Atiyah--Singer Index theorem} (CRC Press, Boca Raton, 1994).}
  \ref{BandY}{Berndt,B.C. and Yeap,B.P. {\it Adv. Appl. Math.}
  {\bf29} (2002) 358.}
  \ref{HandR}{Hanson,A.J. and R\"omer,H. \pl{80B}{1978}{58}.}
  \ref{Hill}{Hill,M.J.M. {\it Trans.Camb.Phil.Soc.} {\bf 13} (1883) 36.}
  \ref{Cayley}{Cayley,A. {\it Quart.Math.J.} {\bf 7} (1866) 304.}
  \ref{Seade}{Seade,J.A. {\it Anal.Inst.Mat.Univ.Nac.Aut\'on
  M\'exico} {\bf 21} (1981) 129.}
  \ref{CM}{Cisneros--Molina,J.L. {\it Geom.Dedicata} {\bf84} (2001)
  \ref{Goette1}{Goette,S. \jram {526} {2000} 181.}
  207.}
  \ref{NandO}{Nash,C. and O'Connor,D--J, \jmp {36}{1995}{1462}.}
  \ref{Dows}{Dowker,J.S. \aop{71}{1972}{577}; Dowker,J.S. and Pettengill,D.F.
  \jpa{7}{1974}{1527}; J.S.Dowker in {\it Quantum Gravity}, edited by
  S. C. Christensen (Hilger,Bristol,1984)}
  \ref{Jadhav2}{Jadhav,S.P. \prD{43}{1991}{2656}.}
  \ref{Dow11}{Dowker,J.S. \cqg{21}{2004}4247.}
  \ref{Dow12}{Dowker,J.S. \cqg{21}{2004}4977.}
  \ref{Dow13}{Dowker,J.S. \jpa{38}{2005}1049.}
  \ref{Zagier}{Zagier,D. \ma{202}{1973}{149}}
  \ref{RandG}{Rademacher, H. and Grosswald,E. {\it Dedekind Sums},
  (Carus, MAA, 1972).}
  \ref{Berndt7}{Berndt,B, \aim{23}{1977}{285}.}
  \ref{HKMM}{Harvey,J.A., Kutasov,D., Martinec,E.J. and Moore,G.
  {\it Localised Tachyons and RG Flows}, hep-th/0111154.}
  \ref{Beck}{Beck,M., {\it Dedekind Cotangent Sums}, {\it Acta Arithmetica}
  {\bf 109} (2003) 109-139 ; math.NT/0112077.}
  \ref{McInnes}{McInnes,B. {\it APS instability and the topology of the brane
  world}, hep-th/0401035.}
  \ref{BHS}{Brevik,I, Herikstad,R. and Skriudalen,S. {\it Entropy Bound for the
  TM Electromagnetic Field in the Half Einstein Universe}; hep-th/0508123.}
  \ref{BandO}{Brevik,I. and Owe,C.  \prD{55}{4689}{1997}.}
  \ref{Kenn}{Kennedy,G. Thesis. University of Manchester 1978.}
  \ref{KandU}{Kennedy,G. and Unwin S. \jpa{12}{L253}{1980}.}
  \ref{BandO1}{Bayin,S.S.and Ozcan,M.
  \prD{48}{2806}{1993}; \prD{49}{5313}{1994}.}
  \ref{Chang}{Chang, P. Thesis. University of Manchester 1993.}
  \ref{Barnesa}{Barnes,E.W. {\it Trans. Camb. Phil. Soc.} {\bf 19} (1903) 374.}
  \ref{Barnesb}{Barnes,E.W. {\it Trans. Camb. Phil. Soc.}
  {\bf 19} (1903) 426.}
  \ref{Stanley1}{Stanley,R.P. \joa {49}{1977}{134}.}
  \ref{Stanley}{Stanley,R.P. \bams {1}{1979}{475}.}
  \ref{Hurley}{Hurley,A.C. \pcps {47}{1951}{51}.}
  \ref{IandK}{Iwasaki,I. and Katase,K. {\it Proc.Japan Acad. Ser} {\bf A55}
  (1979) 141.}
  \ref{IandT}{Ikeda,A. and Taniguchi,Y. {\it Osaka J. Math.} {\bf 15} (1978)
  515.}
  \ref{GandM}{Gallot,S. and Meyer,D. \jmpa{54}{1975}{259}.}
  \ref{Flatto}{Flatto,L. {\it Enseign. Math.} {\bf 24} (1978) 237.}
  \ref{OandT}{Orlik,P and Terao,H. {\it Arrangements of Hyperplanes},
  Grundlehren der Math. Wiss. {\bf 300}, (Springer--Verlag, 1992).}
  \ref{Shepler}{Shepler,A.V. \joa{220}{1999}{314}.}
  \ref{SandT}{Solomon,L. and Terao,H. \cmh {73}{1998}{237}.}
  \ref{Vass}{Vassilevich, D.V. \plb {348}{1995}39.}
  \ref{Vass2}{Vassilevich, D.V. \jmp {36}{1995}3174.}
  \ref{CandH}{Camporesi,R. and Higuchi,A. {\it J.Geom. and Physics}
  {\bf 15} (1994) 57.}
  \ref{Solomon2}{Solomon,L. \tams{113}{1964}{274}.}
  \ref{Solomon}{Solomon,L. {\it Nagoya Math. J.} {\bf 22} (1963) 57.}
  \ref{Obukhov}{Obukhov,Yu.N. \pl{109B}{1982}{195}.}
  \ref{BGH}{Bernasconi,F., Graf,G.M. and Hasler,D. {\it The heat kernel
  expansion for the electromagnetic field in a cavity}; math-ph/0302035.}
  \ref{Baltes}{Baltes,H.P. \prA {6}{1972}{2252}.}
  \ref{BaandH}{Baltes.H.P and Hilf,E.R. {\it Spectra of Finite Systems}
  (Bibliographisches Institut, Mannheim, 1976).}
  \ref{Ray}{Ray,D.B. \aim{4}{1970}{109}.}
  \ref{Hirzebruch}{Hirzebruch,F. {\it Topological methods in algebraic
  geometry} (Springer-- Verlag,\break  Berlin, 1978). }
  \ref{BBG}{Bla\v{z}i\'c,N., Bokan,N. and Gilkey, P.B. {\it Ind.J.Pure and
  Appl.Math.} {\bf 23} (1992) 103.}
  \ref{WandWi}{Weck,N. and Witsch,K.J. {\it Math.Meth.Appl.Sci.} {\bf 17}
  (1994) 1017.}
  \ref{Norlund}{N\"orlund,N.E. \am{43}{1922}{121}.}
  \ref{Duff}{Duff,G.F.D. \aom{56}{1952}{115}.}
  \ref{DandS}{Duff,G.F.D. and Spencer,D.C. \aom{45}{1951}{128}.}
  \ref{BGM}{Berger,M., Gauduchon,P. and Mazet,E. {\it Lect.Notes.Math.}
  {\bf 194} (1971) 1. }
  \ref{Patodi}{Patodi,V.K. \jdg{5}{1971}{233}.}
  \ref{GandS}{G\"unther,P. and Schimming,R. \jdg{12}{1977}{599}.}
  \ref{MandS}{McKean,H.P. and Singer,I.M. \jdg{1}{1967}{43}.}
  \ref{Conner}{Conner,P.E. {\it Mem.Am.Math.Soc.} {\bf 20} (1956).}
  \ref{Gilkey2}{Gilkey,P.B. \aim {15}{1975}{334}.}
  \ref{MandP}{Moss,I.G. and Poletti,S.J. \plb{333}{1994}{326}.}
  \ref{BKD}{Bordag,M., Kirsten,K. and Dowker,J.S. \cmp{182}{1996}{371}.}
  \ref{RandO}{Rubin,M.A. and Ordonez,C. \jmp{25}{1984}{2888}.}
  \ref{BaandD}{Balian,R. and Duplantier,B. \aop {112}{1978}{165}.}
  \ref{Kennedy2}{Kennedy,G. \aop{138}{1982}{353}.}
  \ref{DandKi2}{Dowker,J.S. and Kirsten, K. {\it Analysis and Appl.}
 {\bf 3} (2005) 45.}
 \ref{Watsons}{Watson,S. Thesis Univ. Bristol 1998.}
   \ref{DandA2}{Dowker,J.S. and Apps,J.S.  {\it Int.J.Mod.Phys.D} {\bf 5}
   (1996) 799,}
   \ref{RandTo}{Russell,I.H. and Toms,D.J. \cqg{4}{1987}{1357}.}
   \ref{DandCh}{Dowker,J.S. and Chang,Peter, \prD{46}{1992}{3458}.}
   \ref{Dowl}{Dowker,J.S. {\it Functional determinants on M\"obius corners};
    Proceedings, `Quantum field theory under
    the influence of external conditions', 111-121,Leipzig 1995.}
    \ref{DandCa}{Deutsch,D. and Candelas,P. \prD{20}{1979}{3063}.}
    \ref{Paquet}{Paquet,L. {\it Bull.Sci.Math.} {\bf 105} (1981) 85.}
\end{putreferences}

\bye